\def \<{\langle}
\def \>{\rangle}

\documentclass[a4paper,11pt]{article}  

\usepackage{jcappub,graphicx,epsfig,natbib,color,times,bm,amsmath,multirow,hyperref}
\usepackage{makecell}
\usepackage{comment}
\usepackage{tabularx}
\usepackage{subcaption}

\usepackage[flushleft]{threeparttable}

\title{Mitigating point-source contamination in CMB polarization: a Generalized Point Spread Function fitting approach}

\author[a,b]{Yi-Ming Wang,}\emailAdd{wangyiming@ihep.ac.cn}
\author[a,b]{Wen-Zheng Chen,}\emailAdd{chenwz@ihep.ac.cn}
\author[a]{Yang Liu,}\emailAdd{liuy92@ihep.ac.cn}
\author[*,a]{Si-Yu Li,}\emailAdd{lisy@ihep.ac.cn}
\author[*,a, b]{Hong Li}\emailAdd{hongli@ihep.ac.cn}

\affiliation[a]{Key Laboratory of Particle Astrophysics, Institute of High Energy Physics, Chinese Academy of Sciences, 19B Yuquan Road, Beijing, P.R. China}
\affiliation[b]{University of Chinese Academy of Sciences, 19A Yuquan Road, Beijing, P.R. China}

\begingroup
 
\footnotetext{\mbox{$^*$ Corresponding author.}} 
\endgroup

\begin{document}

\abstract{
Observations of Cosmic Microwave Background (CMB) B-mode polarization provide a way to probe primordial gravitational waves and test inflationary predictions. Extragalactic point sources become a major source of contamination after foreground cleaning and can bias estimates of the tensor-to-scalar ratio $r$ at the $10^{-3}$ level. We introduce Generalized Point Spread Function Fitting (GPSF), a method for removing point-source contamination in polarization maps. GPSF uses the full pixel-domain covariance, including off-diagonal terms, and models overlapping sources. This allows accurate flux estimation under realistic conditions, particularly for small-aperture telescopes with large beams that are more susceptible to source blending. We test GPSF on simulated sky maps, apply foreground cleaning using the Needlet Internal Linear Combination (NILC) method, and compare its performance with standard masking and inpainting. The results show GPSF reduces point-source contamination without significantly affecting the background signal, as seen in both the maps and their power spectra. For the constraint on $r$, GPSF reduces the bias from $1.67 \times 10^{-3}$ to $2.9 \times 10^{-4}$, with only a 2\% increase in standard deviation. Compared to inpainting and masking, GPSF yields lower bias while maintaining comparable variance. This suggests that it may serve as a promising method for future CMB experiments targeting measurements of $r \sim 10^{-3}$.

}

\maketitle

\section{Introduction}\label{sec:intro}
Inflation explains the flatness and homogeneity of the early Universe, as well as the origin of primordial perturbations, through a short period of accelerated expansion~\cite{Lyth:1998xn}. In addition to scalar (density) perturbations, inflation also predicts tensor perturbations—primordial gravitational waves—that produce a characteristic B-mode pattern in the polarization of the Cosmic Microwave Background (CMB). The relative amplitude of tensor to scalar modes is described by the tensor-to-scalar ratio $r$, which is directly related to the energy scale of inflation~\cite{PhysRevLett.78.2054,PhysRevLett.78.2058}. Current measurements constrain $r_{0.05} < 0.036$ at 95\% confidence level~\cite{BICEP_2021}, with no detection yet achieved. Upcoming CMB experiments aim to measure $r$ at the $10^{-3}$ level~\cite{SimonsObservatory:2018koc,LiteBIRD:2020khw}, which would allow much tighter constraints on the inflationary energy scale and the amplitude of primordial gravitational waves.

As experimental sensitivity advances, precise foreground mitigation becomes essential for detecting primordial B modes. A major known source is polarized diffuse Galactic emission, including synchrotron and thermal dust, which dominates the large-scale $B$-mode signal and has been extensively studied and mitigated using multi-frequency component separation techniques~\cite{2016MNRAS.458.2032R,2018MNRAS.479.5577P}.
In addition, extragalactic point sources are emerging as a significant contaminant, especially after foreground cleaning~\cite{Tucci:2004zy, Puglisi:2017lpn}. 
These sources present challenges: they not only generate spurious $B$-mode power that can mimic the primordial tensor signal (with apparent $r \sim 10^{-3}$) at low multipoles~\cite{BICEP_2021, 2013MNRAS.432..728C,BICEP2:2014owc,2020A&A...642A.232L,Trombetti:2017kim}, but also significantly complicate foreground mitigation efforts.
In methods like the Needlet Internal Linear Combination (NILC), the weights are partially assigned to point sources, which reduces the effectiveness of cleaning diffuse Galactic foregrounds, especially around bright sources~\cite{CORE2018}. 
The development of robust point-source mitigation techniques is therefore a crucial challenge for next-generation CMB polarization experiments aiming to achieve sensitivities at the $r \sim 10^{-3}$ level.

Traditional strategies for mitigating compact point-source contamination in CMB maps generally fall into three categories~\cite{Bajkova:2008ygr,Sureau:2014lea}: masking~\cite{WMAP:2012fli,2011ApJS..192...16L}, inpainting~\cite{Abrial:2008mz, 2012MNRAS.424.1694B, planck18_IV, 2011MNRAS.410.2481R}, and point-source removal methods~\cite{2020JCAP...12..045C}. 
However, the efficacy of these methods in current and next-generation experiments designed to measure the tensor-to-scalar ratio $r$ with high precision warrants thorough investigation. One major challenge comes from the relatively large beams of small-aperture telescopes (SATs), which cause the contamination from point sources to spread over a larger area of the sky.
The masking approach, which simply discards contaminated pixels, suffers from two significant drawbacks. First, the large full-width-at-half-maximum (FWHM) results in substantial loss of usable sky area, thereby reducing the available cosmological information~\cite{Scodeller:2012sw}. Second, the mask boundaries create artificial discontinuities that systematically bias subsequent analyses. These effects are particularly pronounced in needlet-transformation, where edge artifacts propagate several degrees beyond the mask boundaries, predominantly affecting large angular scales ($\ell < 300$)~\cite{Scodeller:2010mp}.
Inpainting techniques attempt to fill in masked regions using prior information (such as the power spectrum) or by interpolating from nearby pixels. However, such reconstructions are inherently limited in their accuracy for estimating the true CMB signal in inpainted regions~\cite{Sureau:2014lea}.
The removal method involves modeling each point source individually and subtracting the fitted model from the observed maps. It makes use of both the data in the point-source regions and prior knowledge of the point spread function, which helps recover the underlying signal more accurately in affected areas. Conventional photometric techniques such as matched filter fitting and point spread function fitting have been widely used in previous CMB experiments~\cite{ACT:2023wxl,Planck:2013qym, Planck:2015bin}. However, achieving the level of precision needed to measure $r$ at the $10^{-3}$ level after foreground cleaning requires improved methods that can better utilize the available data. 

In this context, we present the \textit{Generalized Point Spread Function} (GPSF) fitting method—a novel approach that incorporates the full pixel-to-pixel covariance matrix, including off-diagonal contributions from the CMB and diffuse foregrounds, while explicitly accounting for overlapping sources. GPSF enables accurate flux extraction at the map level and is specifically designed to mitigate point-source contamination in polarization maps relevant for tensor-to-scalar ratio ($r$) measurements. 
Its performance is assessed in both the map and power spectrum domains across individual frequency bands, as well as after NILC for foreground cleaning and constraining the tensor-to-scalar ratio $r$. The results are compared with those obtained using conventional masking and inpainting techniques.

The remainder of this paper is organized as follows. 
Section~\ref{sec:method} introduces the GPSF methodology, beginning with the single-source fitting framework and extending to the multi-source case. 
Section~\ref{sec:simu} presents the end-to-end simulation, detailing the sky model, instrumental specifications, and the complete GPSF analysis pipeline, along with performance evaluation setup. 
Section~\ref{sec:result} presents a comparative evaluation of GPSF and other approaches, assessing their performance at both the map and power spectrum domains, before and after NILC, and analysing the resulting constraints on the tensor-to-scalar ratio~$r$.
Finally, Section~\ref{sec:conclusions} summarizes our main findings and discusses future prospects.

\section{Methodology}\label{sec:method}
In this section, we describe the GPSF fitting method for mitigating the contribution of point sources in polarization maps. We first introduce the methodology for the single-source case, and then extend it to the more general case involving multiple point sources.

\subsection{Single-source fitting}\label{sec:single_fit}
The core of GPSF is the direct fitting of a physically motivated point-source model to the observed sky maps using pixel-domain information and a statistically consistent background model. The observed map in a Stokes component $X \in \{I, Q, U\}$ is modeled as:
\begin{align}\label{eq:obs}
m_X^{\text{obs}}(\mathbf{n}) = m_X^{\text{src}}(\mathbf{n}) + m_X^{\text{bkg}}(\mathbf{n}),
\end{align}
where $m_X^{\text{src}}$ is the contribution from point sources, and $m_X^{\text{bkg}}$ represents the background, typically including the CMB, diffuse foregrounds, instrumental noise, etc.

To construct the source component $m_X^{\text{src}}$, we model each point source individually using a parameterized beam-convolved template $\mathcal{T}_X$:
\begin{align}\label{eq:single_model}
m_X^{\text{src}}(\mathbf{n}) \equiv \mathcal{T}_X(\mathbf{n} \mid \hat{\mathbf{n}}_s, \beta_s),
\end{align}
where $\hat{\mathbf{n}}_s$ is the known position of the source, derived either from point-source detection applied to the observed data or from external catalogs, and $\beta_s = \{ I_s, Q_s, U_s \}$ denotes the set of source parameters to be fitted.  

In this work, we adopt a symmetric Gaussian model for the PSF, a common approximation in CMB analyses due to its analytical simplicity and its ability to accurately capture the main lobe of the instrumental beam~\cite{White:1994ru}. A detailed discussion of how GPSF addresses instrumental systematics and atmospheric filtering is provided in appendix~\ref{app:systematics}. For intensity maps ($X = I$), the symmetric Gaussian template is:
\begin{align}
\mathcal{T}_I(\mathbf{n} \mid \hat{\mathbf{n}}_s, \beta_s)
= \frac{I_s}{2\pi\sigma^2} \exp\left( -\frac{\theta^2}{2\sigma^2} \right),
\end{align}
where $\theta = \cos^{-1}(\hat{\mathbf{n}}_s \cdot \mathbf{n})$ is the angular separation between the source and the pixel, and $I_s$ is the intensity amplitude component of the parameter vector $\beta_s$.

For polarization maps ($X \in \{Q, U\}$), the Stokes parameters are coordinate-dependent quantities, and their templates require additional treatment. We adopt the standard \textit{co-polar approximation}~\cite{Planck:2018yye,Hivon:2016qyw}, under which the instrument’s cross-polar response is negligible and the polarized beam is described by a scalar PSF modulated by a spin-2 phase factor. This approximation is fully compatible with the polarization basis used in \texttt{HEALPix}~\cite{2005ApJ...622..759G}, which follows the Ludwig-III convention~\cite{1140406,Carretti:2004zq}, where the polarization components $Q$ and $U$ are defined in the local spherical basis $(\hat{e}_\theta, \hat{e}_\phi)$. In \texttt{HEALPix}, the azimuthal coordinate $\phi$ corresponds to the pixel longitude.

Operationally, we remove the coordinate dependence of the complex polarization field $P_s = Q_s + i U_s$ by multiplying by $e^{+2i\phi_s}$, thereby aligning the pixel to a common reference direction. We then convolve the resulting field with the Gaussian PSF and rotate it back by $e^{-2i\phi(\mathbf{n})}$. The real and imaginary parts of the final field yield the $Q$ and $U$ templates, respectively:
\begin{align}
\mathcal{T}_{Q}(\mathbf{n} \mid \hat{\mathbf{n}}_s, \beta_s) &= 
\Re \Bigg[ e^{-2i\phi(\mathbf{n})}
\left( e^{+2i\phi_s}(Q_s + iU_s)
\frac{1}{2\pi\sigma^2} 
e^{ -\frac{\theta^2}{2\sigma^2}} \right) \Bigg], \\
\mathcal{T}_{U}(\mathbf{n} \mid \hat{\mathbf{n}}_s, \beta_s) &= 
\Im \Bigg[ e^{-2i\phi(\mathbf{n})}
\left( e^{+2i\phi_s}(Q_s + iU_s)
\frac{1}{2\pi\sigma^2} 
e^{ -\frac{\theta^2}{2\sigma^2}} \right) \Bigg],
\end{align}
where $Q_s$ and $U_s$ are the polarization amplitude components of the parameter vector $\beta_s$~, and $\phi_s$ denotes the azimuthal angle of the source center.



To estimate the parameters $\beta_s = \{I_s, Q_s, U_s\}$ of a single point source $s$, we minimize the pixel-domain $\chi^2$ jointly across the above Stokes components within a circular fitting region of radius $R_\mathrm{fit}$, centered at $\hat{\mathbf{n}}_s$, where the point-source beam response dominates over the background signal. The $\chi^2$ is given by:
\begin{align}\label{chisquare}
\chi^2(\beta_s) = 
\left[ \mathbf{m}^{\text{obs}} - \mathbf{m}^{\text{src}}(\beta_s) \right]^\top
\mathbf{C}^{-1}
\left[ \mathbf{m}^{\text{obs}} - \mathbf{m}^{\text{src}}(\beta_s) \right],
\end{align}
where $\mathbf{m}^{\text{obs}}$ and $\mathbf{m}^{\text{src}}(\beta_s)$ are the stacked data and model vectors across all pixels in the fitting region and all Stokes components ($I$, $Q$, and $U$). 

The covariance matrix $\mathbf{C}$ accounts for uncertainties from background components $m_X^{bkg}$. This can be computed using the analytic expression~\cite{Zaldarriaga_1998}:
\begin{align}\label{formula:map_cov}
    \mathbf{C}_{i, j} &= \mathbb{R}(\alpha) \mathbf{M}(r_{i} \cdot r_{j}) \mathbb{R}(\alpha)^{\top},
\end{align}
$\mathbf{M}$ is the covariance matrix in the natural coordinate system:
\begin{align}\label{formula:cov_in_nature}
    \mathbf{M}(r_{i} \cdot r_{j}) = \begin{bmatrix}
        \langle I_{i} I_{j} \rangle & \langle I_{i} Q_{j} \rangle & \langle I_{i} U_{j} \rangle \\
        \langle Q_{i} I_{j} \rangle & \langle Q_{i} Q_{j} \rangle & \langle Q_{i} U_{j} \rangle \\
        \langle U_{i} I_{j} \rangle & \langle U_{i} Q_{j} \rangle & \langle U_{i} U_{j} \rangle
    \end{bmatrix},
\end{align}
where:
\begin{align}\label{formula:corrlation_func}
\left\langle I_{i} I_{j}\right\rangle &= \sum_{l}\left(\frac{2 l+1}{4 \pi}\right) P_{l}(z) C_{l}^{I I}, \\
\left\langle I_{i} Q_{j}\right\rangle &= -\sum_{l}\left(\frac{2 l+1}{4 \pi}\right) F_{l}^{1,0}(z) C_{l}^{I E}, \\
\left\langle I_{i} U_{j}\right\rangle &= -\sum_{l}\left(\frac{2 l+1}{4 \pi}\right) F_{l}^{1,0}(z) C_{l}^{I B}, \\
\left\langle Q_{i} Q_{j}\right\rangle &= \sum_{l}\left(\frac{2 l+1}{4 \pi}\right)\left[F_{l}^{1,2}(z) C_{l}^{E E}-F_{l}^{2,2}(z) C_{l}^{B B}\right], \\
\left\langle U_{i} U_{j}\right\rangle &= \sum_{l}\left(\frac{2 l+1}{4 \pi}\right)\left[F_{l}^{1,2}(z) C_{l}^{B B}-F_{l}^{2,2}(z) C_{l}^{E E}\right], \\
\left\langle Q_{i} U_{j}\right\rangle &= \sum_{l}\left(\frac{2 l+1}{4 \pi}\right)\left[F_{l}^{1,2}(z)+F_{l}^{2,2}(z)\right] C_{l}^{E B},
\end{align}
where $z = r_i \cdot r_j = \cos \theta$ is the cosine of the angle between the two pixels, $i$ and $j$. All the power spectra $C_l$ will be computed from observed maps.
$F_l^{1,0}$, $F_l^{1,1}$, $F_l^{2,2}$ can be written as functions of $z$ as:
\begin{align}
F_{l}^{1,0}(z) &= 2 \frac{\left(\frac{l z}{1-z^{2}}\right) P_{l-1}(z)-\left(\frac{l}{1-z^{2}}+\frac{l(l-1)}{2}\right) P_{l}(z)}{[(l-1) l(l+1)(l+2)]^{1 / 2}}, \\
F_{l}^{1,2}(z) &= 2 \frac{\left(\frac{(l+2) z}{1-z^{2}}\right) P_{l-1}^{2}(z)-\left(\frac{l-4}{1-z^{2}}+\frac{l(l-1)}{2}\right) P_{l}^{2}(z)}{(l-1) l(l+1)(l+2)}, \\
F_{l}^{2,2}(z) &= 4 \frac{(l+2) P_{l-1}^{2}(z)-(l-1) z P_{l}^{2}(z)}{(l-1) l(l+1)(l+2)\left(1-z^{2}\right)}.
\end{align}
Here $P_l$ and $P_l^2$ denote Legendre polynomials $P_l^m$ for the cases $m = 0$ and $m = 2$. The rotation matrix $\mathbb{R}(\alpha)$ is applied to rotate from the nature frame to a global frame where $Q$ and $U$ are referenced to the North-South meridians. The angle between the great circle connecting any two points and the global frame is given by the parameter $\alpha$. Note that here we have used the \texttt{HEALPix} convention~\cite{2005ApJ...622..759G}. The expression of the rotation matrix is:
\begin{align}
    \mathbb{R}(\alpha) &= \begin{bmatrix}
        1 & 0 & 0 \\
        0 & \cos 2 \alpha & -\sin 2 \alpha \\
        0 & \sin 2 \alpha & \cos 2 \alpha
    \end{bmatrix}.
\end{align}


In practical implementation, we construct the pixel–pixel covariance in two steps. First, we compute noise-debiased angular power spectra for each frequency band and convert them into pixel-space covariances using the analytic formalism in equation~\ref{formula:cov_in_nature}, following Ref.~\cite{Tegmark:2001zv}. Second, the noise contribution is taken from noise-only simulations and used as a diagonal noise prior, which we add to the diagonal of the full covariance~$\mathbf{C}$. Based on this covariance, we estimate the point-source parameters by minimizing the $\chi^2$ defined in equation~\ref{chisquare}. The resulting best-fit values are used to construct point-source templates, which are subtracted from the observation maps before subsequent analysis steps.

\subsection{Extension to multi-source fitting}\label{sec:multi_fit}

To account for the blending of nearby sources and improve the accuracy of flux recovery in crowded regions, we extend the GPSF framework to simultaneously fit multiple point sources within a given area. Let $n_s$ denote the number of sources inside the fitting region. The model now becomes a superposition of individual PSF:
\begin{align}\label{eq:multi_model}
m_X^{\text{src}}(\mathbf{n}) = \sum_{s=1}^{n_s} \mathcal{T}_X(\mathbf{n} \mid \hat{\mathbf{n}}_s, \beta_s),
\end{align}
where $X \in \{I, Q, U\}$, $\hat{\mathbf{n}}_s$ is the known position of source $s$, and $\mathcal{T}_{X}$ is the PSF template of that source in component $X$.

Each source contributes three free parameters—one for intensity ($I_s$) and two for polarization ($Q_s$, $U_s$)—yielding a total of $3n_s$ parameters in the model. 
These parameters are estimated using generalized least squares by minimizing the $\chi^2$ statistic with the model substituted by equation~\ref{eq:multi_model}, using the covariance matrix described in section~\ref{sec:single_fit}.

To improve robustness and computational efficiency in polarization analysis, we implement a two-step fitting procedure that targets only the polarized Stokes components $Q$ and $U$:
\begin{itemize}
\item \textbf{Step 1: Identification of significant point sources}

  This step selects bright sources that exceed a polarization detection threshold. Starting from a given catalog (either from existing point source catalogs or constructed using a point-source detection algorithm), we sort all candidate point sources by their total polarization flux density. For each source, we define a local circular fitting region of radius $R_\mathrm{fit}$, as in the single-source case (section~\ref{sec:single_fit}).

  The fitting is performed source-by-source: for each central target, we solve for its $Q_s$   and $U_s$ while accounting for the flux contributions of neighboring sources. All neighboring catalog sources located within $R_\mathrm{neigh}$—the angular distance within which the PSF tails of two sources overlap significantly—are included in the joint model with fixed positions and free flux densities. Only the best-fit parameters and uncertainties of the central source are retained. This process is repeated across the full sorted list.

  We then compute the polarized flux density for each source as $P_s = \sqrt{Q_s^2 + U_s^2}$ and its uncertainty via error propagation:  
\begin{align}
\sigma_{P_s} = \sqrt{
\left( \frac{Q_s}{\sqrt{Q_s^2 + U_s^2}} \cdot \sigma_{Q_s} \right)^2 
+ \left( \frac{U_s}{\sqrt{Q_s^2 + U_s^2}} \cdot \sigma_{U_s} \right)^2 }.
\end{align}

Sources satisfying $P_s > k_\mathrm{sig}\,\sigma_{P_s}$ are flagged as significant and passed to the next step for final subtraction. In practice, $k_\mathrm{sig}$ is typically chosen to be in the range 3–5.

\item \textbf{Step 2: Group-wise simultaneous fitting and subtraction}

In the second step, we re-fit only those sources identified as significant in Step 1. For each source, we search for other significant sources within $R_\mathrm{neigh}$. If any are found, we further check whether they have other nearby sources within the same radius, and repeat this process iteratively until no more are added. All sources found in this way are treated as a group and fit simultaneously. The fitting region is defined as the union of $R_\mathrm{fit}$ disks centered on each source in the group. This approach ensures that blended sources are jointly modeled. Each significant source is fitted exactly once in this process, and all resulting parameters are used for point-source removal.

The refined best-fit values of $Q_s$ and $U_s$ are then used to reconstruct $Q/U$ template maps for each source. These templates are subtracted from the observed maps to obtain source-cleaned polarization maps for further analysis.
\end{itemize}

The covariance matrix $\mathbf{C}$ used in both steps is constructed in the same way as described in Section~\ref{sec:single_fit}.
This two-step strategy ensures robust identification and removal of significant point sources, even in regions with spatial overlap or strong blending.

\paragraph{Computational considerations.}
The dominant computational cost of GPSF is associated with the symmetric positive–definite pixel covariance matrix for each fitting patch. For a patch of size $p$ pixels, the direct Cholesky–based inversion exhibits the standard cubic scaling, $\mathcal{O}(p^{3})$, while constructing the covariance from the estimated signal and noise power spectra scales as $\mathcal{O}(p^{2})$. For the beam sizes considered in this work, the effective patch size ranges from a few hundred pixels (high–frequency channels) up to $\mathcal{O}(10^{4})$ pixels (low–frequency channels). The corresponding wall–clock time for a single patch spans from a few seconds for the smallest patches to $\mathcal{O}(10)$ minutes on a modern CPU. In regions with high source density, the total computational cost increases approximately linearly with the number of sources, since the inverted covariance can be reused and only additional GLS solves are required. Future applications may further reduce this cost by precomputing point–source templates at multiple HEALPix resolutions and explicitly accounting for pixelization effects. For comparison, the alternative point–source mitigation methods used in this work—\textit{masking} and \textit{inpainting} (see Section~\ref{sec:ps_mitigation})—do not require any per–source local operations and therefore incur a much smaller total computational cost. In practice, the masking step requires only a few minutes, while the MCA–based inpainting applied to each frequency band is more computationally demanding and typically takes several to ten minutes.

\section{Simulations and pipelines}\label{sec:simu}

 
In this section, we present the end-to-end simulation developed to evaluate the performance of our GPSF-based point-source mitigation pipeline. 
The simulations incorporate a realistic, multi-component sky model that includes the CMB signal, diffuse foregrounds, and point sources, together with an instrumental model emulating the specifications of a future ground-based SAT CMB experiment. 
We then describe the complete GPSF analysis pipeline and introduce alternative point-source mitigation strategies for direct comparison.

\subsection{Sky model}\label{sec:sky_model}
We use a sky model that aims to closely reflect the true microwave sky. It includes the Cosmic Microwave Background (CMB) signal, diffuse foreground emissions from synchrotron and thermal dust, and extragalactic point sources, which are the main focus of this study. The polarized foreground maps are simulated using the Planck Sky Model~\cite{2013A&A...553A..96D} (PSM). The details of the sky model are as follows:

\begin{enumerate}
    \item \textbf{Extragalactic radio sources:} Compact radio sources are modeled using data from several GHz-band surveys, including NVSS~\cite{NVSS}, SUMSS~\cite{SUMSS}, GB6~\cite{GB6}, and PMN (with subcatalogs PMNE~\cite{PMNE}, PMNT~\cite{PMNT}, PMNZ~\cite{PMNT}, and PMNS~\cite{PMNS}). Flux densities at Planck frequencies are extrapolated from lower frequencies using a power-law spectrum:

    \begin{equation}
        S_\nu = S_{\nu_0} \left( \frac{\nu}{\nu_0} \right)^{-\alpha},
    \end{equation}

    where $S_{\nu_0}$ is the measured flux at a reference frequency $\nu_0$, and $\alpha$ is the spectral index, determined from available multi-frequency data. Different populations (e.g., flat- vs. steep-spectrum sources) are assigned $\alpha$ values from distinct empirical distributions. Polarization fractions are drawn from measured distributions and assumed to be constant with frequency, with random polarization angles.

    \item \textbf{Extragalactic far-infrared sources:} Bright infrared galaxies from the IRAS Point Source Catalog~\cite{1988iras} are extrapolated to millimeter wavelengths using a modified blackbody spectral energy distribution (SED):

    \begin{equation}
        S_\nu = A\, \nu^{\beta} B_\nu(T).
    \end{equation}

    Here, $A$ is a normalization factor derived from the IRAS fluxes, and $\beta$ is the dust emissivity spectral index (typically $\sim 1.3$). $B_\nu(T)$ is the Planck function at dust temperature $T$,

    \begin{equation}
    B_\nu(T) = \frac{2h\nu^3}{c^2} \frac{1}{e^{h\nu / kT} - 1}. 
    \end{equation}

    If only one IRAS band is available, default values for $T$ and $\beta$ are assumed; otherwise, $T$ is fitted using the 60 and 100\,$\mu$m bands. Polarization is modeled by assigning a small fraction (mean $\sim$1\%) drawn from a Rayleigh-like distribution and a random polarization angle.

    \item \textbf{Diffuse synchrotron:} The synchrotron intensity template is derived from the 408~MHz all-sky map by Haslam et al.~\cite{Haslam:1982zz}, extrapolated to microwave frequencies using a fixed spectral index $\beta_s = -3.0$. Polarization information is obtained from the WMAP 7-year $Q$ and $U$ maps at 23~GHz~\cite{Gold:2010fm}, which are used to reconstruct the polarization angle $\gamma_s(p)$ and depolarization factor $g_s(p)$:
    \begin{align}
        \gamma_s(p) &= \frac{1}{2} \tan^{-1}\left(-\frac{U_{23}(p)}{Q_{23}(p)}\right), \\
        g_s(p) &= \frac{\sqrt{Q_{23}^2 + U_{23}^2}}{f_s I_{408}(p) \left(\frac{23}{0.408}\right)^{-\beta_s}},
    \end{align}
    where $f_s = 0.75$ is the intrinsic polarization fraction. The polarized synchrotron emission at frequency $\nu$ is then synthesized using:
    \begin{align}
        Q_\nu(p) &= f_s \, g_s(p) \, I_{408}(p) \left( \frac{\nu}{0.408} \right)^{\beta_s} \cos(2\gamma_s(p)), \\
        U_\nu(p) &= f_s \, g_s(p) \, I_{408}(p) \left( \frac{\nu}{0.408} \right)^{\beta_s} \sin(2\gamma_s(p)).
    \end{align}
    
    \item \textbf{Diffuse thermal dust:} The total intensity of thermal dust is modeled using the two-component modified blackbody model (Model~7) from Finkbeiner et al.~\cite{Finkbeiner:1999aq}, with spatially varying temperature and spectral index. Polarized emission is synthesized from the same intensity template $I_\nu(p)$, combined with a Galactic magnetic field model to compute the polarization angle $\gamma_d(p)$ and geometric depolarization factor $g_d(p)$. The Stokes parameters are given by:
    \begin{align}
        Q_\nu(p) &= f_d \, g_d(p) \, I_\nu(p) \cos(2\gamma_d(p)), \\
        U_\nu(p) &= f_d \, g_d(p) \, I_\nu(p) \sin(2\gamma_d(p)),
    \end{align}
    where the intrinsic polarization fraction is set to $f_d = 0.15$.
    
    \item \textbf{CMB:} 
    

    The primordial CMB maps are Gaussian realizations generated from the power spectra computed using the Boltzmann code \texttt{CAMB}~\cite{lewis2011camb} package\footnote{\url{https://github.com/cmbant/CAMB}}, based on the Planck 2018 best-fit cosmological parameters and assuming $r = 0$. The CMB dipole was excluded from the simulations.
    
    Additionally, the effects of CMB lensing are incorporated into the simulations. Weak gravitational lensing by large-scale structure (LSS) remaps the CMB polarization field on the sphere, converting part of the primordial $E$-mode signal into $B$-modes—commonly referred to as lensing $B$-modes. To simulate this effect, we employ the pixel rearrangement method implemented in the \texttt{lenspyx} \cite{reinecke2023improved} package\footnote{\url{https://github.com/carronj/lenspyx}}, following the algorithm described in Ref.~\cite{2005PhRvD..71h3008L}. This approach applies distortions to the primordial signal using a realization of the lensing potential map generated from the theoretical lensing power spectrum, $C_{\ell}^{\phi\phi}$.
    
\end{enumerate}

\subsection{Instrumental configuration}

\begin{figure}[h]
    \centering
    \includegraphics[width=0.3\textwidth]{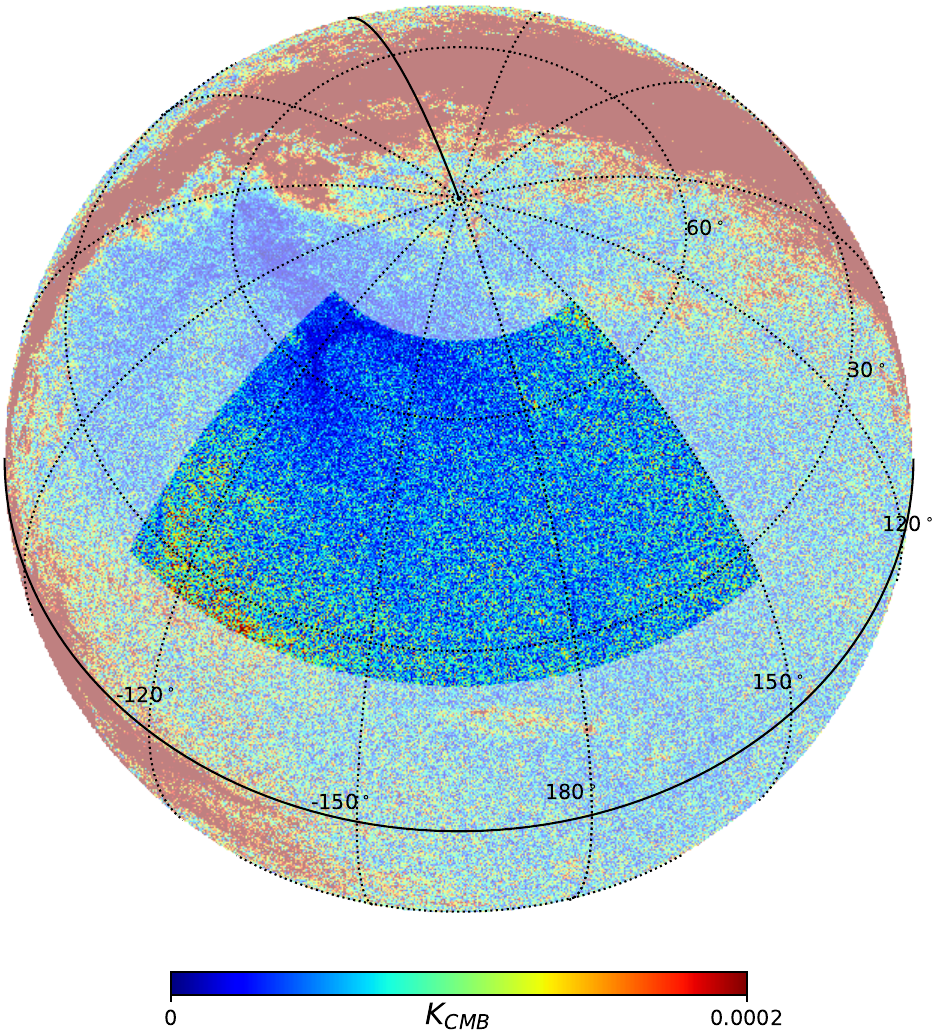}
    \caption{Orthogonal projection of the northern sky patch and its corresponding masks in Equatorial coordinates. The patch is centered at $(RA = 197^\circ, Dec = 55^\circ)$, extending from $RA = 150^\circ$ to $RA = 250^\circ$ and from $Dec = 25^\circ$ to $Dec = 70^\circ$. The background represents the polarization intensity map, $P = \sqrt{Q^2 + U^2}$, observed by \textit{Planck} at 353 GHz.}
    \label{fig:sky_map}
\end{figure}

The configuration of the simulation is designed to emulate future ground-based CMB experiments. For the detection of primordial gravitational waves, it is essential to conduct a deep survey on a clean sky patch with sufficient scanning. 
Following this principle, we selected a clean patch in the northern hemisphere, centered at ($\text{RA}\ 13.13\,\text{h}$, $\text{Dec}\ 55^\circ$) and covering approximately $7\%$ of the sky.
The region has been identified as the patch with the lowest galactic foreground contamination based on the Planck 353 GHz polarization intensity map. The sky patches, along with the Planck 353 GHz polarization emission intensity map, are illustrated in figure~\ref{fig:sky_map}.

Due to the atmospheric emission and absorption effects, ground-based experiments are limited to a few available frequency bands. The selected frequency bands and their corresponding beam resolutions are detailed in table~\ref{tab:setup}. The noise distribution across the patch is assumed to be homogeneous, and the map depth for each frequency band is also provided in the same table. 

The final observed multi-frequency maps were generated by combining the noise map, realized from the map depth, with the simulated sky maps at each frequency. All maps are pixelized in the \texttt{HEALPix} format at $N_{\text{side}}=2048$. 
The noise map realizations are also utilized to estimate the noise-induced biases at the power spectrum level, which are taken into account during the likelihood analysis presented in section~\ref{sec:likelihood_analysis}.

\begin{table}[bthp]
\caption{Instrumental specifications for the northern patch. The map-depth and beam-size are taken from the reference~\cite{2017PhRvD..95d3504A} with $f_{sky}\simeq7\%$.} 
    \centering
    \begin{threeparttable}
        \begin{tabular}{c|c|c|c}
            \hline\hline
            Frequency & Map-depth T & Map-depth P & Beam size \\
            (GHz) & ($\mu$K$\cdot$arcmin) & ($\mu$K$\cdot$arcmin) & (arcmin) \\
            \hline
            $30$          & $7.49$ & $10.60$ & $67.0$ \\
            $95$          & $0.96$ & $1.35$  & $30.0$ \\
            $155$         & $0.87$ & $1.24$  & $17.0$ \\
            $215$         & $1.34$ & $1.90$  & $11.0$ \\
            $270$         & $1.16$ & $1.64$  & $9.0$  \\
            \hline\hline
        \end{tabular}
        \end{threeparttable}
    \label{tab:setup}
\end{table}

\subsection{Full analysis pipeline using GPSF}\label{sec:pipe_GPSF}
In this section, we outline the full analysis pipeline using the GPSF method. The pipeline involves the following steps: initially, we apply GPSF to remove the resolved point sources from the observed maps. Subsequently, we perform map preprocessing to smooth the maps to a common beam and correct for EB leakage before proceeding with foreground cleaning. Next, we employ the NILC algorithm to isolate the CMB signal. Finally, we conduct power spectrum estimation and likelihood analysis to constrain the tensor-to-scalar ratio 
$r$.
\subsubsection{Point-source mitigation}\label{sec: ps_mitigation}
As the first step of the GPSF-based pipeline, we apply the point-source fitting and subtraction procedure described in Section~\ref{sec:multi_fit} to all frequency maps (see Table~\ref{tab:setup}). In this analysis, we adopt the PSM catalog as the input source list and use fixed parameters for the fitting: $R_\mathrm{fit}=1.5\,\theta_\mathrm{FWHM}$ for the local fitting radius, $R_\mathrm{neigh}=3.0\,\theta_\mathrm{FWHM}$ for the neighboring-source radius, and a detection threshold of $k_\mathrm{sig}=3$ for polarized flux significance. A detailed justification of the chosen hyperparameters—$R_\mathrm{fit}$, $R_\mathrm{neigh}$, and $k_\mathrm{sig}$—including their physical motivation and sensitivity tests, is provided in appendix~\ref{app:hyperparams}. Note that we construct the pixel-domain covariance matrix using only the $Q$ and $U$ components, as the fitting targets polarized point sources in the polarization maps. The correlation between intensity and polarization is ignored, as it has a negligible impact on the estimation of polarized source parameters. Accordingly, the intensity component $I$ is excluded from the $\chi^2$ minimization.

Table~\ref{tab:pol_src_det} summarizes the performance of the GPSF fitting stage in recovering polarized point-source flux densities. For each frequency band, we report: (i) the characteristic flux uncertainty $\sigma_{\rm pol}$ obtained from the fitted $(Q_s, U_s)$ parameters, (ii) the number of sources whose fitted polarized flux $P_s=\sqrt{Q_s^2+U_s^2}$ exceeds $3\sigma_{\rm pol}$, and (iii) the number of blended sources identified within the neighbor-inclusion radius $R_{\rm neigh}$. Bands with narrower beams yield smaller flux uncertainties and thus a larger number of measured polarized sources, while blending affects a small subset of closely spaced sources.

\begin{table}[htbp]
    \centering
    \renewcommand{\arraystretch}{1.3}
    \caption{Summary of the GPSF fitting performance for each frequency band. For each band, we list the characteristic polarized flux uncertainty $\sigma_{\mathrm{pol}}$ derived from the fitted $(Q_s,U_s)$ parameters, the number of sources with polarized flux $P_s > 3\sigma_{\mathrm{pol}}$, and the number of blended sources found within the neighbor-inclusion radius $R_{\mathrm{neigh}}$.}
    \label{tab:pol_src_det}
    \begin{tabular}{c|ccc}
    \hline\hline
    Frequency (GHz) 
      & $\sigma_{\mathrm{pol}}$ (mJy) 
      & $N_{\mathrm{det}}(>3\sigma_{\mathrm{pol}})$ 
      & $N_{\mathrm{blend}}$ \\
    \hline
     30  & 10.0 & 21 & 3 \\
     95 & 4.4 & 79 & 10 \\
     155 & 3.9 &  83 & 4 \\
     215 & 3.8 &  83 & 4 \\
     270 & 2.4 &  133 & 6 \\
    \hline\hline
    \end{tabular}
    \label{tab:pol_src_det}
\end{table}

\subsubsection{Map preprocessing}\label{sec:pre-process}
Prior to foreground cleaning, all $QU$ maps are first smoothed to a common angular resolution of 17 arcmin FWHM, and then corrected for $EB$ leakage using the recycling method~\cite{Liu:2018kut}. These preprocessing steps serve two main purposes: (1) ensuring that the CMB signal exhibits the same spatial pattern across all frequency channels, which is essential for NILC; and (2) mitigating $EB$ leakage, which would otherwise bias the $B$-mode power spectrum and thus the estimation of the tensor-to-scalar ratio $r$. Further details of the smoothing and leakage correction procedures are provided in appendix~\ref{appendix:map_preprocess}.

\subsubsection{Foreground cleaning}\label{sec:fg_cln} 
We apply the NILC method to the smoothed polarization maps at five frequencies in order to extract the CMB signal, producing a cleaned map as the output.
NILC minimizes the local variance in the needlet domain under a CMB preservation constraint. Details are described in appendix~\ref{appendix:nilc}.

\subsubsection{Power spectrum estimation and likelihood analysis}\label{sec:likelihood_analysis}
After foreground cleaning, we estimate the angular power spectra of the cleaned maps to assess the residual contamination and ultimately constrain the tensor-to-scalar ratio $r$. The pseudo-$C_\ell$ spectra are computed using the \texttt{PyMaster}\footnote{\url{https://github.com/LSSTDESC/NaMaster}}  package~\cite{10.1093/mnras/stz093}, which accounts for the effects of partial sky coverage. 
The $BB$ spectrum is calculated using the polarization mode of \texttt{PyMaster} when the input is $QU$ polarization maps, and the scalar mode when the input is B-mode maps.
The choice ensures consistency with the map format and accurate estimation of the $BB$ spectrum. We use a piecewise binning scheme with fixed width $\Delta\ell = 40$ for $\ell < 400$, and variable width $\Delta\ell = \max(40, \left[ 0.1\ell \right])$ at higher multipoles.

The noise bias is estimated using simulated noise maps that undergo the same pipeline as the observed maps.
The resulting noise power spectra are then subtracted from the measured spectra of the observed maps.

Then we constrain the tensor-to-scalar ratio $r$ using the noise debiased\footnote{We only debias the noise component, as it is more straightforward to model in real experiments. In contrast, foreground components are difficult to accurately characterize through simulations. For unresolved point sources, the modeling can introduce biases in cosmological inference, as discussed in Ref.~\cite{Planck:2018yye}. We include a plot of the associated biases in the appendix~\ref{app:bias} for reference.} $BB$ spectrum in multipole range $\ell \in [40, 320]$, where we adopt a Gaussian likelihood of the form
\begin{align}
-2 \ln \mathcal{L} = \left(\hat{C}_{\ell_b}^{BB} - C_{\ell_b}^{BB}\right)^T \mathrm{Cov}^{-1} \left(\hat{C}_{\ell_b}^{BB} - C_{\ell_b}^{BB}\right) + \text{const.}
\end{align}
This approximation is well justified in the high-$\ell$ regime, where the number of modes per band power is sufficiently large for the central limit theorem to render the distribution of estimated spectra approximately Gaussian~\cite{Hamimeche:2008ai,Planck:2019nip}. 
Here, $\hat{C}_{\ell_b}^{BB}$ denotes the mean of the noise-debiased B-mode power spectra obtained after point-source mitigation, used to evaluate the effectiveness of different mitigation strategies. In contrast, $C_{\ell_b}^{BB}$ refers to the theoretical model spectrum computed from the assumed cosmological parameters. 
The covariance is estimated from the spectra of 200 end-to-end simulated maps that include CMB and instrumental noise and are propagated through the full analysis pipeline.
In the analysis, we fix all cosmological parameters to their \textit{Planck} 2018 best-fit values~\cite{Planck:2018vyg}, except for $r$ and the lensing amplitude $A_L$, which are treated as free parameters. We perform Markov Chain Monte Carlo sampling using the \texttt{Cobaya}\footnote{\url{https://github.com/CobayaSampler/cobaya}} package~\cite{Torrado_2021}, and analyze the posterior distributions with \texttt{GetDist}\footnote{\url{https://github.com/cmbant/getdist}}~\cite{lewis2019}.

\subsection{Performance evaluation setup}\label{sec:ps_mitigation}

To evaluate the effectiveness of our GPSF-based analysis pipeline, we test five distinct cases, each defined by a specific combination of simulation inputs and processing pipeline.  
All cases use the same underlying CMB, diffuse foregrounds, and instrumental noise realizations, but differ in:  
(i) whether point sources are present in the input simulation;  
(ii) the point-source mitigation method applied (e.g., GPSF, inpainting, masking, or none); and  
(iii) the ordering of processing steps (e.g., EB leakage correction, smoothing, foreground cleaning).  
The details of each case are as follows:

\begin{enumerate}
    \item \textbf{\textit{No PS}}\\
    \begin{tabularx}{\linewidth}{@{}lX@{}}
    (i) & No point sources are included in the simulation input (CMB + diffuse foregrounds + instrumental noise only). \\
    (ii) & No point-source mitigation is applied. \\
    (iii) & The maps are processed using the full GPSF pipeline, except for the point-source removal step. \\
    \end{tabularx}

    \item \textbf{\textit{PS unmitigated}}\\
    \begin{tabularx}{\linewidth}{@{}lX@{}}
    (i) & Point sources are added to the same input as in \textit{No PS}. \\
    (ii) & No point-source mitigation is applied. \\
    (iii) & The maps are processed identically to those in \textit{No PS}. \\
    \end{tabularx}

    \item \textbf{\textit{GPSF}}\\
    \begin{tabularx}{\linewidth}{@{}lX@{}}
    (i) & Same simulation input as \textit{PS unmitigated}. \\
    (ii) & Point sources are removed using the GPSF method described in Section~\ref{sec:method}. \\
    (iii) & The resulting maps are processed through the GPSF pipeline outlined in Section~\ref{sec:pipe_GPSF}. \\
    \end{tabularx}

    \item \textbf{\textit{Inpainting}}\\
    \begin{tabularx}{\linewidth}{@{}lX@{}}
    (i) & Same simulation input as \textit{PS unmitigated}. \\
    (ii) & In each frequency band, bright $B$-mode point sources, identified in Step~1 of Section~\ref{sec:ps_mitigation}, are masked with circular disks of radius $2.5\,\theta_{\text{FWHM}}$ and filled via MCA-based inpainting using the \texttt{iSAP} package\footnote{\url{https://www.cosmostat.org/software/isap}}, \texttt{m3} configuration. \\
    (iii) & EB leakage correction is performed first, followed by inpainting, then smoothing the inpainted maps to $17'$ FWHM, and finally applying NILC. \\
    \end{tabularx}

    \item \textbf{\textit{Masking}}\\
    \begin{tabularx}{\linewidth}{@{}lX@{}}
    (i) & Same simulation input as \textit{PS unmitigated}. \\
    (ii) & The 20 brightest point sources are masked with circular disks of radius $2.5\,\theta_{\text{FWHM}}$ (corresponding to the beam size at 30\,GHz). No inpainting is applied. \\
    (iii) & Smoothing to $17'$ FWHM is performed first, followed by EB leakage correction, then applying NILC, and finally masking. \\
    \end{tabularx}
\end{enumerate}

Although all five cases share the same underlying survey footprint, the usable sky area entering the power-spectrum estimation is not strictly identical across mitigation methods. The \textit{Masking} case applies circular point-source masks after NILC, reducing the effective sky fraction of the final map. For reference, the effective sky fraction without point-source masks is $f_{\mathrm{sky}}^{\mathrm{eff}} = 3.66\%$, while applying the point-source mask reduces it to $f_{\mathrm{sky}}^{\mathrm{eff}} = 3.10\%$. We compute  $f_{\mathrm{sky}}^{\mathrm{eff}}$ using the standard pseudo-$C_\ell$ definition $f_{\mathrm{sky}}^{\mathrm{eff}} = N_{\mathrm{pix}}^{-1}\sum_i W_i^2$.

\section{Results and discussions}\label{sec:result}
We present the results of the performance evaluation cases both before and after NILC. Prior to NILC, we compare the cases by examining the map morphology and B-mode power spectra at 30\,GHz and 155\,GHz. These two frequencies are chosen for specific reasons: 30\,GHz has a relatively large beam, which may lead to more prominent residual effects, while 155\,GHz is representative of the higher-frequency channels. Results at 95, 215, and 270\,GHz are similar to those at 155\,GHz and are included in Appendix~\ref{app:95215270} for completeness. We also assess the NILC results at both the map and power spectrum levels. These NILC-processed results are then used to evaluate the effectiveness of point-source mitigation in CMB reconstruction and its impact on the inferred constraint on the tensor-to-scalar ratio~$r$.

\subsection{Performance at 30 and 155\,GHz frequency bands}\label{sec:each_freq}

\begin{figure}[htbp]
    \centering
    
    \includegraphics[width=0.8\textwidth]{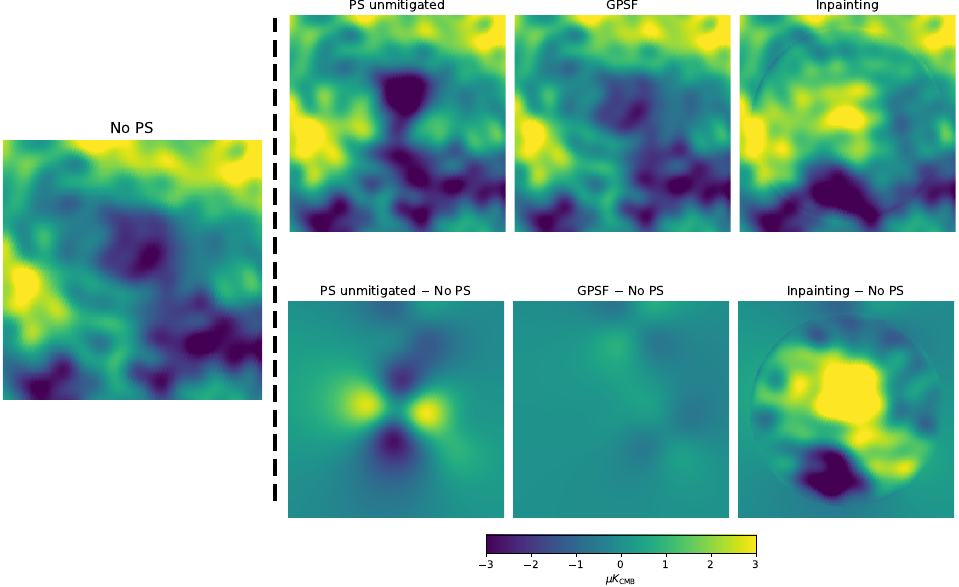}

    \caption{B-mode polarization maps for a representative point-source region at 30~GHz. The leftmost panel shows the \textit{No~PS} case, which serves as the reference for all comparisons. The upper row displays the reconstructed maps for the \textit{PS unmitigated}, \textit{GPSF}, and \textit{Inpainting} cases, with the dashed line separating the reference from the reconstructed maps. The lower row shows the corresponding difference maps relative to the \textit{No~PS} case.}
    \label{fig:freq30_maps}
\end{figure}

\begin{figure}[htbp]
    \centering

    \includegraphics[width=0.8\textwidth]{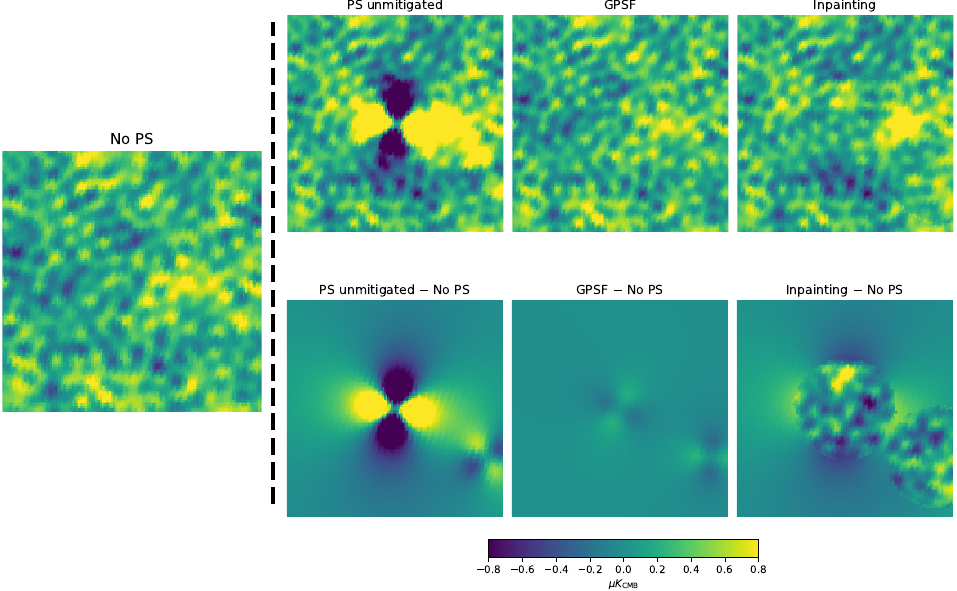}
    \caption{Same as Figure~\ref{fig:freq30_maps} but for 155~GHz. 
    }
    \label{fig:freq155_maps}
\end{figure}

Figures~\ref{fig:freq30_maps} and \ref{fig:freq155_maps} present the B-mode maps in the region of a representative point source at 30\,GHz and 155\,GHz, respectively. Each figure includes the maps for the \textit{No~PS}, \textit{PS unmitigated}, \textit{GPSF}, and \textit{Inpainting} cases, together with their corresponding difference maps relative to the \textit{No~PS} case. At both frequencies, the impact of point sources is most visible when comparing the \textit{PS unmitigated} map with the \textit{No~PS} case, with the effect being substantially stronger at 155\,GHz due to its lower noise levels and foreground contamination. The difference maps highlight these localized residuals and clearly show that the \textit{PS unmitigated} case produces large deviations from the case without point sources. The characteristic butterfly-shaped signature of point sources in $B$-mode maps is also evident~\cite{Diego-Palazuelos_2021}. The \textit{Inpainting} method reduces the point-source signatures but introduces non-negligible distortions within the inpainted region, which are apparent in both the maps and the difference panels. In contrast, the \textit{GPSF} reconstruction yields maps that more closely resemble the \textit{No~PS} case, with significantly smaller residuals in the difference maps. These results demonstrate that GPSF effectively mitigates point-source contamination while preserving the underlying signal structure.

\begin{figure}[h]
    \centering
    \subfloat{\includegraphics[width=0.47\textwidth]{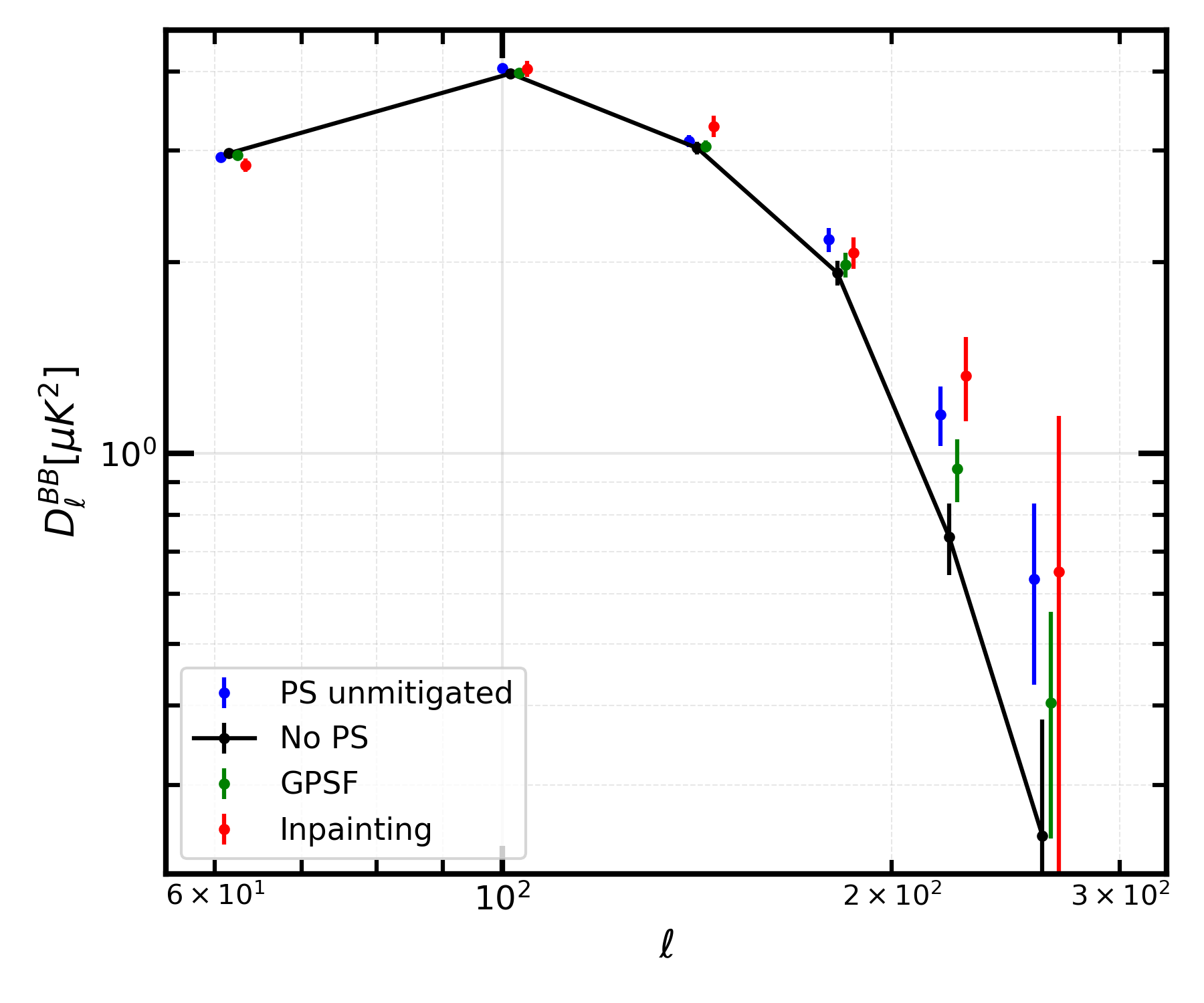}}\hfill
    \subfloat{\includegraphics[width=0.47\textwidth]{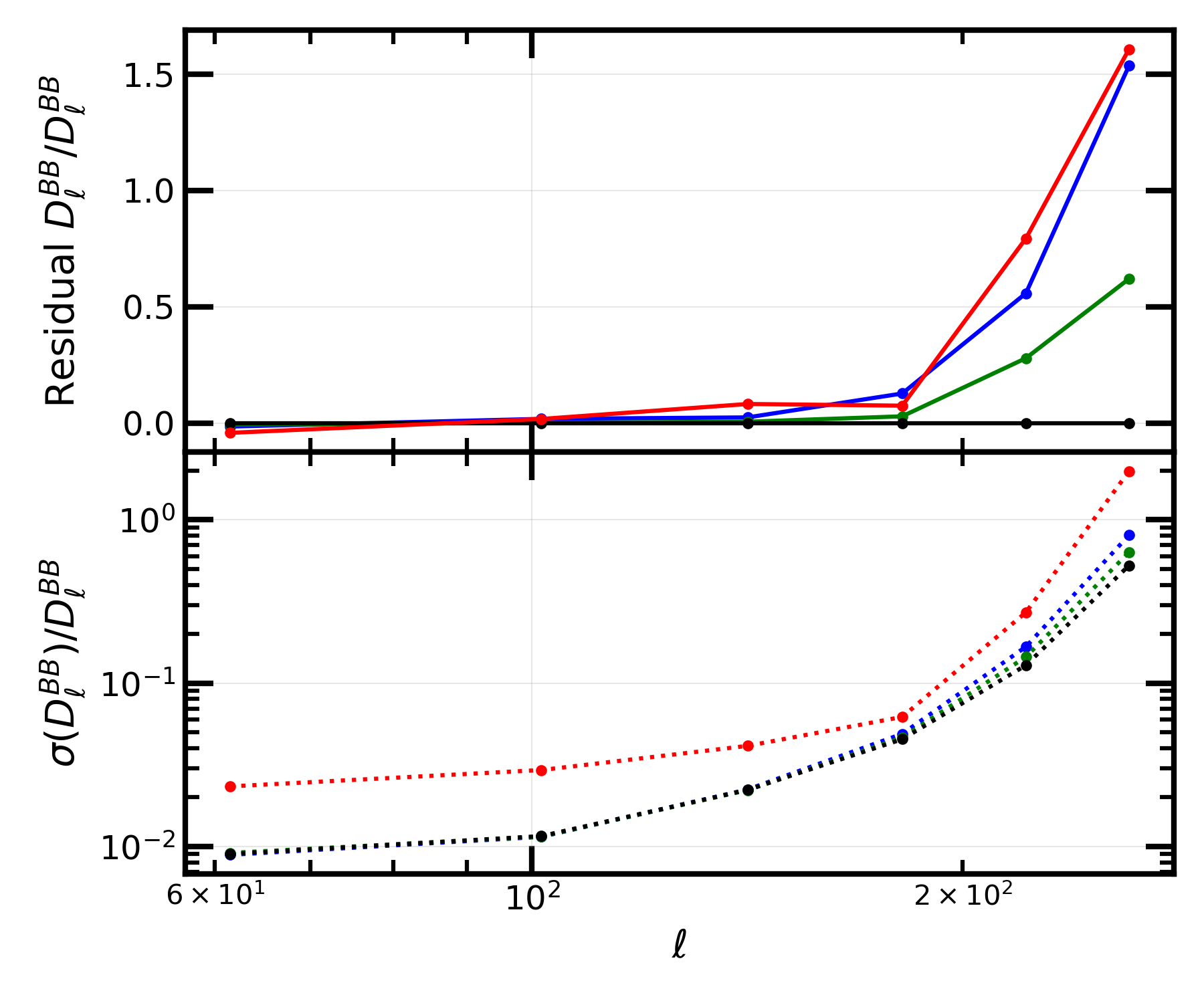}}\\[0.5ex]
    \subfloat{\includegraphics[width=0.47\textwidth]{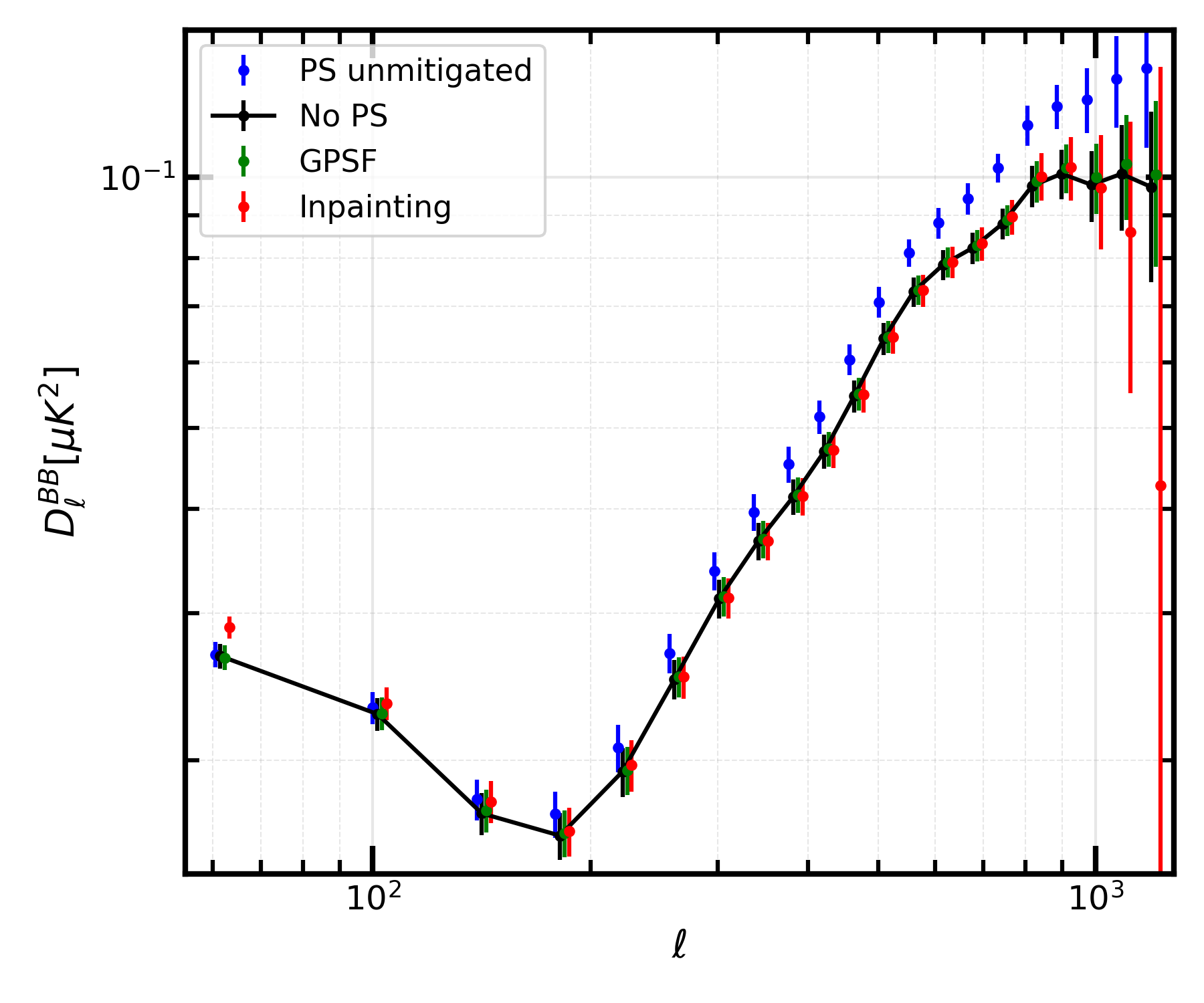}}\hfill
    \subfloat{\includegraphics[width=0.47\textwidth]{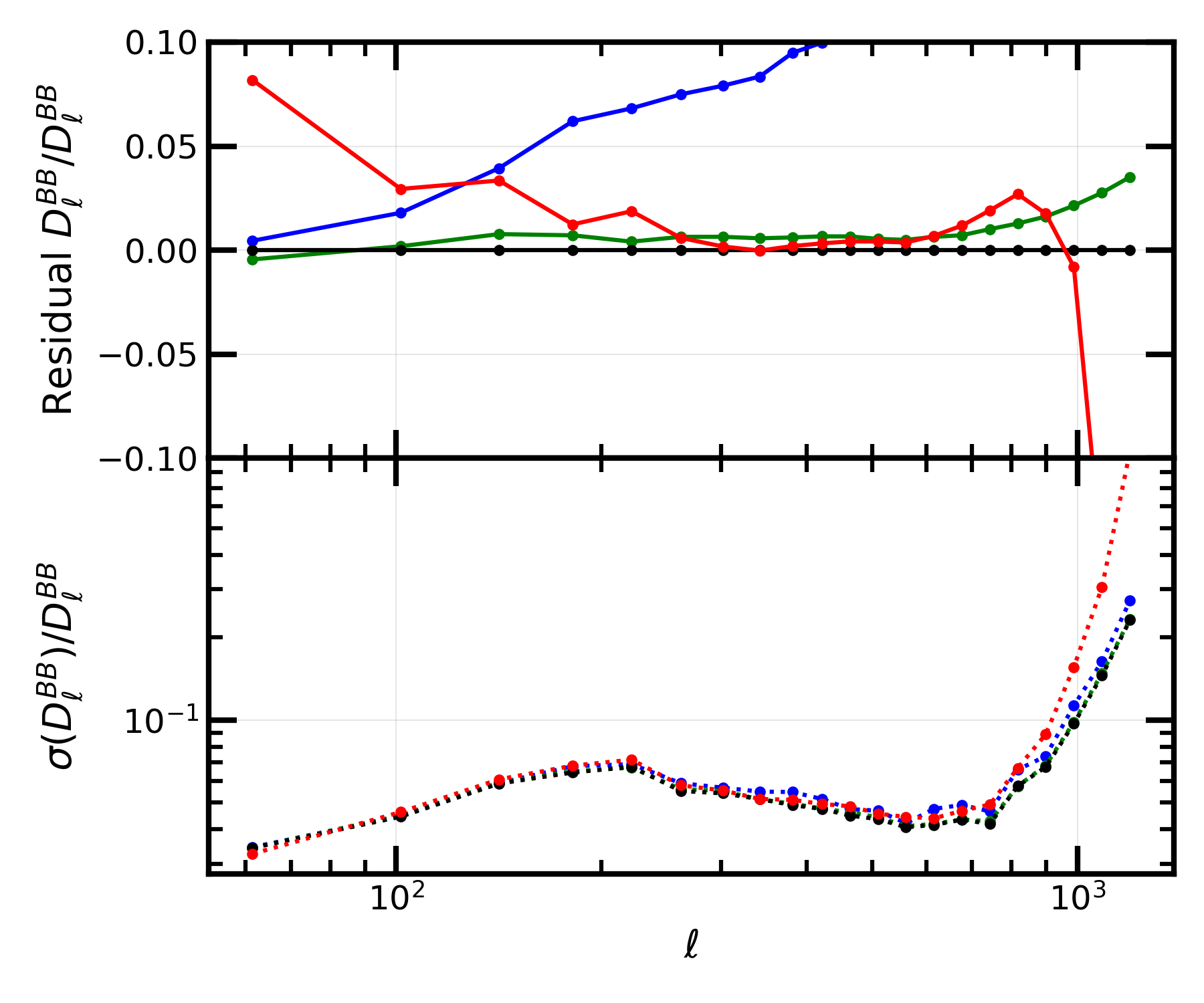}}
    \caption{Noise-debiased $B$-mode power spectra (left) and their relative residuals and relative standard deviations (right) for different point-source mitigation methods. The upper row shows results at 30\,GHz, and the lower row at 155\,GHz. In each row, the left panel presents the noise-debiased B-mode power spectrum with standard deviations estimated from 200 realizations (accounting for CMB and noise), while the right panel displays the relative residuals and relative standard deviations. The \textit{No PS} serves as the reference.
    The relative residual is defined as the difference between the mean power spectrum of each method and that of the reference, normalized by the reference mean. The relative standard deviation is computed as the standard deviation of each method divided by its own mean.}
    \label{fig:each_freq_cl}
\end{figure}

The B-mode noise-debiased power spectra for 30 and 155\,GHz are shown in the left panel of figure~\ref{fig:each_freq_cl}, along with the relative residual and relative standard deviation in the right panel. 

For 30\,GHz, some interesting features emerge. 
It is evident that the diffuse Galactic synchrotron emission dominates at large scales, as indicated by the \textit{No PS} case.
Additionally, strong contributions from point sources are visible even at low multipoles $100 < \ell < 300$, as shown by the difference between the \textit{PS unmitigated} case and \textit{No PS} case.
This is primarily attributed to the intense synchrotron emission from radio galaxies at this frequency. The relative residuals and relative standard deviations in the right panel offer a straightforward comparison of the mitigation performance between \textit{GPSF} and \textit{Inpainting}.
As shown in the upper-right panels, the \textit{Inpainting} method does not effectively recover the background signal: its relative residual and standard deviation are generally higher than those of the \textit{PS unmitigated} case across most multipoles.  
This degradation likely results from the limited available sky area around bright sources, which restricts the amount of information available for \textit{Inpainting} to infer the background signal.
The \textit{GPSF} method, by contrast, shows reduced residual power—about 50\% lower at $\ell \simeq 200$—and a standard deviation that remains close to that of the \textit{No PS}.
These results indicate that \textit{GPSF} can partially mitigate point-source contamination while preserving the statistical properties of the underlying signal, even in the presence of a large beam.

The 155\,GHz noise-debiased B-mode spectrum reveals the effectiveness of various point-source mitigation methods.
The difference between \textit{No PS} and \textit{PS unmitigated} cases increases with multipole $\ell$, as point source contributions become more significant at higher $\ell$. The effectiveness of mitigation strategies varies across the multipole range as shown in lower right panel.
The \textit{Inpainting} method does reduce the point-source contribution on intermediate multipole ranges. However, it exhibits larger residuals at both low and high $\ell$, along with a increased standard deviation at high $\ell$.
In contrast, the \textit{GPSF} method yields lower residuals across the full multipole range and maintains standard deviations comparable to the reference case.
These results suggest that GPSF is effective for power spectrum analysis at moderate beam sizes, where it can reduce most of bias on power spectra to a low level with minimal change in variance.

\subsection{Impact on NILC maps and power spectra}\label{sec:nilc_res}
\begin{figure}[h]
    \centering
    \includegraphics[width=0.9\textwidth]{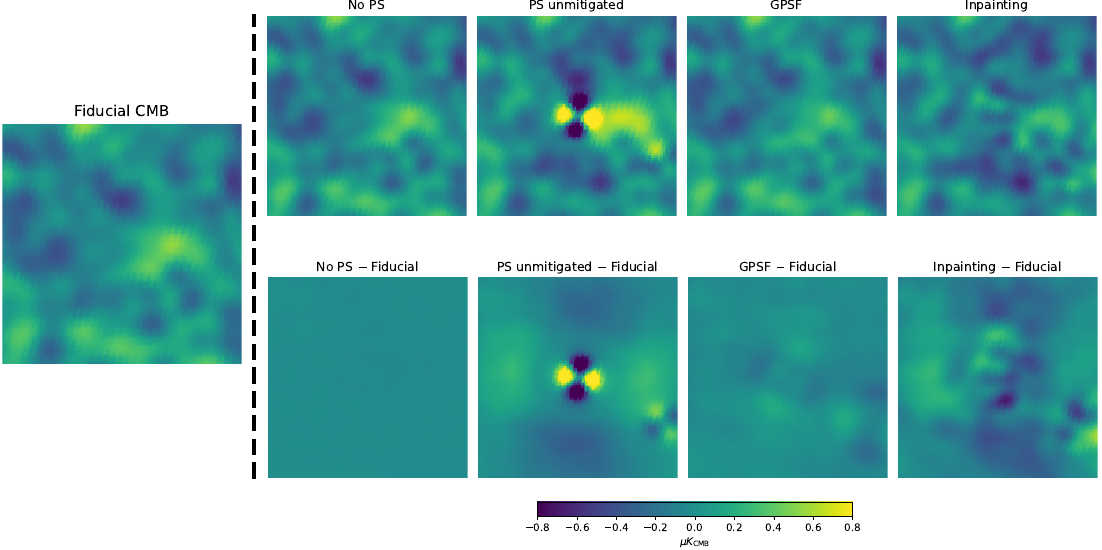}
    \caption{NILC B-mode polarization maps in a representative sky region.
The leftmost panel shows the fiducial (input) CMB realization used as
the reference. The upper row displays the reconstructed maps for the
\textit{No~PS}, \textit{PS unmitigated}, \textit{GPSF}, and
\textit{Inpainting} cases, with the dashed line separating the reference
from the reconstructed maps. The lower row shows the corresponding
difference maps relative to the fiducial CMB. }
    \label{fig:nilc_map_res}
\end{figure}

Figure~\ref{fig:nilc_map_res} shows the NILC B-mode maps at the location of a representative source for different cases: the fiducial CMB, \textit{No PS}, \textit{PS unmitigated}, \textit{Inpainting}, and \textit{GPSF} case, as well as their corresponding difference maps relative to the fiducial CMB.
The similarity between the \textit{No PS} and the fiducial CMB indicates that the diffuse foregrounds in the former have been effectively removed through the NILC operation.
However, noticeable residuals from point sources are observed in the \textit{PS unmitigated} case. This justifies the fact that standard NILC has limited effectiveness in mitigating compact sources, consistent with findings in Ref.~\cite{Zhang:2020ltv}. Therefore, specific point-source mitigation techniques should be incorporated. 
Among the point-source mitigation strategies, the \textit{Inpainting} method visually removes compact source features from the map.
However, the resulting reconstruction exhibits noticeable deviations from the fiducial CMB, suggesting that the underlying background signal may not be accurately recovered in the inpainted regions.
In contrast, the map produced with the \textit{GPSF} method shows closer visual agreement with the fiducial CMB, indicating that GPSF can suppress point-source contamination while preserving the large-scale structure of the CMB signal in NILC framework.

\begin{figure}[h]
    \centering
    \subfloat{\includegraphics[width=0.47\textwidth]{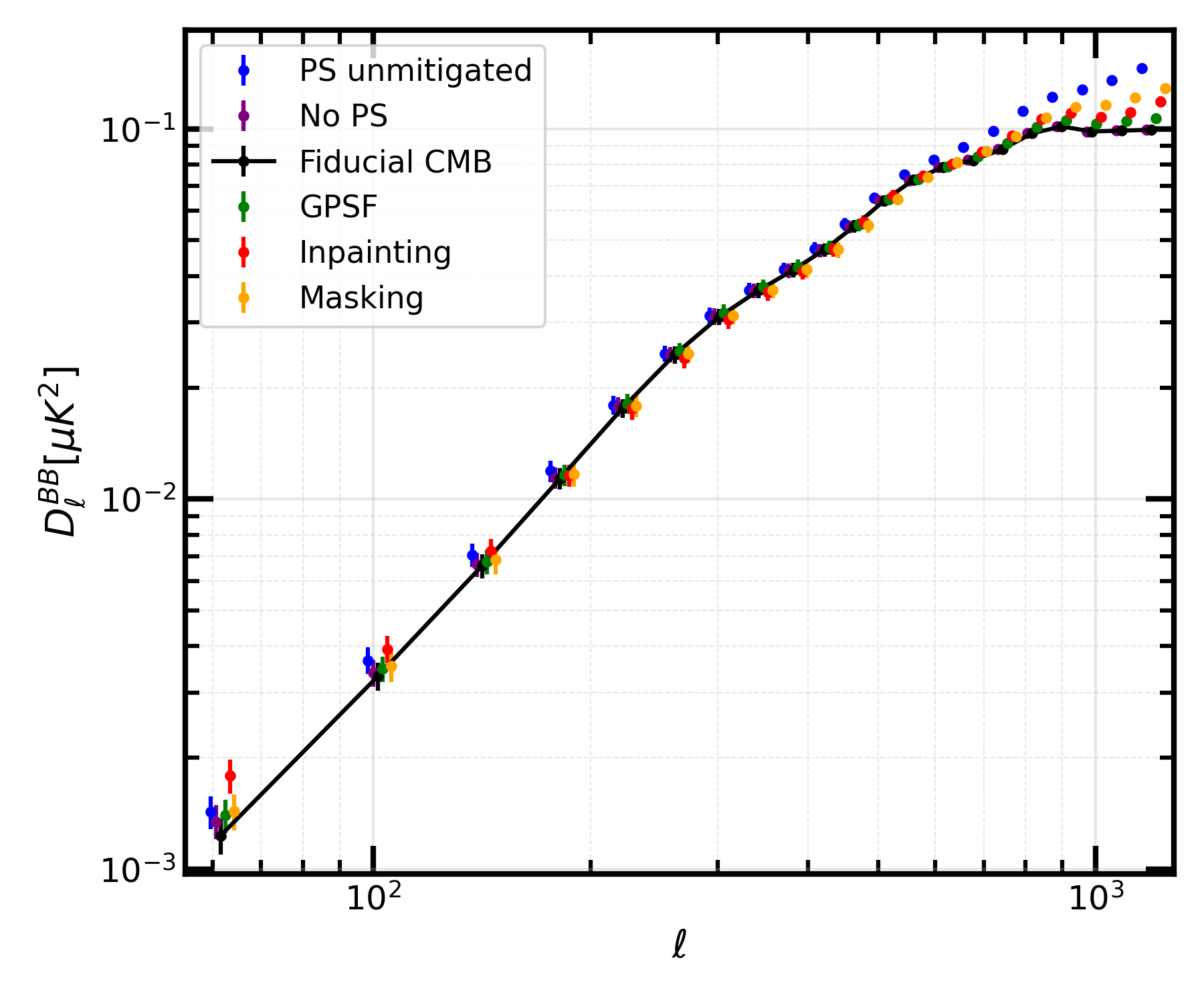}}\hfill
    \subfloat{\includegraphics[width=0.47\textwidth]{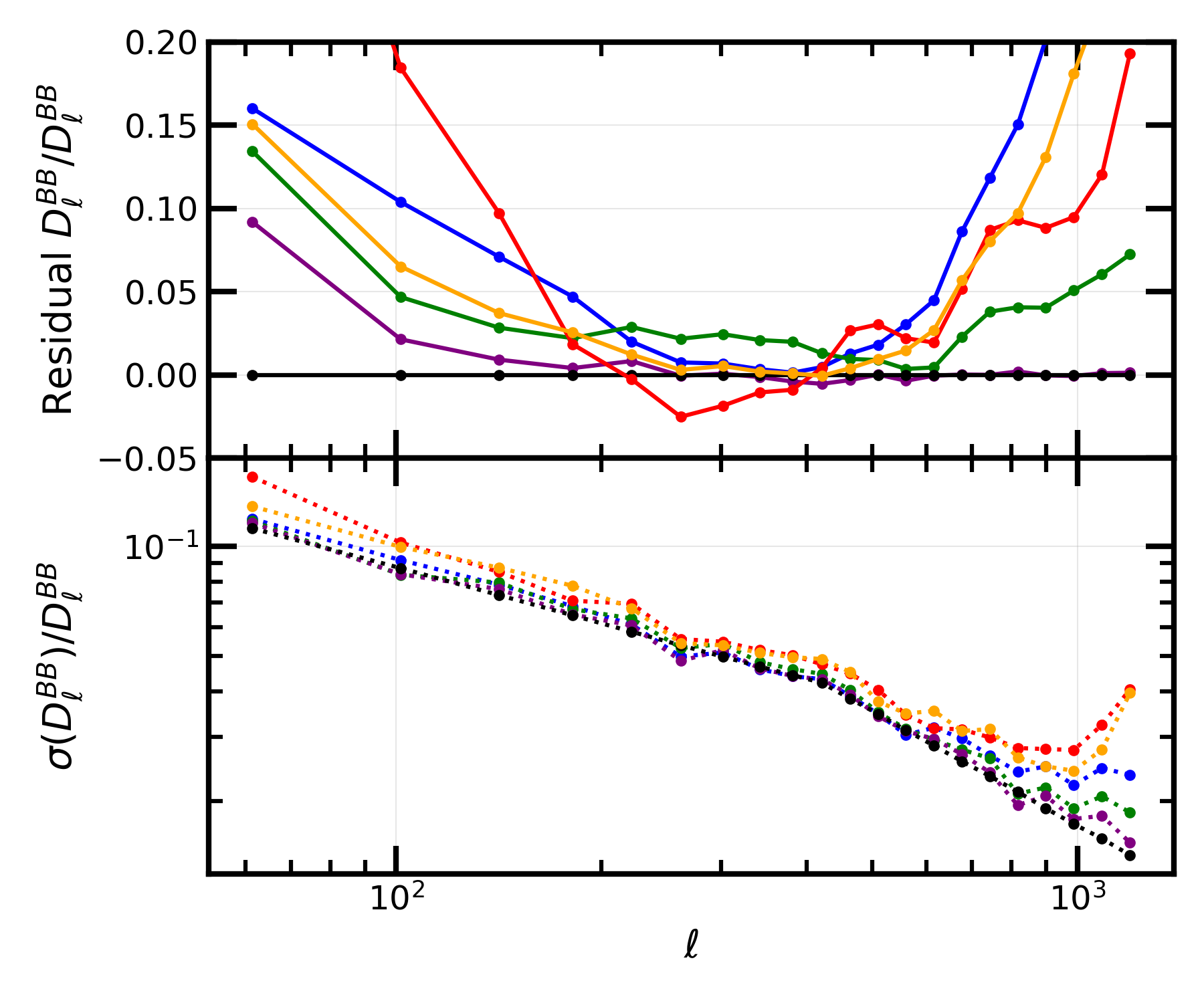}}\\
    \caption{Noise‑debiased $B$‑mode power spectra (left) and their relative residual and relative standard deviation (right) for different point-source mitigation methods after NILC. The left panel presents the noise-debiased B-mode power spectrum with standard deviations estimated from 200 realizations (accounting for CMB and noise), while the right panel displays the relative residuals and relative standard deviations. The fiducial CMB serves as the reference. The relative residual is defined as the difference between the mean power spectrum of each method and that of the reference, normalized by the reference mean. The relative standard deviation is computed as the standard deviation of each method divided by its own mean.}
    \label{fig:nilc_res}
\end{figure}

The NILC B-mode noise-debiased power spectra are shown in the left panel of figure~\ref{fig:nilc_res}, along with the relative residual and relative standard deviation in the right panel.

As shown in the left panel, the NILC method is effective in suppressing diffuse foreground contamination, particularly at large angular scales ($\ell < 300$) where foreground emission is most prominent. In the \textit{No PS} case, only a small foreground-induced bias appears in the lowest multipole bin, and the spectrum closely follows the fiducial CMB at higher multipoles. In comparison, the \textit{PS unmitigated} case exhibits a significant excess at $\ell > 550$, even after NILC processing. This contrast highlights the limited ability of NILC to remove point-source contamination at smaller angular scales~\cite{Zhang:2020ltv}.

From the right panel, the results show that the \textit{Inpainting} method reduces the bias of point sources compared to the \textit{PS unmitigated} at $\ell > 600$.
However, the residuals exhibit greater variability at $\ell < 600$, and \textit{Inpainting} leads to more standard deviation than the \textit{No PS} case across all multipoles.
The \textit{Masking} method reduces the bias from the point source across scales. However, the increased standard deviation resulting from the reduction in the usable sky area represents a significant drawback. We have tried to mask more point sources and the increased standard deviation makes the results even worse.
In contrast, \textit{GPSF} yields lower residuals than the \textit{PS unmitigated} case at $\ell < 240$ and $\ell > 500$, but demonstrates reduced efficiency in the intermediate range $240 < \ell < 500$. In this regime, the method shows limited effectiveness in point-source removal, and the residual point-source contamination significantly complicates the NILC procedure. Meanwhile, GPSF maintains standard deviations comparable to \textit{No PS} case. The particularly good performance at $\ell < 240$ demonstrates its ability to better constrain the tensor-to-scalar ratio $r$.

\begin{table}[htbp]
    \centering
    \renewcommand{\arraystretch}{1.3} 
    \caption{Mean and standard deviation on $r$ for different point-source mitigation cases.}
    \label{tab:r_mean_std}
    \begin{tabular}{c|ccccc}
    \hline\hline
    $r$ & \makecell{No PS} & \makecell{PS unmitigated} & Inpainting & Masking & GPSF \\
    \hline
    Mean & $1.89 \times 10^{-3}$ & $3.56 \times 10^{-3}$ & $1.05 \times 10^{-2}$ & $3.08 \times 10^{-3}$ & $2.18 \times 10^{-3}$ \\
    Std. dev & $2.40 \times 10^{-3}$ & $2.52 \times 10^{-3}$ & $3.01 \times 10^{-3}$ & $2.59 \times 10^{-3}$ & $2.45 \times 10^{-3}$ \\
    \hline\hline
    \end{tabular}
\end{table}

\subsection{Tensor-to-scalar ratio $r$ constraint}
Considering the scientific objective of detecting primordial gravitational waves, we ultimately evaluate the performance of the point-source mitigation approach based on its ability to improve the constraints on the tensor-to-scalar ratio $r$. The mean and standard deviation on $r$ obtained from 200 simulations for different point-source mitigation methods are summarized in table~\ref{tab:r_mean_std}.

In this work, we define the bias as the shift in the mean of $r$ relative to the \textit{No PS}. Physically, this bias quantifies the systematic error introduced by imperfect point-source mitigation and foreground removal. A smaller bias implies a more faithful recovery of the underlying CMB $B$-mode signal and thus a more reliable estimate of $r$.
The presence of point sources introduces a bias of $1.67 \times 10^{-3}$ of $r$, consistent with the findings reported in Ref.~\cite{Zhang:2020ltv}.
A comparative analysis of the mitigation approaches shows that the \textit{Inpainting} method yields a residual bias and standard deviation that remain higher than those of the \textit{No PS} case. The \textit{Masking} method achieves a $28\%$ reduction in bias ($1.19 \times 10^{-3}$) at the cost of an $8\%$ elevation in standard deviation ($2.59 \times 10^{-3}$ versus $2.40 \times 10^{-3}$), attributable to decreased sky coverage.
The \textit{GPSF} method yields a residual bias of $2.9 \times 10^{-4}$, representing an 82\% reduction relative to the \textit{PS unmitigated} case, while the standard deviation increases by only 2\%, remaining close to that of the ideal simulation.
These results indicate that \textit{GPSF} is capable of suppressing point-source contamination with minimal impact on the statistical properties of the CMB signal, supporting precise estimation of the tensor-to-scalar ratio~$r$.

\subsection{Discussion}
In the previous section, we examined how the GPSF method addresses point-source contamination in CMB B-mode polarization analysis, both before and after NILC.
In the following, we provide a detailed analysis of these results.
 
At the map level, we find the \textit{GPSF} leads to the least point-source residual while preserving the most reliable CMB signal.
This success stems from the use of an explicit point-source modeling, which subtracts localized flux while preserving the background CMB signal.
In contrast, the \textit{inpainting} method reconstructs masked regions based on surrounding structures, exhibiting strong dependence on local morphological context. This leads to distinct artifacts in point-source region, as evidenced by the 30 and 155\,GHz maps and the NILC maps.
The \textit{masking} approach adopts a more radical solution by completely discarding contaminated regions, which inevitably reduces both the usable sky coverage and the resulting statistical significance of the analysis.

At the power spectrum level, \textit{GPSF} achieves low residuals at $\ell < 240$. Numerical results in appendix~\ref{app:bias} indicate that, in this multipole range, GPSF effectively reduces point-source contributions by approximately 50\% while not interfering with the diffuse foreground removal, which in turn lowers the overall residual level.
For other methods, although \textit{Inpainting} is computationally efficient and can reconstruct CMB at some level, our results show that its recovering effectiveness is limited for large beams.
\textit{Masking} reduces bias but, by lowering sky coverage, increases the standard deviation; for SATs at low-frequency bands, this coverage loss significantly weakens sensitivity to $r$.

Our analysis suggests that the \textit{GPSF} approach demonstrates considerable potential as an effective point-source mitigation strategy for CMB $B$-mode studies, especially when implemented within the NILC framework. This method achieves an optimal trade-off between minimizing residual contamination and maintaining robustness of recovering the CMB signal.

Finally, thanks to the strengths of GPSF—particularly its ability to suppress residuals at high multipoles ($\ell > 500$)—we anticipate that it may have broader applicability. For example, CMB lensing reconstruction is known to be sensitive to biases from compact sources on small angular scales \cite{osborne2014extragalactic,sailer2023foreground}, which can be substantially mitigated through a prior point-source removal step. A detailed investigation of this potential application is left for future work.

\section{Conclusions}\label{sec:conclusions}
For CMB polarization experiments targeting the detection of primordial gravitational waves, polarized point sources represent a significant source of contamination.  
In this work, we introduce a new method—\textit{Generalized Point Spread Function Fitting} (GPSF)—alongside a complete analysis pipeline incorporating NILC component separation, designed to resolve point-source contamination in polarization maps.

We conducted a comprehensive evaluation of point-source mitigation strategies performance across multiple frequency channels, examining their impact both before and after NILC. The evaluation considered three aspects: single-frequency maps and power spectra (with focus on 30\,GHz and 155\,GHz), NILC maps and power spectra, and the resulting constraints on the tensor-to-scalar ratio~$r$. 

At the single-frequency level, the \textit{GPSF} method shows improved preservation of CMB signal in the map domain. Correspondingly, in the spectrum domain, it yields lower residuals across angular scales, albeit with a marginal increase in standard deviation. These results suggest that GPSF can help mitigate point-source contamination while preserving statistical properties. 
In contrast, the \textit{Inpainting} method displays somewhat reduced effectiveness for background signal reconstruction within masked regions, with more pronounced challenges under large beam conditions as observed at 30\,GHz.

After NILC, residual contamination from point sources remains evident in the \textit{PS unmitigated} case, especially at small angular scales.
In contrast, the \textit{GPSF} method improves agreement with the fiducial CMB both in map morphology and in the power spectrum. At the map level, point-source residuals are visibly reduced, and the large-scale structure is more consistent with the reference. For the power spectrum, GPSF yields low residuals across multipoles, with standard deviations remaining comparable to the ideal case. These results suggest that applying GPSF prior to NILC can help reduce point-source contamination without introducing significant additional variance.
While the \textit{Inpainting} method also suppresses compact source features in the maps, though some residual differences remain. For the power spectrum, it exhibits higher residuals and increased standard deviations, particularly at low and high multipoles, compared to the unmitigated case.
The \textit{Masking} method reduces residual bias in the power spectrum by discarding contaminated regions, but this comes at the cost of increased variance due to the reduced sky coverage.

For the constraints on $r$, the presence of point sources introduces a bias of $1.67 \times 10^{-3}$ in $r$, consistent with previous findings~\cite{Zhang:2020ltv}.  
The \textit{Inpainting} method yields higher bias and standard deviation compared to the case with point sources unmitigated. This may be due to the relatively large masks around point sources and the limited surrounding area available for reconstruction, which together can lead to deviations from the true background in the recovered signal and increased uncertainty in the masked regions.  
\textit{Masking} reduces the bias by 28\%, but increases the standard deviation by 8\%, reflecting the trade-off between contamination removal and sky coverage.  
The \textit{GPSF} method reduces the bias by 82\%—to $2.9 \times 10^{-4}$—while increasing the standard deviation by only 2\%, remaining close to the ideal case.

In summary, our results demonstrate that the \textit{GPSF} method effectively mitigates point-source contamination while robustly preserving the key features of the CMB signal—both in the map domain and power spectrum—before and after NILC processing. 
This performance leads to improved stability in estimating the tensor-to-scalar ratio, $r$, positioning \textit{GPSF} as a promising strategy for future CMB polarization analyses targeting primordial gravitational wave detection.

\acknowledgments
We thank Zirui Zhang, Yongping Li, Yepeng Yan, Zixuan Zhang, Qian Chen, Xingjian Lv, Luya Lou for useful discussion. This study is supported by the National Key R\&D Program of China No.2020YFC2201601, National Natural Science Foundation of China No.12403005.
We acknowledge the use of packages: \texttt{PSM}~\cite{2013A&A...553A..96D}, \texttt{lenspyx} \cite{reinecke2023improved}, \texttt{HEALPix}~\cite{2005ApJ...622..759G}, \texttt{healpy}~\cite{Zonca2019}, \texttt{iminuit}~\cite{iminuit}, \texttt{CAMB}~\cite{lewis2011camb}, \texttt{PyMaster}~\cite{10.1093/mnras/stz093}, \texttt{Cobaya}~\cite{Torrado_2021}, \texttt{GetDist}~\cite{lewis2019}.

\bibliography{main}

@article{reinecke2023improved,
  title={Improved cosmic microwave background (de-) lensing using general spherical harmonic transforms},
  author={Reinecke, Martin and Belkner, Sebastian and Carron, Julien},
  journal={Astronomy \& Astrophysics},
  volume={678},
  pages={A165},
  year={2023},
  publisher={EDP Sciences}
}

@article{lewis2011camb,
  title={Camb: Code for anisotropies in the microwave background},
  author={Lewis, Antony and Challinor, Anthony},
  journal={Astrophysics source code library},
  pages={ascl--1102},
  year={2011}
}

@article{Planck:2018yye,
    author = "Akrami, Y. and others",
    collaboration = "Planck",
    title = "{Planck 2018 results. IV. Diffuse component separation}",
    eprint = "1807.06208",
    archivePrefix = "arXiv",
    primaryClass = "astro-ph.CO",
    doi = "10.1051/0004-6361/201833881",
    journal = "Astron. Astrophys.",
    volume = "641",
    pages = "A4",
    year = "2020"
}

@article{Planck:2015bin,
    author = "Ade, P. A. R. and others",
    collaboration = "Planck",
    title = "{Planck 2015 results. XXVI. The Second Planck Catalogue of Compact Sources}",
    eprint = "1507.02058",
    archivePrefix = "arXiv",
    primaryClass = "astro-ph.CO",
    doi = "10.1051/0004-6361/201526914",
    journal = "Astron. Astrophys.",
    volume = "594",
    pages = "A26",
    year = "2016"
}

@article{ACT:2023wxl,
    author = "Vargas, Cristian and others",
    collaboration = "ACT",
    title = "{The Atacama Cosmology Telescope: Extragalactic Point Sources in the Southern Surveys at 150, 220 and 280 GHz observed between 2008-2010}",
    eprint = "2310.17535",
    archivePrefix = "arXiv",
    primaryClass = "astro-ph.GA",
    month = "10",
    year = "2023"
}

@inproceedings{Bajkova:2008ygr,
    author = "Bajkova, Anisa T.",
    title = "{High-accuracy method for the removal of point sources from maps of the cosmic microwave background}",
    booktitle = "{International Conference on Problems of Practical Cosmology}",
    volume = "2",
    pages = "220--227",
    year = "2008"
}

@article{Scodeller:2012sw,
    author = "Scodeller, Sandro and Hansen, Frode K.",
    title = "{Masking versus removing point sources in CMB data: the source corrected WMAP power spectrum from new extended catalogue}",
    eprint = "1207.2315",
    archivePrefix = "arXiv",
    primaryClass = "astro-ph.CO",
    doi = "10.1088/0004-637X/761/2/119",
    journal = "Astrophys. J.",
    volume = "761",
    pages = "119",
    year = "2012"
}

@article{White:1994ru,
    author = "White, Martin J. and Srednicki, Mark",
    title = "{Window functions for CMB experiments}",
    eprint = "astro-ph/9402037",
    archivePrefix = "arXiv",
    reportNumber = "CFPA-TH-94-11, CFPA-94-TH-11, UCSBTH-94-03",
    doi = "10.1086/175497",
    journal = "Astrophys. J.",
    volume = "443",
    pages = "6",
    year = "1995"
}

@article{Zhang:2020ltv,
    author = "Zhang, Zirui and Liu, Yang and Li, Si-Yu and Wu, De-Liang and Li, Haifeng and Li, Hong",
    title = "{Efficient ILC analysis on polarization maps after EB leakage correction}",
    eprint = "2109.12619",
    archivePrefix = "arXiv",
    primaryClass = "astro-ph.CO",
    doi = "10.1088/1475-7516/2022/07/044",
    journal = "JCAP",
    volume = "22",
    pages = "044",
    year = "2020"
}

@article{Abrial:2008mz,
    author = "Abrial, P. and Moudden, Y. and Starck, J. -L. and Fadili, J. and Delabrouille, J. and Nguyen, M. K.",
    title = "{CMB data analysis and sparsity}",
    eprint = "0804.1295",
    archivePrefix = "arXiv",
    primaryClass = "astro-ph",
    doi = "10.1016/j.stamet.2007.11.005",
    journal = "Stat. Meth.",
    volume = "5",
    pages = "289",
    year = "2008"
}

@article{Hamimeche:2008ai,
    author = "Hamimeche, Samira and Lewis, Antony",
    title = "{Likelihood Analysis of CMB Temperature and Polarization Power Spectra}",
    eprint = "0801.0554",
    archivePrefix = "arXiv",
    primaryClass = "astro-ph",
    doi = "10.1103/PhysRevD.77.103013",
    journal = "Phys. Rev. D",
    volume = "77",
    pages = "103013",
    year = "2008"
}

@article{Planck:2019nip,
    author = "Aghanim, N. and others",
    collaboration = "Planck",
    title = "{Planck 2018 results. V. CMB power spectra and likelihoods}",
    eprint = "1907.12875",
    archivePrefix = "arXiv",
    primaryClass = "astro-ph.CO",
    doi = "10.1051/0004-6361/201936386",
    journal = "Astron. Astrophys.",
    volume = "641",
    pages = "A5",
    year = "2020"
}

@article{Planck:2018vyg,
    author = "Aghanim, N. and others",
    collaboration = "Planck",
    title = "{Planck 2018 results. VI. Cosmological parameters}",
    eprint = "1807.06209",
    archivePrefix = "arXiv",
    primaryClass = "astro-ph.CO",
    doi = "10.1051/0004-6361/201833910",
    journal = "Astron. Astrophys.",
    volume = "641",
    pages = "A6",
    year = "2020",
    note = "[Erratum: Astron.Astrophys. 652, C4 (2021)]"
}

@article{Torrado_2021,
doi = {10.1088/1475-7516/2021/05/057},
url = {https://dx.doi.org/10.1088/1475-7516/2021/05/057},
year = {2021},
month = {may},
publisher = {IOP Publishing},
volume = {2021},
number = {05},
pages = {057},
author = {Torrado, Jesús and Lewis, Antony},
title = {Cobaya: code for Bayesian analysis of hierarchical physical models},
journal = {Journal of Cosmology and Astroparticle Physics}
}

@misc{lewis2019,
      title={GetDist: a Python package for analysing Monte Carlo samples}, 
      author={Antony Lewis},
      year={2019},
      eprint={1910.13970},
      archivePrefix={arXiv},
      primaryClass={astro-ph.IM},
      url={https://arxiv.org/abs/1910.13970}, 
}

@article{Finkbeiner:1999aq,
    author = "Finkbeiner, Douglas P. and Davis, Marc and Schlegel, David J.",
    title = "{Extrapolation of galactic dust emission at 100 microns to CMBR frequencies using FIRAS}",
    eprint = "astro-ph/9905128",
    archivePrefix = "arXiv",
    doi = "10.1086/307852",
    journal = "Astrophys. J.",
    volume = "524",
    pages = "867--886",
    year = "1999"
}

@article{Gold:2010fm,
    author = "Gold, B. and others",
    title = "{Seven-Year Wilkinson Microwave Anisotropy Probe (WMAP) Observations: Galactic Foreground Emission}",
    eprint = "1001.4555",
    archivePrefix = "arXiv",
    primaryClass = "astro-ph.GA",
    doi = "10.1088/0067-0049/192/2/15",
    journal = "Astrophys. J. Suppl.",
    volume = "192",
    pages = "15",
    year = "2011"
}

@article{Haslam:1982zz,
    author = "Haslam, C. G. T. and Salter, C. J. and Stoffel, H. and Wilson, W. E.",
    title = "{A 408 MHz all-sky continuum survey. II. The atlas of contour maps}",
    journal = "Astron. Astrophys. Suppl. Ser.",
    volume = "47",
    pages = "1--142",
    year = "1982"
}

@PROCEEDINGS{1988iras,
        title = "{Infrared Astronomical Satellite (IRAS) Catalogs and Atlases.Volume 1: Explanatory Supplement.}",
    booktitle = {Infrared astronomical satellite (IRAS) catalogs and atlases. Volume 1: Explanatory supplement},
         year = 1988,
       editor = {{Beichman}, C.~A. and {Neugebauer}, G. and {Habing}, H.~J. and {Clegg}, P.~E. and {Chester}, Thomas J.},
       volume = {1},
        month = jan,
       adsurl = {https://ui.adsabs.harvard.edu/abs/1988iras....1.....B}
}

@ARTICLE{GB6,
       author = {{Gregory}, P.~C. and {Scott}, W.~K. and {Douglas}, K. and {Condon}, J.~J.},
        title = "{The GB6 Catalog of Radio Sources}",
      journal = {\apjs},
         year = 1996,
        month = apr,
       volume = {103},
        pages = {427},
          doi = {10.1086/192282},
       adsurl = {https://ui.adsabs.harvard.edu/abs/1996ApJS..103..427G}
}

@ARTICLE{PMNS,
       author = {{Wright}, Alan E. and {Griffith}, Mark R. and {Burke}, B.~F. and {Ekers}, R.~D.},
        title = "{The Parkes-MIT-NRAO (PMN) Surveys. II. Source Catalog for the Southern Survey (-87 degrees -4pt.5 < delta < -37 degrees )}",
      journal = {\apjs},
         year = 1994,
        month = mar,
       volume = {91},
        pages = {111},
          doi = {10.1086/191939},
       adsurl = {https://ui.adsabs.harvard.edu/abs/1994ApJS...91..111W}
}

@ARTICLE{PMNT,
       author = {{Griffith}, Mark R. and {Wright}, Alan E. and {Burke}, B.~F. and {Ekers}, R.~D.},
        title = "{The Parkes-MIT-NRAO (PMN) Surveys. III. Source Catalog for the Tropical Survey (-29 degrees < delta < -9 degrees -3pt.5)}",
      journal = {\apjs},
         year = 1994,
        month = jan,
       volume = {90},
        pages = {179},
          doi = {10.1086/191863},
       adsurl = {https://ui.adsabs.harvard.edu/abs/1994ApJS...90..179G}
}

@ARTICLE{PMNE,
       author = {{Griffith}, Mark R. and {Wright}, Alan E. and {Burke}, B.~F. and {Ekers}, R.~D.},
        title = "{The Parkes-MIT-NRAO (PMN) Surveys. VI. Source Catalog for the Equatorial Survey (-9.5 degrees < delta < +10.0 degrees )}",
      journal = {\apjs},
         year = 1995,
        month = apr,
       volume = {97},
        pages = {347},
          doi = {10.1086/192146},
       adsurl = {https://ui.adsabs.harvard.edu/abs/1995ApJS...97..347G}
}

@ARTICLE{NVSS,
       author = {{Condon}, J.~J. and {Cotton}, W.~D. and {Greisen}, E.~W. and {Yin}, Q.~F. and {Perley}, R.~A. and {Taylor}, G.~B. and {Broderick}, J.~J.},
        title = "{The NRAO VLA Sky Survey}",
      journal = {\aj},
         year = 1998,
        month = may,
       volume = {115},
       number = {5},
        pages = {1693-1716},
          doi = {10.1086/300337},
       adsurl = {https://ui.adsabs.harvard.edu/abs/1998AJ....115.1693C}
}

@ARTICLE{SUMSS,
       author = {{Mauch}, T. and {Murphy}, T. and {Buttery}, H.~J. and {Curran}, J. and {Hunstead}, R.~W. and {Piestrzynski}, B. and {Robertson}, J.~G. and {Sadler}, E.~M.},
        title = "{SUMSS: a wide-field radio imaging survey of the southern sky - II. The source catalogue}",
      journal = {\mnras},
         year = 2003,
        month = jul,
       volume = {342},
       number = {4},
        pages = {1117-1130},
          doi = {10.1046/j.1365-8711.2003.06605.x},
archivePrefix = {arXiv},
       eprint = {astro-ph/0303188},
 primaryClass = {astro-ph},
       adsurl = {https://ui.adsabs.harvard.edu/abs/2003MNRAS.342.1117M}
}

@ARTICLE{2013A&A...553A..96D,
       author = {{Delabrouille}, J. and {Betoule}, M. and {Melin}, J. -B. and {Miville-Desch{\^e}nes}, M. -A. and {Gonzalez-Nuevo}, J. and {Le Jeune}, M. and {Castex}, G. and {de Zotti}, G. and {Basak}, S. and {Ashdown}, M. and {Aumont}, J. and {Baccigalupi}, C. and {Banday}, A.~J. and {Bernard}, J. -P. and {Bouchet}, F.~R. and {Clements}, D.~L. and {da Silva}, A. and {Dickinson}, C. and {Dodu}, F. and {Dolag}, K. and {Elsner}, F. and {Fauvet}, L. and {Fa{\"y}}, G. and {Giardino}, G. and {Leach}, S. and {Lesgourgues}, J. and {Liguori}, M. and {Mac{\'\i}as-P{\'e}rez}, J.~F. and {Massardi}, M. and {Matarrese}, S. and {Mazzotta}, P. and {Montier}, L. and {Mottet}, S. and {Paladini}, R. and {Partridge}, B. and {Piffaretti}, R. and {Prezeau}, G. and {Prunet}, S. and {Ricciardi}, S. and {Roman}, M. and {Schaefer}, B. and {Toffolatti}, L.},
        title = "{The pre-launch Planck Sky Model: a model of sky emission at submillimetre to centimetre wavelengths}",
      journal = {\aap},
         year = 2013,
        month = may,
       volume = {553},
          eid = {A96},
        pages = {A96},
          doi = {10.1051/0004-6361/201220019},
archivePrefix = {arXiv},
       eprint = {1207.3675},
 primaryClass = {astro-ph.CO},
       adsurl = {https://ui.adsabs.harvard.edu/abs/2013A&A...553A..96D}
}

@ARTICLE{2005PhRvD..71h3008L,
       author = {{Lewis}, Antony},
        title = "{Lensed CMB simulation and parameter estimation}",
      journal = {\prd},
         year = 2005,
        month = apr,
       volume = {71},
       number = {8},
          eid = {083008},
        pages = {083008},
          doi = {10.1103/PhysRevD.71.083008},
archivePrefix = {arXiv},
       eprint = {astro-ph/0502469},
 primaryClass = {astro-ph},
       adsurl = {https://ui.adsabs.harvard.edu/abs/2005PhRvD..71h3008L}
}

@ARTICLE{2017PhRvD..95d3504A,
       author = {{Alonso}, David and {Dunkley}, Joanna and {Thorne}, Ben and {N{\ae}ss}, Sigurd},
        title = "{Simulated forecasts for primordial B -mode searches in ground-based experiments}",
      journal = {\prd},
         year = 2017,
        month = feb,
       volume = {95},
       number = {4},
          eid = {043504},
        pages = {043504},
          doi = {10.1103/PhysRevD.95.043504},
archivePrefix = {arXiv},
       eprint = {1608.00551},
 primaryClass = {astro-ph.CO},
       adsurl = {https://ui.adsabs.harvard.edu/abs/2017PhRvD..95d3504A}
}

@ARTICLE{2019JCAP...04..046L,
       author = {{Liu}, Hao and {Creswell}, James and {Dachlythra}, Konstantina},
        title = "{Blind correction of the EB-leakage in the pixel domain}",
      journal = {\jcap},
         year = 2019,
        month = apr,
       volume = {2019},
       number = {4},
          eid = {046},
        pages = {046},
          doi = {10.1088/1475-7516/2019/04/046},
archivePrefix = {arXiv},
       eprint = {1904.00451},
 primaryClass = {astro-ph.CO},
       adsurl = {https://ui.adsabs.harvard.edu/abs/2019JCAP...04..046L}
}

@article{Liu:2018kut,
    author = "Liu, Hao and Creswell, James and von Hausegger, Sebastian and Naselsky, Pavel",
    title = "{Methods for pixel domain correction of EB leakage}",
    eprint = "1811.04691",
    archivePrefix = "arXiv",
    primaryClass = "astro-ph.CO",
    doi = "10.1103/PhysRevD.100.023538",
    journal = "Phys. Rev. D",
    volume = "100",
    number = "2",
    pages = "023538",
    year = "2019"
}

@article{Diego-Palazuelos_2021,
doi = {10.1088/1475-7516/2021/03/048},
url = {https://dx.doi.org/10.1088/1475-7516/2021/03/048},
year = {2021},
month = {mar},
publisher = {IOP Publishing},
volume = {2021},
number = {03},
pages = {048},
author = {Diego-Palazuelos, P. and Vielva, P. and Herranz, D.},
title = {Characterization of extragalactic point-sources on E- and B-mode maps of the CMB polarization},
journal = {Journal of Cosmology and Astroparticle Physics}
}

@ARTICLE{2020A&A...642A.232L,
       author = {{Lagache}, G. and {B{\'e}thermin}, M. and {Montier}, L. and {Serra}, P. and {Tucci}, M.},
        title = "{Impact of polarised extragalactic sources on the measurement of CMB B-mode anisotropies}",
      journal = {\aap},
         year = 2020,
        month = oct,
       volume = {642},
          eid = {A232},
        pages = {A232},
          doi = {10.1051/0004-6361/201937147},
archivePrefix = {arXiv},
       eprint = {1911.09466},
 primaryClass = {astro-ph.CO},
       adsurl = {https://ui.adsabs.harvard.edu/abs/2020A&A...642A.232L}
}

@article{WMAP:2012fli,
    author = "Bennett, C. L. and others",
    collaboration = "WMAP",
    title = "{Nine-Year Wilkinson Microwave Anisotropy Probe (WMAP) Observations: Final Maps and Results}",
    eprint = "1212.5225",
    archivePrefix = "arXiv",
    primaryClass = "astro-ph.CO",
    doi = "10.1088/0067-0049/208/2/20",
    journal = "Astrophys. J. Suppl.",
    volume = "208",
    pages = "20",
    year = "2013"
}

@ARTICLE{CORE2018,
       author = {{Remazeilles}, M. and {Banday}, A.~J. and {Baccigalupi}, C. and {Basak}, S. and {Bonaldi}, A. and {De Zotti}, G. and {Delabrouille}, J. and {Dickinson}, C. and {Eriksen}, H.~K. and {Errard}, J. and {Fernandez-Cobos}, R. and {Fuskeland}, U. and {Herv{\'\i}as-Caimapo}, C. and {L{\'o}pez-Caniego}, M. and {Martinez-Gonz{\'a}lez}, E. and {Roman}, M. and {Vielva}, P. and {Wehus}, I. and {Achucarro}, A. and {Ade}, P. and {Allison}, R. and {Ashdown}, M. and {Ballardini}, M. and {Banerji}, R. and {Bartlett}, J. and {Bartolo}, N. and {Baumann}, D. and {Bersanelli}, M. and {Bonato}, M. and {Borrill}, J. and {Bouchet}, F. and {Boulanger}, F. and {Brinckmann}, T. and {Bucher}, M. and {Burigana}, C. and {Buzzelli}, A. and {Cai}, Z. -Y. and {Calvo}, M. and {Carvalho}, C. -S. and {Castellano}, G. and {Challinor}, A. and {Chluba}, J. and {Clesse}, S. and {Colantoni}, I. and {Coppolecchia}, A. and {Crook}, M. and {D'Alessandro}, G. and {de Bernardis}, P. and {de Gasperis}, G. and {Diego}, J. -M. and {Di Valentino}, E. and {Feeney}, S. and {Ferraro}, S. and {Finelli}, F. and {Forastieri}, F. and {Galli}, S. and {Genova-Santos}, R. and {Gerbino}, M. and {Gonz{\'a}lez-Nuevo}, J. and {Grandis}, S. and {Greenslade}, J. and {Hagstotz}, S. and {Hanany}, S. and {Handley}, W. and {Hernandez-Monteagudo}, C. and {Hills}, M. and {Hivon}, E. and {Kiiveri}, K. and {Kisner}, T. and {Kitching}, T. and {Kunz}, M. and {Kurki-Suonio}, H. and {Lamagna}, L. and {Lasenby}, A. and {Lattanzi}, M. and {Lesgourgues}, J. and {Lewis}, A. and {Liguori}, M. and {Lindholm}, V. and {Luzzi}, G. and {Maffei}, B. and {Martins}, C.~J.~A.~P. and {Masi}, S. and {Matarrese}, S. and {McCarthy}, D. and {Melin}, J. -B. and {Melchiorri}, A. and {Molinari}, D. and {Monfardini}, A. and {Natoli}, P. and {Negrello}, M. and {Notari}, A. and {Paiella}, A. and {Paoletti}, D. and {Patanchon}, G. and {Piat}, M. and {Pisano}, G. and {Polastri}, L. and {Polenta}, G. and {Pollo}, A. and {Poulin}, V. and {Quartin}, M. and {Rubino-Martin}, J. -A. and {Salvati}, L. and {Tartari}, A. and {Tomasi}, M. and {Tramonte}, D. and {Trappe}, N. and {Trombetti}, T. and {Tucker}, C. and {Valiviita}, J. and {Van de Weijgaert}, R. and {van Tent}, B. and {Vennin}, V. and {Vittorio}, N. and {Young}, K. and {Zannoni}, M.},
        title = "{Exploring cosmic origins with CORE: B-mode component separation}",
      journal = {\jcap},
         year = 2018,
        month = apr,
       volume = {2018},
       number = {4},
          eid = {023},
        pages = {023},
          doi = {10.1088/1475-7516/2018/04/023},
archivePrefix = {arXiv},
       eprint = {1704.04501},
 primaryClass = {astro-ph.CO},
       adsurl = {https://ui.adsabs.harvard.edu/abs/2018JCAP...04..023R}
}

@article{Zaldarriaga_1998,
doi = {10.1086/305987},
url = {https://dx.doi.org/10.1086/305987},
year = {1998},
month = {aug},
publisher = {},
volume = {503},
number = {1},
pages = {1},
author = {Zaldarriaga, Matias},
title = {Cosmic Microwave Background Polarization Experiments},
journal = {The Astrophysical Journal}
}

@article{Tegmark:2001zv,
    author = "Tegmark, Max and de Oliveira-Costa, Angelica",
    title = "{How to measure CMB polarization power spectra without losing information}",
    eprint = "astro-ph/0012120",
    archivePrefix = "arXiv",
    doi = "10.1103/PhysRevD.64.063001",
    journal = "Phys. Rev. D",
    volume = "64",
    pages = "063001",
    year = "2001"
}

@article{10.1093/mnras/stz093,
    author = {Alonso, David and Sanchez, Javier and Slosar, Anže and LSST Dark Energy Science Collaboration},
    title = {A unified pseudo-Cl framework},
    journal = {Monthly Notices of the Royal Astronomical Society},
    volume = {484},
    number = {3},
    pages = {4127-4151},
    year = {2019},
    month = {01},
    issn = {0035-8711},
    doi = {10.1093/mnras/stz093},
    url = {https://doi.org/10.1093/mnras/stz093},
    eprint = {https://academic.oup.com/mnras/article-pdf/484/3/4127/27747342/stz093.pdf},
}

@article{PhysRevLett.78.2054,
  title = {Signature of Gravity Waves in the Polarization of the Microwave Background},
  author = {Seljak, Uros\ifmmode \breve{}\else \u{}\fi{} and Zaldarriaga, Matias},
  journal = {Phys. Rev. Lett.},
  volume = {78},
  issue = {11},
  pages = {2054--2057},
  numpages = {0},
  year = {1997},
  month = {Mar},
  publisher = {American Physical Society},
  doi = {10.1103/PhysRevLett.78.2054},
  url = {https://link.aps.org/doi/10.1103/PhysRevLett.78.2054}
}

@article{PhysRevLett.78.2058,
  title = {A Probe of Primordial Gravity Waves and Vorticity},
  author = {Kamionkowski, Marc and Kosowsky, Arthur and Stebbins, Albert},
  journal = {Phys. Rev. Lett.},
  volume = {78},
  issue = {11},
  pages = {2058--2061},
  numpages = {0},
  year = {1997},
  month = {Mar},
  publisher = {American Physical Society},
  doi = {10.1103/PhysRevLett.78.2058},
  url = {https://link.aps.org/doi/10.1103/PhysRevLett.78.2058}
}

@ARTICLE{planck18_IV,
       author = {{Planck Collaboration} and {Akrami}, Y. and {Ashdown}, M. and {Aumont}, J. and {Baccigalupi}, C. and {Ballardini}, M. and {Banday}, A.~J. and {Barreiro}, R.~B. and {Bartolo}, N. and {Basak}, S. and {Benabed}, K. and {Bersanelli}, M. and {Bielewicz}, P. and {Bond}, J.~R. and {Borrill}, J. and {Bouchet}, F.~R. and {Boulanger}, F. and {Bucher}, M. and {Burigana}, C. and {Calabrese}, E. and {Cardoso}, J. -F. and {Carron}, J. and {Casaponsa}, B. and {Challinor}, A. and {Colombo}, L.~P.~L. and {Combet}, C. and {Crill}, B.~P. and {Cuttaia}, F. and {de Bernardis}, P. and {de Rosa}, A. and {de Zotti}, G. and {Delabrouille}, J. and {Delouis}, J. -M. and {Di Valentino}, E. and {Dickinson}, C. and {Diego}, J.~M. and {Donzelli}, S. and {Dor{\'e}}, O. and {Ducout}, A. and {Dupac}, X. and {Efstathiou}, G. and {Elsner}, F. and {En{\ss}lin}, T.~A. and {Eriksen}, H.~K. and {Falgarone}, E. and {Fernandez-Cobos}, R. and {Finelli}, F. and {Forastieri}, F. and {Frailis}, M. and {Fraisse}, A.~A. and {Franceschi}, E. and {Frolov}, A. and {Galeotta}, S. and {Galli}, S. and {Ganga}, K. and {G{\'e}nova-Santos}, R.~T. and {Gerbino}, M. and {Ghosh}, T. and {Gonz{\'a}lez-Nuevo}, J. and {G{\'o}rski}, K.~M. and {Gratton}, S. and {Gruppuso}, A. and {Gudmundsson}, J.~E. and {Handley}, W. and {Hansen}, F.~K. and {Helou}, G. and {Herranz}, D. and {Hildebrandt}, S.~R. and {Huang}, Z. and {Jaffe}, A.~H. and {Karakci}, A. and {Keih{\"a}nen}, E. and {Keskitalo}, R. and {Kiiveri}, K. and {Kim}, J. and {Kisner}, T.~S. and {Krachmalnicoff}, N. and {Kunz}, M. and {Kurki-Suonio}, H. and {Lagache}, G. and {Lamarre}, J. -M. and {Lasenby}, A. and {Lattanzi}, M. and {Lawrence}, C.~R. and {Le Jeune}, M. and {Levrier}, F. and {Liguori}, M. and {Lilje}, P.~B. and {Lindholm}, V. and {L{\'o}pez-Caniego}, M. and {Lubin}, P.~M. and {Ma}, Y. -Z. and {Mac{\'\i}as-P{\'e}rez}, J.~F. and {Maggio}, G. and {Maino}, D. and {Mandolesi}, N. and {Mangilli}, A. and {Marcos-Caballero}, A. and {Maris}, M. and {Martin}, P.~G. and {Mart{\'\i}nez-Gonz{\'a}lez}, E. and {Matarrese}, S. and {Mauri}, N. and {McEwen}, J.~D. and {Meinhold}, P.~R. and {Melchiorri}, A. and {Mennella}, A. and {Migliaccio}, M. and {Miville-Desch{\^e}nes}, M. -A. and {Molinari}, D. and {Moneti}, A. and {Montier}, L. and {Morgante}, G. and {Natoli}, P. and {Oppizzi}, F. and {Pagano}, L. and {Paoletti}, D. and {Partridge}, B. and {Peel}, M. and {Pettorino}, V. and {Piacentini}, F. and {Polenta}, G. and {Puget}, J. -L. and {Rachen}, J.~P. and {Reinecke}, M. and {Remazeilles}, M. and {Renzi}, A. and {Rocha}, G. and {Roudier}, G. and {Rubi{\~n}o-Mart{\'\i}n}, J.~A. and {Ruiz-Granados}, B. and {Salvati}, L. and {Sandri}, M. and {Savelainen}, M. and {Scott}, D. and {Seljebotn}, D.~S. and {Sirignano}, C. and {Spencer}, L.~D. and {Suur-Uski}, A. -S. and {Tauber}, J.~A. and {Tavagnacco}, D. and {Tenti}, M. and {Thommesen}, H. and {Toffolatti}, L. and {Tomasi}, M. and {Trombetti}, T. and {Valiviita}, J. and {Van Tent}, B. and {Vielva}, P. and {Villa}, F. and {Vittorio}, N. and {Wandelt}, B.~D. and {Wehus}, I.~K. and {Zacchei}, A. and {Zonca}, A.},
        title = "{Planck 2018 results. IV. Diffuse component separation}",
      journal = {\aap},
         year = 2020,
        month = sep,
       volume = {641},
          eid = {A4},
        pages = {A4},
          doi = {10.1051/0004-6361/201833881},
archivePrefix = {arXiv},
       eprint = {1807.06208},
 primaryClass = {astro-ph.CO},
       adsurl = {https://ui.adsabs.harvard.edu/abs/2020A&A...641A...4P}
}

@ARTICLE{2011ApJS..192...16L,
       author = {{Larson}, D. and {Dunkley}, J. and {Hinshaw}, G. and {Komatsu}, E. and {Nolta}, M.~R. and {Bennett}, C.~L. and {Gold}, B. and {Halpern}, M. and {Hill}, R.~S. and {Jarosik}, N. and {Kogut}, A. and {Limon}, M. and {Meyer}, S.~S. and {Odegard}, N. and {Page}, L. and {Smith}, K.~M. and {Spergel}, D.~N. and {Tucker}, G.~S. and {Weiland}, J.~L. and {Wollack}, E. and {Wright}, E.~L.},
        title = "{Seven-year Wilkinson Microwave Anisotropy Probe (WMAP) Observations: Power Spectra and WMAP-derived Parameters}",
      journal = {\apjs},
         year = 2011,
        month = feb,
       volume = {192},
       number = {2},
          eid = {16},
        pages = {16},
          doi = {10.1088/0067-0049/192/2/16},
archivePrefix = {arXiv},
       eprint = {1001.4635},
 primaryClass = {astro-ph.CO},
       adsurl = {https://ui.adsabs.harvard.edu/abs/2011ApJS..192...16L}
}

@article{Scodeller:2010mp,
    author = "Scodeller, S. and Rudjord, O. and Hansen, F. K. and Marinucci, D. and Geller, D. and Mayeli, A.",
    title = "{Introducing Mexican needlets for CMB analysis: Issues for practical applications and comparison with standard needlets}",
    eprint = "1004.5576",
    archivePrefix = "arXiv",
    primaryClass = "astro-ph.CO",
    doi = "10.1088/0004-637X/733/2/121",
    journal = "Astrophys. J.",
    volume = "733",
    pages = "121",
    year = "2011"
}

@ARTICLE{BICEP_2021,
       author = {{Ade}, P.~A.~R. and {Ahmed}, Z. and {Amiri}, M. and {Barkats}, D. and {Thakur}, R. Basu and {Bischoff}, C.~A. and {Beck}, D. and {Bock}, J.~J. and {Boenish}, H. and {Bullock}, E. and {Buza}, V. and {Cheshire}, J.~R. and {Connors}, J. and {Cornelison}, J. and {Crumrine}, M. and {Cukierman}, A. and {Denison}, E.~V. and {Dierickx}, M. and {Duband}, L. and {Eiben}, M. and {Fatigoni}, S. and {Filippini}, J.~P. and {Fliescher}, S. and {Goeckner-Wald}, N. and {Goldfinger}, D.~C. and {Grayson}, J. and {Grimes}, P. and {Hall}, G. and {Halal}, G. and {Halpern}, M. and {Hand}, E. and {Harrison}, S. and {Henderson}, S. and {Hildebrandt}, S.~R. and {Hilton}, G.~C. and {Hubmayr}, J. and {Hui}, H. and {Irwin}, K.~D. and {Kang}, J. and {Karkare}, K.~S. and {Karpel}, E. and {Kefeli}, S. and {Kernasovskiy}, S.~A. and {Kovac}, J.~M. and {Kuo}, C.~L. and {Lau}, K. and {Leitch}, E.~M. and {Lennox}, A. and {Megerian}, K.~G. and {Minutolo}, L. and {Moncelsi}, L. and {Nakato}, Y. and {Namikawa}, T. and {Nguyen}, H.~T. and {O'Brient}, R. and {Ogburn}, R.~W. and {Palladino}, S. and {Prouve}, T. and {Pryke}, C. and {Racine}, B. and {Reintsema}, C.~D. and {Richter}, S. and {Schillaci}, A. and {Schwarz}, R. and {Schmitt}, B.~L. and {Sheehy}, C.~D. and {Soliman}, A. and {Germaine}, T. St. and {Steinbach}, B. and {Sudiwala}, R.~V. and {Teply}, G.~P. and {Thompson}, K.~L. and {Tolan}, J.~E. and {Tucker}, C. and {Turner}, A.~D. and {Umilt{\`a}}, C. and {Verg{\`e}s}, C. and {Vieregg}, A.~G. and {Wandui}, A. and {Weber}, A.~C. and {Wiebe}, D.~V. and {Willmert}, J. and {Wong}, C.~L. and {Wu}, W.~L.~K. and {Yang}, H. and {Yoon}, K.~W. and {Young}, E. and {Yu}, C. and {Zeng}, L. and {Zhang}, C. and {Zhang}, S. and {Bicep/Keck Collaboration}},
        title = "{Improved Constraints on Primordial Gravitational Waves using Planck, WMAP, and BICEP/Keck Observations through the 2018 Observing Season}",
      journal = {\prl},
         year = 2021,
        month = oct,
       volume = {127},
       number = {15},
          eid = {151301},
        pages = {151301},
          doi = {10.1103/PhysRevLett.127.151301},
archivePrefix = {arXiv},
       eprint = {2110.00483},
 primaryClass = {astro-ph.CO},
       adsurl = {https://ui.adsabs.harvard.edu/abs/2021PhRvL.127o1301A}
}

@ARTICLE{2013MNRAS.432..728C,
       author = {{Curto}, A. and {Tucci}, M. and {Gonz{\'a}lez-Nuevo}, J. and {Toffolatti}, L. and {Mart{\'\i}nez-Gonz{\'a}lez}, E. and {Arg{\"u}eso}, F. and {Lapi}, A. and {L{\'o}pez-Caniego}, M.},
        title = "{Forecasts on the contamination induced by unresolved point sources in primordial non-Gaussianity beyond Planck}",
      journal = {\mnras},
         year = 2013,
        month = jun,
       volume = {432},
       number = {1},
        pages = {728-742},
          doi = {10.1093/mnras/stt511},
archivePrefix = {arXiv},
       eprint = {1301.1544},
 primaryClass = {astro-ph.CO},
       adsurl = {https://ui.adsabs.harvard.edu/abs/2013MNRAS.432..728C}
}

@ARTICLE{2012MNRAS.424.1694B,
       author = {{Bucher}, Martin and {Louis}, Thibaut},
        title = "{Filling in cosmic microwave background map missing data using constrained Gaussian realizations}",
      journal = {\mnras},
         year = 2012,
        month = aug,
       volume = {424},
       number = {3},
        pages = {1694-1713},
          doi = {10.1111/j.1365-2966.2012.21138.x},
archivePrefix = {arXiv},
       eprint = {1109.0286},
 primaryClass = {astro-ph.CO},
       adsurl = {https://ui.adsabs.harvard.edu/abs/2012MNRAS.424.1694B}
}

@ARTICLE{2011MNRAS.410.2481R,
       author = {{Remazeilles}, Mathieu and {Delabrouille}, Jacques and {Cardoso}, Jean-Fran{\c{c}}ois},
        title = "{CMB and SZ effect separation with constrained Internal Linear Combinations}",
      journal = {\mnras},
         year = 2011,
        month = feb,
       volume = {410},
       number = {4},
        pages = {2481-2487},
          doi = {10.1111/j.1365-2966.2010.17624.x},
archivePrefix = {arXiv},
       eprint = {1006.5599},
 primaryClass = {astro-ph.CO},
       adsurl = {https://ui.adsabs.harvard.edu/abs/2011MNRAS.410.2481R}
}

@ARTICLE{2020JCAP...12..045C,
       author = {{Choi}, Steve K. and {Hasselfield}, Matthew and {Ho}, Shuay-Pwu Patty and {Koopman}, Brian and {Lungu}, Marius and {Abitbol}, Maximilian H. and {Addison}, Graeme E. and {Ade}, Peter A.~R. and {Aiola}, Simone and {Alonso}, David and {Amiri}, Mandana and {Amodeo}, Stefania and {Angile}, Elio and {Austermann}, Jason E. and {Baildon}, Taylor and {Battaglia}, Nick and {Beall}, James A. and {Bean}, Rachel and {Becker}, Daniel T. and {Bond}, J. Richard and {Bruno}, Sarah Marie and {Calabrese}, Erminia and {Calafut}, Victoria and {Campusano}, Luis E. and {Carrero}, Felipe and {Chesmore}, Grace E. and {Cho}, Hsiao-mei and {Clark}, Susan E. and {Cothard}, Nicholas F. and {Crichton}, Devin and {Crowley}, Kevin T. and {Darwish}, Omar and {Datta}, Rahul and {Denison}, Edward V. and {Devlin}, Mark J. and {Duell}, Cody J. and {Duff}, Shannon M. and {Duivenvoorden}, Adriaan J. and {Dunkley}, Jo and {D{\"u}nner}, Rolando and {Essinger-Hileman}, Thomas and {Fankhanel}, Max and {Ferraro}, Simone and {Fox}, Anna E. and {Fuzia}, Brittany and {Gallardo}, Patricio A. and {Gluscevic}, Vera and {Golec}, Joseph E. and {Grace}, Emily and {Gralla}, Megan and {Guan}, Yilun and {Hall}, Kirsten and {Halpern}, Mark and {Han}, Dongwon and {Hargrave}, Peter and {Henderson}, Shawn and {Hensley}, Brandon and {Hill}, J. Colin and {Hilton}, Gene C. and {Hilton}, Matt and {Hincks}, Adam D. and {Hlo{\v{z}}ek}, Ren{\'e}e and {Hubmayr}, Johannes and {Huffenberger}, Kevin M. and {Hughes}, John P. and {Infante}, Leopoldo and {Irwin}, Kent and {Jackson}, Rebecca and {Klein}, Jeff and {Knowles}, Kenda and {Kosowsky}, Arthur and {Lakey}, Vincent and {Li}, Dale and {Li}, Yaqiong and {Li}, Zack and {Lokken}, Martine and {Louis}, Thibaut and {MacInnis}, Amanda and {Madhavacheril}, Mathew and {Maldonado}, Felipe and {Mallaby-Kay}, Maya and {Marsden}, Danica and {Maurin}, Lo{\"\i}c and {McMahon}, Jeff and {Menanteau}, Felipe and {Moodley}, Kavilan and {Morton}, Tim and {Naess}, Sigurd and {Namikawa}, Toshiya and {Nati}, Federico and {Newburgh}, Laura and {Nibarger}, John P. and {Nicola}, Andrina and {Niemack}, Michael D. and {Nolta}, Michael R. and {Orlowski-Sherer}, John and {Page}, Lyman A. and {Pappas}, Christine G. and {Partridge}, Bruce and {Phakathi}, Phumlani and {Prince}, Heather and {Puddu}, Roberto and {Qu}, Frank J. and {Rivera}, Jesus and {Robertson}, Naomi and {Rojas}, Felipe and {Salatino}, Maria and {Schaan}, Emmanuel and {Schillaci}, Alessandro and {Schmitt}, Benjamin L. and {Sehgal}, Neelima and {Sherwin}, Blake D. and {Sierra}, Carlos and {Sievers}, Jon and {Sifon}, Cristobal and {Sikhosana}, Precious and {Simon}, Sara and {Spergel}, David N. and {Staggs}, Suzanne T. and {Stevens}, Jason and {Storer}, Emilie and {Sunder}, Dhaneshwar D. and {Switzer}, Eric R. and {Thorne}, Ben and {Thornton}, Robert and {Trac}, Hy and {Treu}, Jesse and {Tucker}, Carole and {Vale}, Leila R. and {Van Engelen}, Alexander and {Van Lanen}, Jeff and {Vavagiakis}, Eve M. and {Wagoner}, Kasey and {Wang}, Yuhan and {Ward}, Jonathan T. and {Wollack}, Edward J. and {Xu}, Zhilei and {Zago}, Fernando and {Zhu}, Ningfeng},
        title = "{The Atacama Cosmology Telescope: a measurement of the Cosmic Microwave Background power spectra at 98 and 150 GHz}",
      journal = {\jcap},
         year = 2020,
        month = dec,
       volume = {2020},
       number = {12},
          eid = {045},
        pages = {045},
          doi = {10.1088/1475-7516/2020/12/045},
archivePrefix = {arXiv},
       eprint = {2007.07289},
 primaryClass = {astro-ph.CO},
       adsurl = {https://ui.adsabs.harvard.edu/abs/2020JCAP...12..045C}
}

@article{Sureau:2014lea,
    author = "Sureau, F. C. and Starck, J. -L. and Bobin, J. and Paykari, P. and Rassat, A.",
    title = "{Sparse point-source removal for full-sky CMB experiments: application to WMAP 9-year data}",
    eprint = "1405.5482",
    archivePrefix = "arXiv",
    primaryClass = "astro-ph.IM",
    doi = "10.1051/0004-6361/201322706",
    journal = "Astron. Astrophys.",
    volume = "566",
    pages = "A100",
    year = "2014"
}

@ARTICLE{2016MNRAS.458.2032R,
       author = {{Remazeilles}, M. and {Dickinson}, C. and {Eriksen}, H.~K.~K. and {Wehus}, I.~K.},
        title = "{Sensitivity and foreground modelling for large-scale cosmic microwave background B-mode polarization satellite missions}",
      journal = {\mnras},
         year = 2016,
        month = may,
       volume = {458},
       number = {2},
        pages = {2032-2050},
          doi = {10.1093/mnras/stw441},
archivePrefix = {arXiv},
       eprint = {1509.04714},
 primaryClass = {astro-ph.CO},
       adsurl = {https://ui.adsabs.harvard.edu/abs/2016MNRAS.458.2032R}
}

@ARTICLE{2018MNRAS.479.5577P,
       author = {{Philcox}, Oliver H.~E. and {Sherwin}, Blake D. and {van Engelen}, Alexander},
        title = "{Detection and removal of B-mode dust foregrounds with signatures of statistical anisotropy}",
      journal = {\mnras},
         year = 2018,
        month = oct,
       volume = {479},
       number = {4},
        pages = {5577-5595},
          doi = {10.1093/mnras/sty1769},
archivePrefix = {arXiv},
       eprint = {1805.09177},
 primaryClass = {astro-ph.CO},
       adsurl = {https://ui.adsabs.harvard.edu/abs/2018MNRAS.479.5577P}
}

@ARTICLE{2005ApJ...622..759G,
   author = {{G{\'o}rski}, K.~M. and {Hivon}, E. and {Banday}, A.~J. and 
	{Wandelt}, B.~D. and {Hansen}, F.~K. and {Reinecke}, M. and 
	{Bartelmann}, M.},
    title = "{HEALPix: A Framework for High-Resolution Discretization and Fast Analysis of Data Distributed on the Sphere}",
  journal = {\apj},
   eprint = {arXiv:astro-ph/0409513},
     year = 2005,
    month = apr,
   volume = 622,
    pages = {759-771},
      doi = {10.1086/427976},
   adsurl = {http://adsabs.harvard.edu/abs/2005ApJ...622..759G}
}

@article{Zonca2019,
  doi = {10.21105/joss.01298},
  url = {https://doi.org/10.21105/joss.01298},
  year = {2019},
  month = mar,
  publisher = {The Open Journal},
  volume = {4},
  number = {35},
  pages = {1298},
  author = {Andrea Zonca and Leo Singer and Daniel Lenz and Martin Reinecke and Cyrille Rosset and Eric Hivon and Krzysztof Gorski},
  title = {healpy: equal area pixelization and spherical harmonics transforms for data on the sphere in Python},
  journal = {Journal of Open Source Software}
}

@misc{iminuit,
	author = {Piti Ongmongkolkul (@piti118) and Christoph Deil (@cdeil) and Hans Dembinski (@HDembinski) and David Men\'endez Hurtado (@Dapid) and Chris Burr (@chrisburr) and Andrew ZP Smith (@energynumbers) and Fabian Rost (@fabianrost84) and Alex Pearce (@alexpearce) and Lukas Geiger (@lgeiger) and Omar Zapata (@omazapa)},
	title = {iminuit - A Python interface to MINUIT},
	year = {2012--},
	url = {https://github.com/iminuit/iminuit},
	note = {[Online; accessed 2018.03.05]}
}

@article{SimonsObservatory:2018koc,
    author = "Ade, Peter and others",
    collaboration = "Simons Observatory",
    title = "{The Simons Observatory: Science goals and forecasts}",
    eprint = "1808.07445",
    archivePrefix = "arXiv",
    primaryClass = "astro-ph.CO",
    doi = "10.1088/1475-7516/2019/02/056",
    journal = "JCAP",
    volume = "02",
    pages = "056",
    year = "2019"
}

@article{Lyth:1998xn,
    author = "Lyth, David H. and Riotto, Antonio",
    title = "{Particle physics models of inflation and the cosmological density perturbation}",
    eprint = "hep-ph/9807278",
    archivePrefix = "arXiv",
    reportNumber = "LANCS-TH-9720, FERMILAB-PUB-97-292-A, CERN-TH-97-383, OUTP-98-39-P",
    doi = "10.1016/S0370-1573(98)00128-8",
    journal = "Phys. Rept.",
    volume = "314",
    pages = "1--146",
    year = "1999"
}

@article{BICEP2:2014owc,
    author = "Ade, P. A. R. and others",
    collaboration = "BICEP2",
    title = "{Detection of $B$-Mode Polarization at Degree Angular Scales by BICEP2}",
    eprint = "1403.3985",
    archivePrefix = "arXiv",
    primaryClass = "astro-ph.CO",
    doi = "10.1103/PhysRevLett.112.241101",
    journal = "Phys. Rev. Lett.",
    volume = "112",
    number = "24",
    pages = "241101",
    year = "2014"
}

@article{Planck:2013qym,
    author = "Ade, P. A. R. and others",
    collaboration = "Planck",
    title = "{Planck 2013 results. XXVIII. The Planck Catalogue of Compact Sources}",
    eprint = "1303.5088",
    archivePrefix = "arXiv",
    primaryClass = "astro-ph.CO",
    reportNumber = "CERN-PH-TH-2013-140",
    doi = "10.1051/0004-6361/201321524",
    journal = "Astron. Astrophys.",
    volume = "571",
    pages = "A28",
    year = "2014"
}

@ARTICLE{1140406,
  author={Ludwig, A.},
  journal={IEEE Transactions on Antennas and Propagation}, 
  title={The definition of cross polarization}, 
  year={1973},
  volume={21},
  number={1},
  pages={116-119},
  doi={10.1109/TAP.1973.1140406}}

@article{Carretti:2004zq,
    author = "Carretti, Ettore and Cortiglioni, Stefano and Sbarra, Carla and Tascone, Riccardo",
    title = "{Antenna instrumental polarization and its effects on E- and B-modes for CMBP observations}",
    eprint = "astro-ph/0403493",
    archivePrefix = "arXiv",
    doi = "10.1051/0004-6361:20035601",
    journal = "Astron. Astrophys.",
    volume = "420",
    pages = "437--445",
    year = "2004"
}

@article{Hivon:2016qyw,
    author = "Hivon, Eric and Mottet, Sylvain and Ponthieu, Nicolas",
    title = "{QuickPol: Fast calculation of effective beam matrices for CMB polarization}",
    eprint = "1608.08833",
    archivePrefix = "arXiv",
    primaryClass = "astro-ph.CO",
    doi = "10.1051/0004-6361/201629626",
    journal = "Astron. Astrophys.",
    volume = "598",
    pages = "A25",
    year = "2017"
}

@article{LiteBIRD:2020khw,
    author = "Hazumi, M. and others",
    collaboration = "LiteBIRD",
    title = "{LiteBIRD: JAXA's new strategic L-class mission for all-sky surveys of cosmic microwave background polarization}",
    eprint = "2101.12449",
    archivePrefix = "arXiv",
    primaryClass = "astro-ph.IM",
    doi = "10.1117/12.2563050",
    journal = "Proc. SPIE Int. Soc. Opt. Eng.",
    volume = "11443",
    pages = "114432F",
    year = "2020"
}

@article{osborne2014extragalactic,
  title={Extragalactic foreground contamination in temperature-based CMB lens reconstruction},
  author={Osborne, Stephen J and Hanson, Duncan and Dor{\'e}, Olivier},
  journal={Journal of Cosmology and Astroparticle Physics},
  volume={2014},
  number={03},
  pages={024},
  year={2014},
  publisher={IOP Publishing}
}

@article{sailer2023foreground,
  title={Foreground-immune CMB lensing reconstruction with polarization},
  author={Sailer, Noah and Ferraro, Simone and Schaan, Emmanuel},
  journal={Physical Review D},
  volume={107},
  number={2},
  pages={023504},
  year={2023},
  publisher={APS}
}

@article{Mitra:2010rt,
    author = "Mitra, S. and Rocha, G. and Gorski, K. M. and Huffenberger, K. M. and Eriksen, H. K. and Ashdown, M. A. J. and Lawrence, C. R.",
    title = "{Fast Pixel Space Convolution for CMB Surveys with Asymmetric Beams and Complex Scan Strategies: FEBeCoP}",
    eprint = "1005.1929",
    archivePrefix = "arXiv",
    primaryClass = "astro-ph.CO",
    doi = "10.1088/0067-0049/193/1/5",
    journal = "Astrophys. J. Suppl.",
    volume = "193",
    pages = "5",
    year = "2011"
}

@article{ACT:2025xdm,
    author = "Naess, Sigurd and others",
    collaboration = "ACT",
    title = "{The Atacama Cosmology Telescope: DR6 Maps}",
    eprint = "2503.14451",
    archivePrefix = "arXiv",
    primaryClass = "astro-ph.CO",
    reportNumber = "FERMILAB-PUB-25-0160-PPD",
    month = "3",
    year = "2025"
}

@article{Naess:2019nrw,
    author = "N{\ae}ss, Sigurd K.",
    title = "{How to avoid X'es around point sources in maximum likelihood CMB maps}",
    eprint = "1906.08030",
    archivePrefix = "arXiv",
    primaryClass = "astro-ph.IM",
    doi = "10.1088/1475-7516/2019/12/060",
    journal = "JCAP",
    volume = "12",
    pages = "060",
    year = "2019"
}

@article{SPT:2023jql,
    author = "Pan, Z. and others",
    collaboration = "SPT",
    title = "{Measurement of gravitational lensing of the cosmic microwave background using SPT-3G 2018 data}",
    eprint = "2308.11608",
    archivePrefix = "arXiv",
    primaryClass = "astro-ph.CO",
    doi = "10.1103/PhysRevD.108.122005",
    journal = "Phys. Rev. D",
    volume = "108",
    number = "12",
    pages = "122005",
    year = "2023"
}

@article{CLASS:2020zyk,
    author = "Harrington, Kathleen and others",
    collaboration = "CLASS",
    title = "{Two Year Cosmology Large Angular Scale Surveyor (CLASS) Observations: Long Timescale Stability Achieved with a Front-end Variable-delay Polarization Modulator at 40 GHz}",
    eprint = "2101.00034",
    archivePrefix = "arXiv",
    primaryClass = "astro-ph.IM",
    doi = "10.3847/1538-4357/ac2235",
    journal = "Astrophys. J.",
    volume = "922",
    number = "2",
    pages = "212",
    year = "2021"
}

@article{Rosenberg:2024geh,
    author = "Rosenberg, Erik and Gratton, Steven and Challinor, Anthony",
    title = "{Filtering in CMB data analysis with application to ACT DR4 and Planck observations}",
    eprint = "2412.13995",
    archivePrefix = "arXiv",
    primaryClass = "astro-ph.CO",
    doi = "10.1103/rvhm-fsdc",
    journal = "Phys. Rev. D",
    volume = "112",
    number = "2",
    pages = "023534",
    year = "2025"
}

@article{Takakura:2017ddx,
    author = "Takakura, Satoru and others",
    title = "{Performance of a continuously rotating half-wave plate on the POLARBEAR telescope}",
    eprint = "1702.07111",
    archivePrefix = "arXiv",
    primaryClass = "astro-ph.IM",
    doi = "10.1088/1475-7516/2017/05/008",
    journal = "JCAP",
    volume = "05",
    pages = "008",
    year = "2017"
}

@article{Tucci:2004zy,
    author = "Tucci, Marco and Martinez-Gonzalez, E. and Vielva, P. and Delabrouille, J.",
    title = "{Limits on the detectability of the CMB B-mode polarization imposed by foregrounds}",
    eprint = "astro-ph/0411567",
    archivePrefix = "arXiv",
    doi = "10.1111/j.1365-2966.2005.09123.x",
    journal = "Mon. Not. Roy. Astron. Soc.",
    volume = "360",
    pages = "935--949",
    year = "2005"
}

@article{Puglisi:2017lpn,
    author = "Puglisi, G. and Galluzzi, V. and Bonavera, L. and Gonzalez-Nuevo, J. and Lapi, A. and Massardi, M. and Perrotta, F. and Baccigalupi, C. and Celotti, A. and Danese, L.",
    title = "{Forecasting the Contribution of Polarized Extragalactic Radio Sources in CMB Observations}",
    eprint = "1712.09639",
    archivePrefix = "arXiv",
    primaryClass = "astro-ph.CO",
    doi = "10.3847/1538-4357/aab3c7",
    journal = "Astrophys. J.",
    volume = "858",
    number = "2",
    pages = "85",
    year = "2018"
}

@article{Trombetti:2017kim,
    author = "Trombetti, T. and Burigana, C. and De Zotti, G. and Galluzzi, V. and Massardi, M.",
    title = "{Average fractional polarization of extragalactic sources at Planck frequencies}",
    eprint = "1712.08412",
    archivePrefix = "arXiv",
    primaryClass = "astro-ph.CO",
    doi = "10.1051/0004-6361/201732342",
    journal = "Astron. Astrophys.",
    volume = "618",
    pages = "A29",
    year = "2018"
}

@article{BICEP2:2019upn,
    author = "Ade, P. A. R. and others",
    collaboration = "BICEP2, Keck Array",
    title = "{BICEP2 / Keck Array XI: Beam Characterization and Temperature-to-Polarization Leakage in the BK15 Dataset}",
    eprint = "1904.01640",
    archivePrefix = "arXiv",
    primaryClass = "astro-ph.IM",
    doi = "10.3847/1538-4357/ab391d",
    journal = "Astrophys. J.",
    volume = "884",
    pages = "114",
    year = "2019"
}
\bibliographystyle{JHEP} 

\appendix
\section{Instrumental systematics and atmospheric filtering relevant for GPSF}\label{app:systematics}
In realistic CMB observations, uncertainties in the beam response (such as ellipticity and sidelobes), instrumental polarization leakage (including cross-polar terms), and the filtering of atmospheric fluctuations can all modify point-source response. We summarize their treatment and the relevance for GPSF below.

{\it (i) Elliptical beam shape.} 
Beam ellipticity and its spatial variation can be introduced by the scan strategy. Nevertheless, it is generally adequate to use an \textit{effective beam} evaluated at each sky location when performing source fitting~\cite{Planck:2015bin}. For realistic experiments such as \textit{Planck}, the difference between maps convolved with the full asymmetric beams (via Level-S simulations) and those using the FEBeCoP effective beams is extremely small: the fractional RMS discrepancy is only a few $\times\, 0.01\%$ in temperature and a few $\times\, 0.1\%$ in polarization~\cite{Mitra:2010rt}. Therefore, in practical applications, adopting the effective beam provides an accurate approximation to the local PSF on the sky and allows point sources to be reliably fitted.

{\it (ii) Sidelobes.}    
Sidelobes, including both near and far components, can bias point-source fitting if they are not properly characterized and mitigated.
Far sidelobes are caused by oblique reflections off flat surfaces surrounding the main mirror. In ACT~\cite{ACT:2025xdm}, these sidelobes are measured using Moon-centered and Sun-centered maps, and the Sun-centered measurements show that the most important sidelobes reach a few mK when sourced by the 6000 K Sun, corresponding to a 60--70 dB suppression compared to the main beam. The ACT analysis states that, since they are so weak, far sidelobes can be ignored unless something extremely bright enters them.
In BICEP~\cite{BICEP2:2019upn}, measurements using an amplified microwave source show that maps made with and without the forebaffle exhibit no perceptible difference in the main-beam region. This is consistent with the small amount of power intercepted by the forebaffle ($\sim 0.7\%$).
Given these results, we therefore do not explicitly include far sidelobes in the current GPSF treatment.

In contrast, the near-sidelobe response redistributes a small fraction of the main-beam flux over extended angular scales, making the apparent source profiles slightly broader and asymmetric, which can lead to small but systematic biases even when fitting with an effective beam model.

To mitigate this effect, we can adopt the two-step approximation approach used in the ACT pipeline~\cite{ACT:2025xdm}.  
Instead of including the sidelobe term explicitly in the mapmaking operator, its contribution is removed from the time-ordered data (TOD) through
\begin{align}
d &= d_{\rm raw} - P_{\rm side} m_{\rm guess},
\end{align}
where \(P_{\rm side}\) is the sidelobe pointing operator and \(m_{\rm guess}\) is the sky estimate obtained from the previous mapmaking pass.  
Since the near-sidelobe amplitude is only an \(O(10^{-3})\) perturbation relative to the main beam, this subtraction efficiently suppresses the leading-order contamination without the need for costly full deconvolution.

Then, \(P_{\rm side}\) is approximated as a sum of a small number (typically about a dozen) of scaled and offset copies of the main beam, each representing one near-sidelobe lobe.  
This compact representation reduces the computational cost to less than 1\% of the total mapmaking expense while effectively removing the dominant sidelobe signal.  
After this correction, the resulting maps can be approximately treated as observations through the main beam only, effectively minimizing the impact of sidelobe leakage on subsequent point-source fits (e.g., using GPSF).

{\it (iii) Instrumental leakage and atmospheric filtering.} 
Non-idealities in the instrument’s polarization response—such as cross-polar leakage, beam mismatch, and other systematic effects—can cause a fraction of the total-intensity signal to leak into the $Q/U$ channels (T-P leakage). This is particularly relevant for atmospheric emission: although the atmosphere is intrinsically unpolarized, its large temperature fluctuations make it a dominant source of intensity contamination. Through T-P leakage, a small portion of this strong atmospheric signal is transferred into the polarization time streams.

Consequently, the polarization time-ordered data still exhibit pronounced low-frequency ($1/f$) noise tails. Filtering the $Q/U$ time streams is therefore required to suppress these residual low-frequency components. However, such filtering not only removes large-scale sky modes but can also introduce cross-shaped artifacts around bright sources, depending on the scan strategy~\cite{Naess:2019nrw, Rosenberg:2024geh}. In addition, filtering may distort the intrinsic point-source profile, potentially biasing the subsequent GPSF fitting.



We perform a simplified simulation to quantify and compare how the recovered point-source amplitude changes after applying the filtering procedure, which directly affects the effectiveness of the subsequent point-source handling pipeline (e.g., GPSF). The simulation includes four components: a point source, the CMB, white noise, and $1/f$ noise. Among all observing frequencies, we choose the 30\,GHz channel for this test, since its large beam size makes it particularly sensitive to atmospheric fluctuations and low-frequency contamination. This channel therefore represents the most challenging case for evaluating the impact of atmospheric filtering on point-source recovery. The detailed setup is as follows:

1. \textbf{Point source.}  
A single Gaussian point source with a flux density of 303.4\,mJy was injected into the Q map, corresponding to a temperature amplitude of $26.84\,\mu{\rm K}$ in CMB units. For reference, typical polarized sources within our survey region exhibit flux densities of only a few tens of mJy. We intentionally adopted a higher flux value to examine how the algorithm performs under stronger signal conditions, particularly in handling the emergence of the characteristic cross-shaped structure and its impact on point-source fitting accuracy.

2. \textbf{CMB.}  
The CMB realization is smoothed with a Gaussian beam of FWHM $67'$, matching the angular resolution of the 30\,GHz channel.

3. \textbf{White noise.}  
The white noise level is set to be consistent with the instrumental noise at 30\,GHz, ensuring that the simulation reflects realistic instrument performance.

4. \textbf{$1/f$ noise.}  
The $1/f$ noise amplitude is calibrated from the intensity (temperature) channel, assuming a T--P leakage fraction of $1/100$. Under this assumption, the effective knee frequency in polarization is $f_{\mathrm{knee}} \approx 0.1$\,Hz. The spectral index is set to $\alpha = -1.6$, following Ref.~\cite{CLASS:2020zyk}.

\begin{figure}[htbp]
  \centering
  \includegraphics[width=0.31\linewidth]{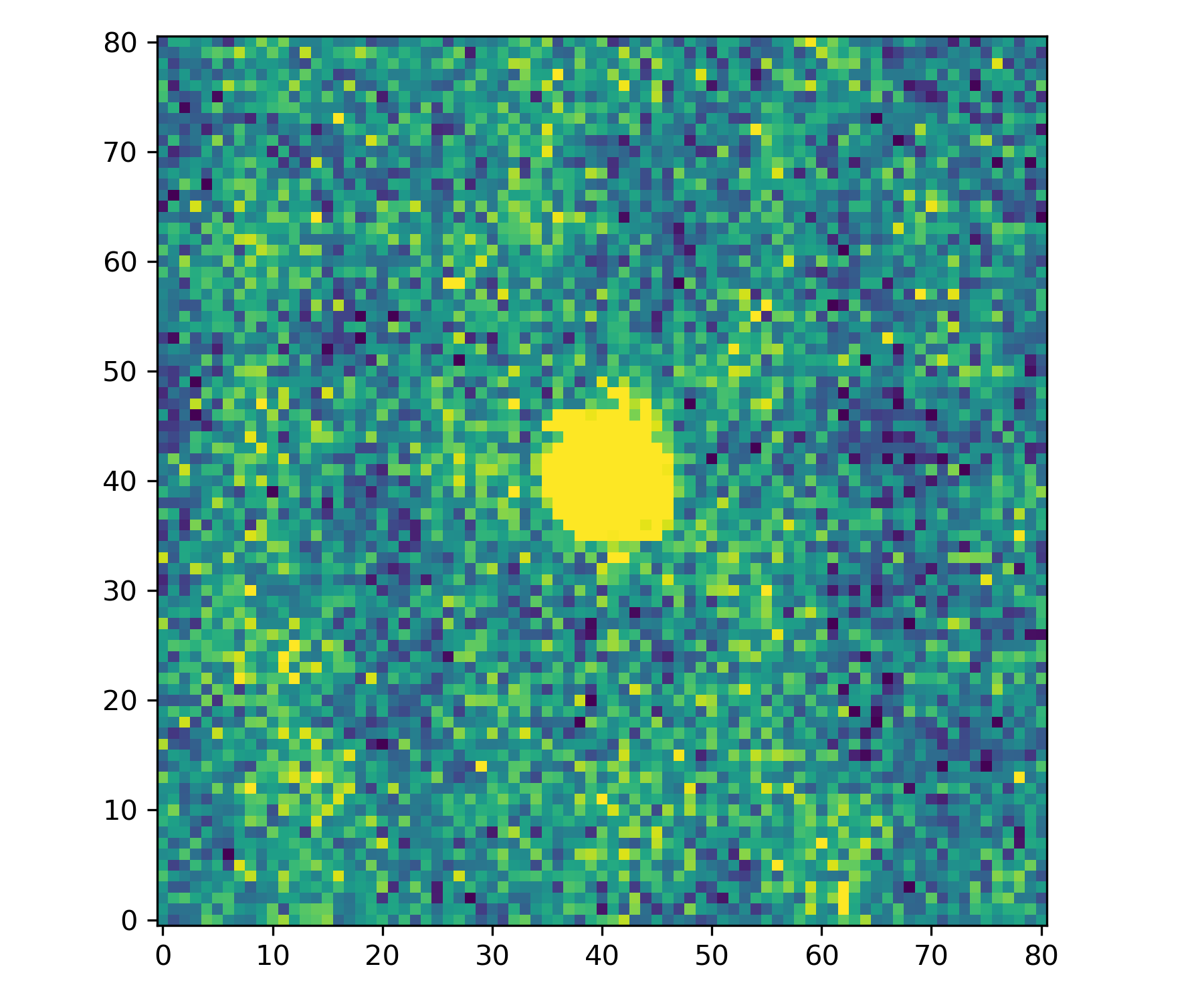}
  \includegraphics[width=0.31\linewidth]{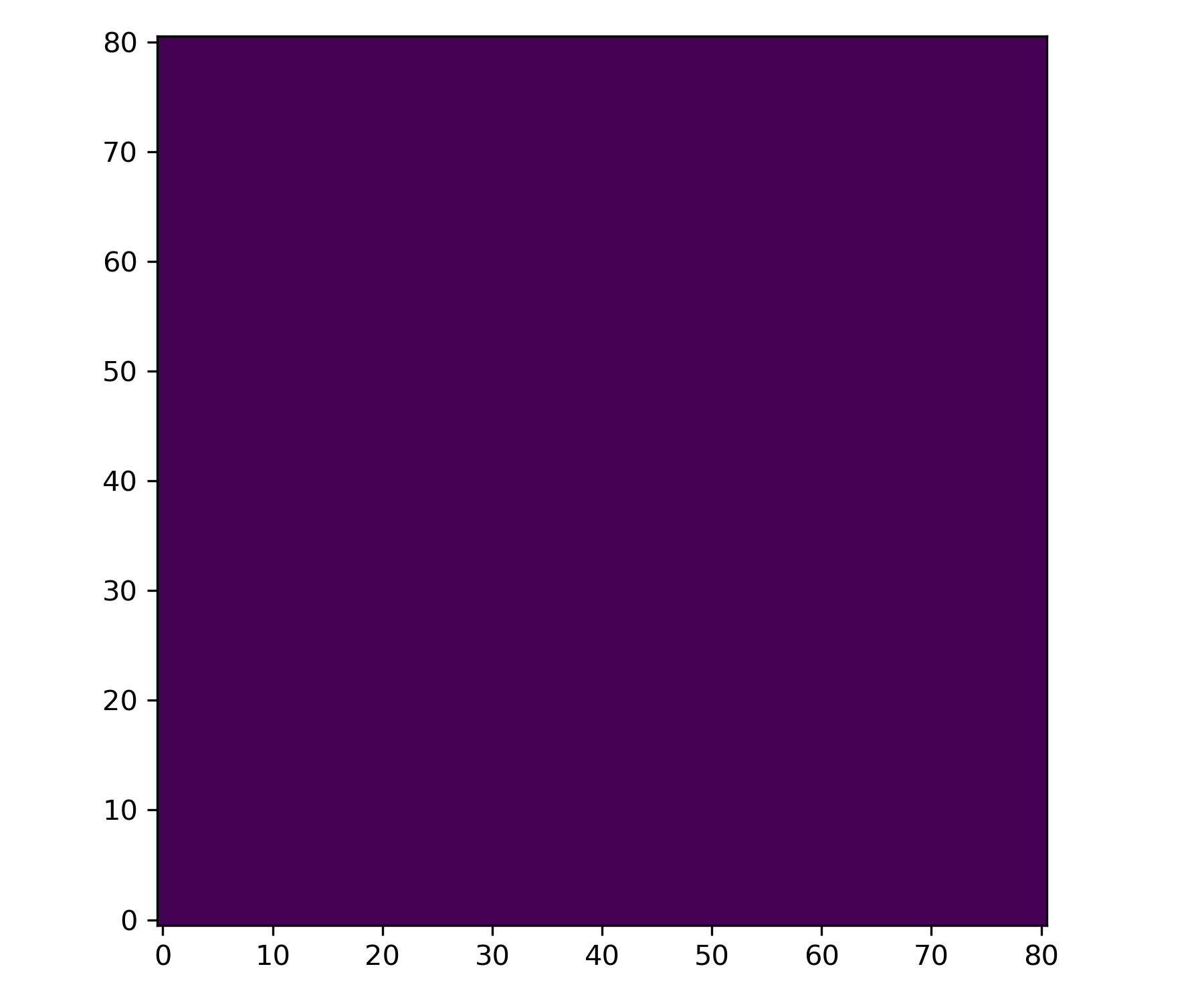}
  \includegraphics[width=0.31\linewidth]{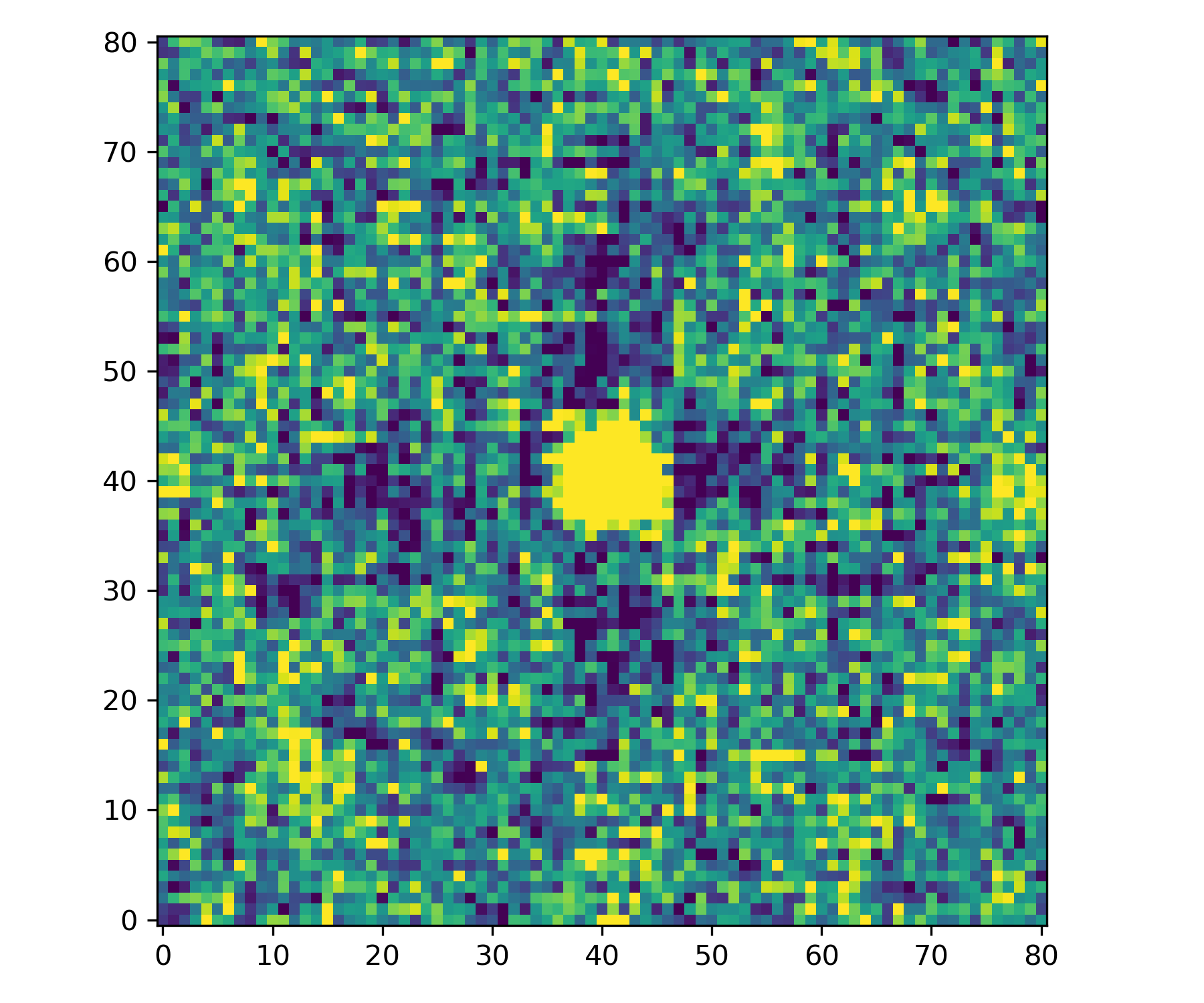}

  \vspace{3pt}
  \parbox[c]{0.31\linewidth}{\centering (a) Source + CMB \\+ White noise}%
  \parbox[c]{0.31\linewidth}{\centering (b) Source + CMB \\+ 1/f noise}%
  \parbox[c]{0.31\linewidth}{\centering (c) Polynomial filter\\(no mask)}%

  \vspace{6pt}

  \includegraphics[width=0.31\linewidth]{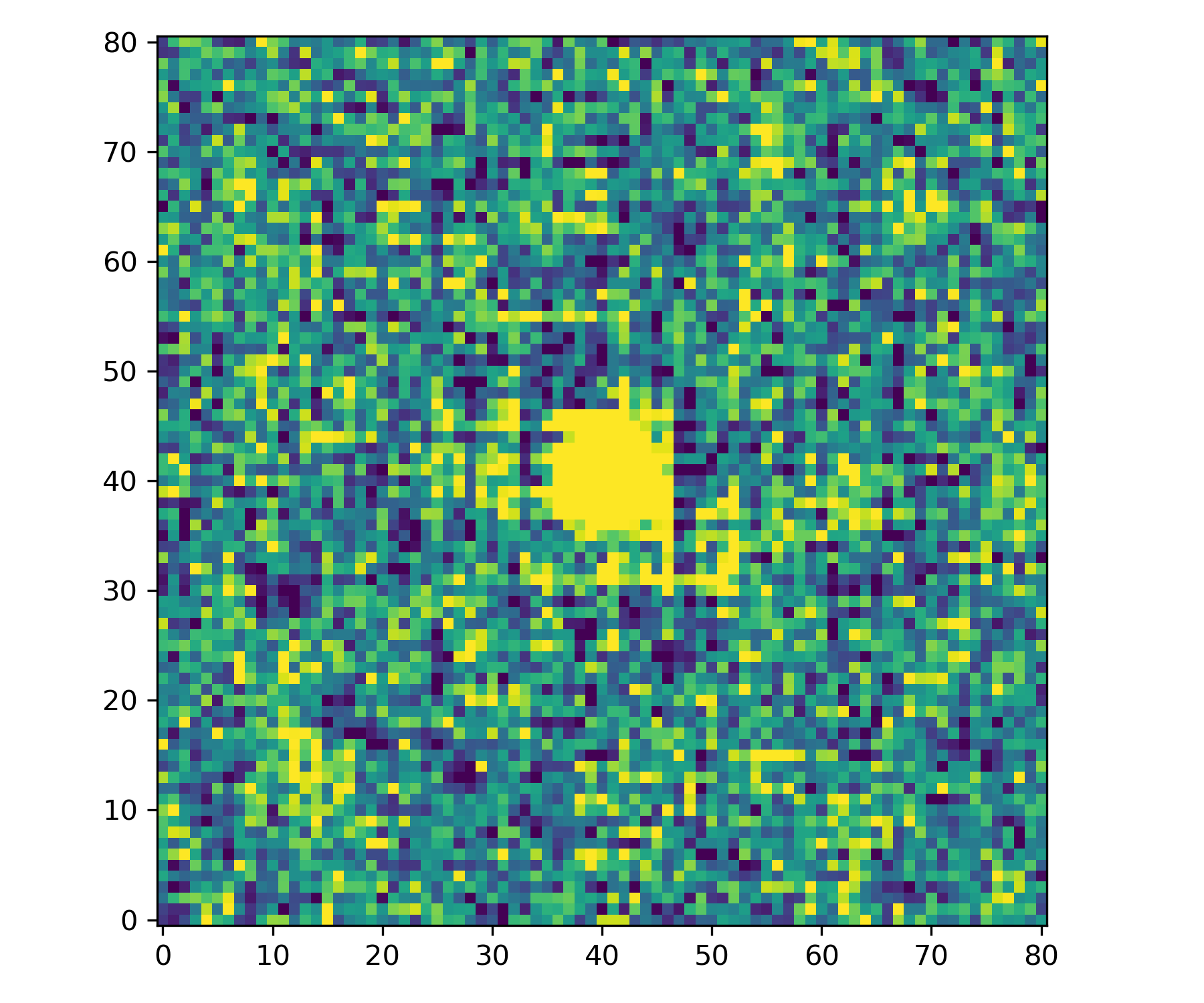}
  \includegraphics[width=0.31\linewidth]{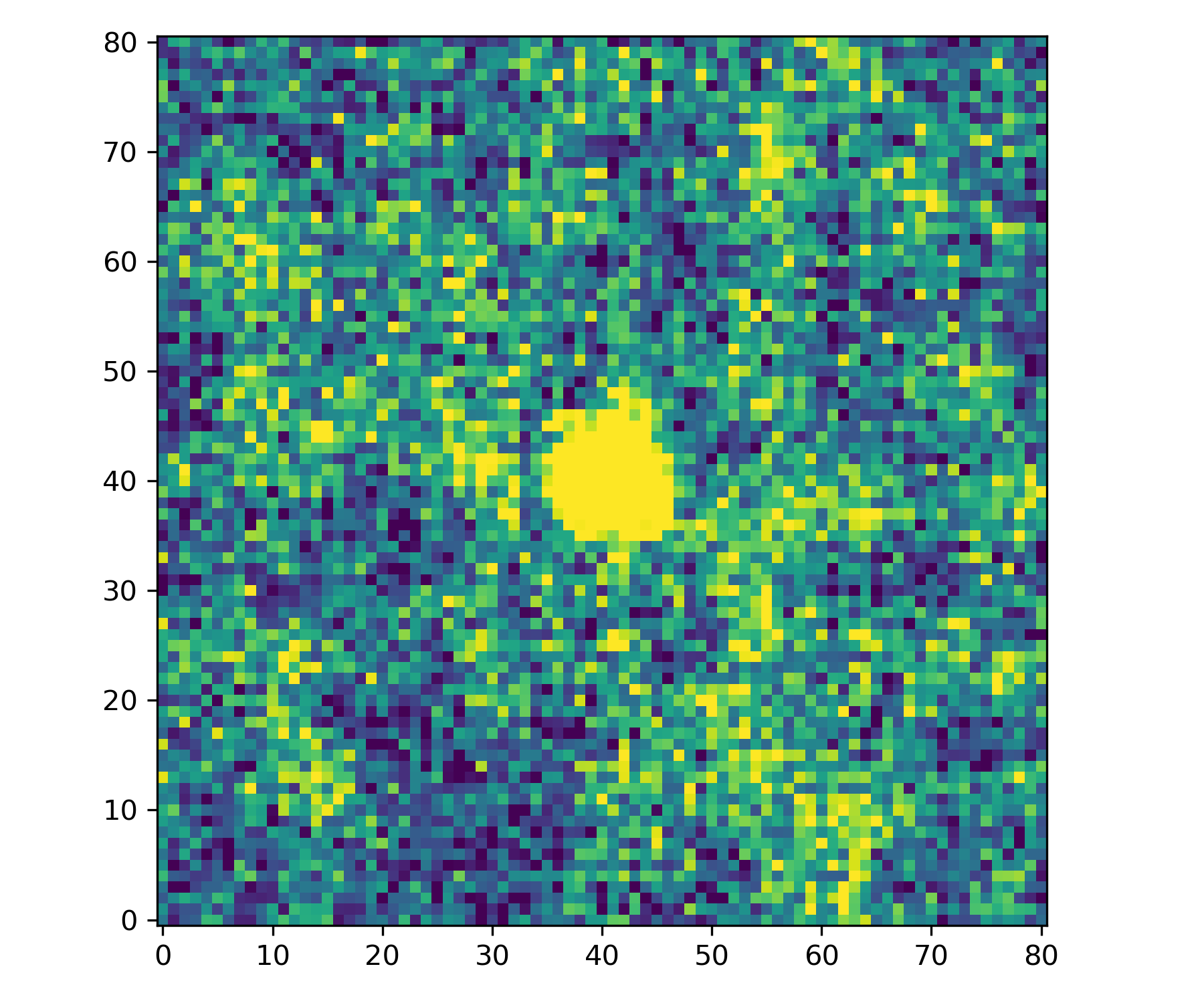}
  \includegraphics[width=0.31\linewidth]{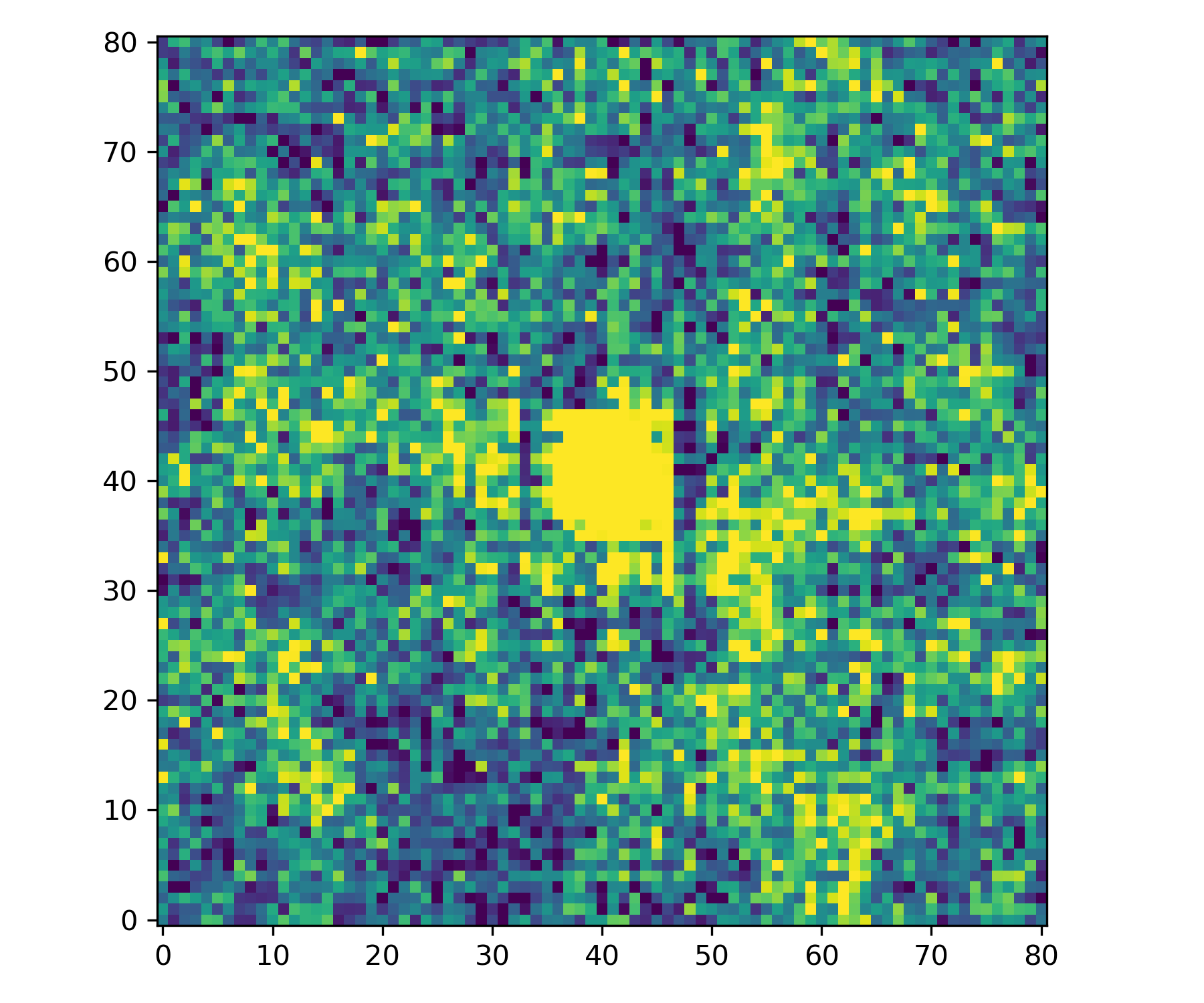}

  \vspace{3pt}
  \parbox[c]{0.31\linewidth}{\centering (d) Polynomial filter\\(with source mask)}%
  \parbox[c]{0.31\linewidth}{\centering (e) Standard maximum-likelihood}%
  \parbox[c]{0.31\linewidth}{\centering (f) Source sub-sampling}%

  \caption{Reconstructed polarization Q maps under six configurations. All maps are shown in units of $\mu K$, with color scales ranging from $-3$ to $3~\mu$K, and have a pixel size of $10'$. In panel (b), the map values are concentrated around a nearly uniform offset at the level of $\sim 1600~\mu$K, which results in a visually uniform appearance when displayed using the same color scale.
}
  \label{fig:map_series}
\end{figure}

Figure~\ref{fig:map_series} presents the reconstructed sky maps under six different configurations:
(a) point sources superimposed on CMB with white noise;
(b) point sources with CMB and $1/f$ atmospheric noise;
(c) 3-order polynomial filtering applied along each row or column;
(d) 3-order polynomial filtering with the source region masked~\cite{SPT:2023jql};
(e) standard maximum-likelihood map-making; and
(f) source sub-sampling map-making, where additional degrees of freedom are introduced within the source region. For detailed introduction, please see Ref.~\cite{Naess:2019nrw}.

After introducing the $1/f$ noise as in panel~(b), the entire map becomes severely contaminated, indicating the necessity of 1/f noise removal. 
Applying polynomial filtering in panel~(c) successfully suppresses a large fraction of the 1/f component; however, a faint cross-shaped artifact appears around the source, introduced by the filter~\cite{Rosenberg:2024geh}. 
This artifact can bias the flux estimation of nearby sources. 
When the source region is masked during filtering, as shown in panel~(d), the cross pattern is effectively eliminated. The standard maximum-likelihood map-making in panel~(e) also achieves a comparable level of atmospheric suppression. 
Although in principle similar cross-shaped residuals could exist, their expected amplitude is on the order of $10^{-4}$~\cite{Naess:2019nrw}, which is negligible in our polarization setup where the point-source signal is relatively weak in our sky patch. 
Finally, the source sub-sampling approach shown in panel~(f) also removes atmospheric contamination effectively while avoiding the cross-shaped filtering artifacts.

Table~\ref{tab:src_fit_stats} quantitatively summarizes the bias (mean - input value) and standard deviation of the recovered point-source amplitudes, obtained from 500 realizations of CMB and instrumental noise for each configuration. \footnote{Note that, for all configurations, the covariance matrix used in the GPSF fitting accounts only for the CMB and white-noise components.}
Cases~(b)--(e) are compared against the atmospheric-free reference case~(a). 
For cases~(b), (c), and (e), the biases already exceeds the standard deviation of case~(a), indicating a significant distortion in the recovered point-source amplitude.
In contrast, cases~(d) and~(f) not only reduce the bias to within the ideal statistical uncertainty but also eliminate the cross-shaped artifacts introduced by atmospheric filtering, albeit at the cost of a slightly increased variance.
In practice, the level of $1/f$ noise can be further reduced by employing a continuously rotating half-wave plate (CWHWP)~\cite{Takakura:2017ddx} or similar modulation techniques.
To test such an improved scenario, we repeat the simulation with a lower knee frequency of $f_{\mathrm{knee}} = 0.02$\,Hz.
In this case, the biases for methods~(d) and~(f) remain negligible, while their standard deviations decrease to 6.70 and 7.18, respectively, leading to a noticeably better accuracy in point-source fitting.
All in all, we have shown that the $1/f$ noise can be properly handled in realistic situations using methods (d) and (f), which preserve the effectiveness of our GPSF pipeline.

\begin{table}[htbp]
  \centering
  \caption{Bias and standard deviation of fitted point-source amplitudes (with GPSF method) for different map-making and filtering methods, based on 500 realizations of CMB and noise. The input source have flux density of 303.4mJy on Q map at 30\,GHz.}
  \label{tab:src_fit_stats}
  \setlength{\tabcolsep}{8pt}
  \renewcommand{\arraystretch}{1.2}
  \begin{tabular}{lcc}
    \hline
    \textbf{Case} & \textbf{Bias [mJy]} & \textbf{Standard Deviation [mJy]} \\
    \hline
    (a) Source + CMB + White noise & -0.01 & 6.24 \\
    (b) Source + CMB + 1/f noise & -319.04 & 12671.62\\
    (c) Polynomial filter (no mask) & -58.20 & 7.75 \\
    (d) Polynomial filter (with source mask) & 0.17 & 11.64 \\
    (e) Standard maximum-likelihood & -8.11 & 8.90 \\
    (f) Source sub-sampling & -0.23 & 11.55 \\
    \hline
  \end{tabular}
\end{table}

In summary, based on the discussion above, we conclude that the GPSF framework should remain effective in realistic situations, provided that effective beams, sidelobe corrections, instrumental leakage, and atmosphere-filtered data are properly accounted for.

\section{GPSF hyperparameter choices}\label{app:hyperparams}

In this appendix, we describe the motivation and validation of the hyperparameters used in the GPSF point-source fitting and subtraction procedure, namely the fitting radius $R_\mathrm{fit}$, the neighboring-source radius $R_\mathrm{neigh}$, and the polarized flux significance threshold $k_\mathrm{sig}$.

\begin{figure}[!htb]
    \centering
    \includegraphics[width=0.95\linewidth]{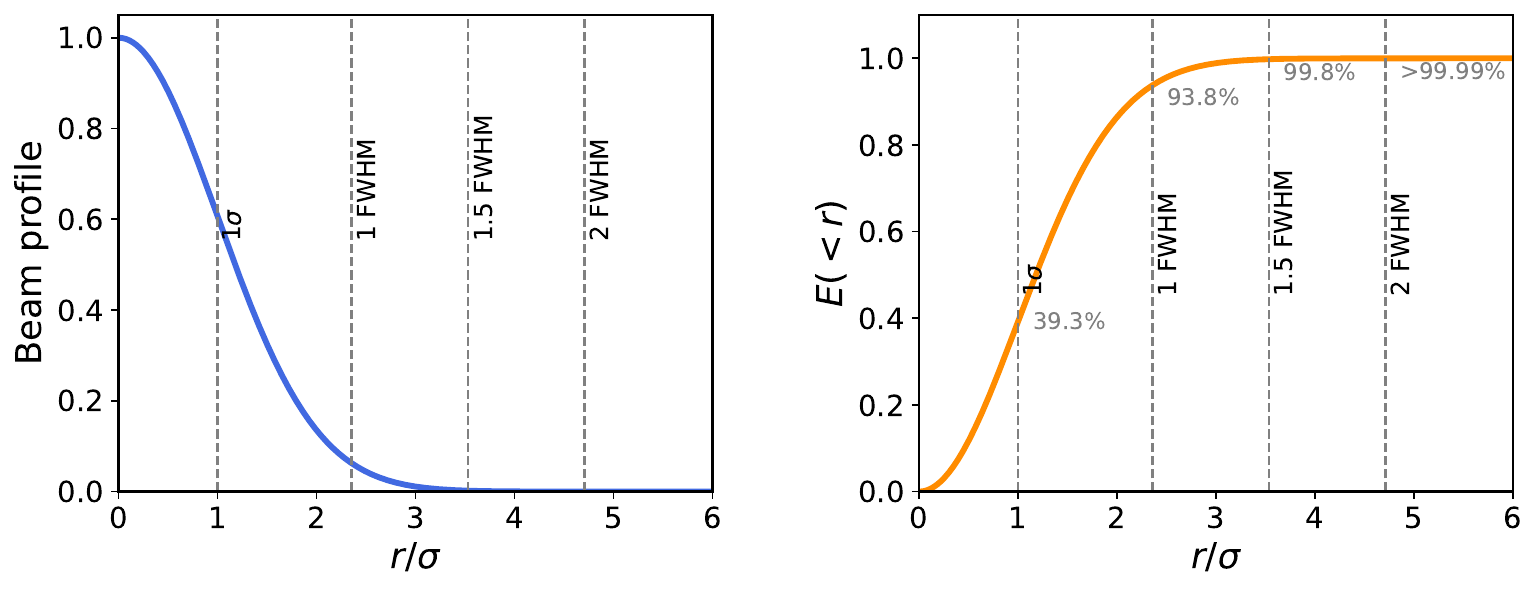}
    \caption{
    Radial profile and encircled energy of a two-dimensional Gaussian beam.
    \textbf{Left:} normalized beam profile as a function of radius in units of $\sigma$.
    \textbf{Right:} encircled energy fraction $E(<r)=1-\exp[-r^{2}/(2\sigma^{2})]$.
    The dashed vertical lines mark radii corresponding to $1\sigma$, 1, 1.5, and $2\,\mathrm{FWHM}$.
    At $1.5\,\mathrm{FWHM}$ ($\simeq3.5\sigma$), the beam encloses $\approx99.8\%$
    of the total energy, while $2\,\mathrm{FWHM}$ encloses $>99.99\%$.
    Extending the fitting window further therefore provides negligible additional signal gain
    but increases the inclusion of correlated noise pixels.
    }
    \label{fig:beam_energy}
\end{figure}

First, the fitting radius $R_{\mathrm{fit}}$ is chosen to be large enough to enclose essentially all of the beam power while avoiding the rapid growth of background noise at larger radii. As shown in figure~\ref{fig:beam_energy}, a radius of $1.5\,\mathrm{FWHM}$ already captures $99.8\%$ of the beam energy; enlarging the fitting window beyond this value yields only negligible additional signal.

\begin{figure}[!htb]
    \centering
    \begin{subfigure}{0.48\textwidth}
        \includegraphics[width=\linewidth]{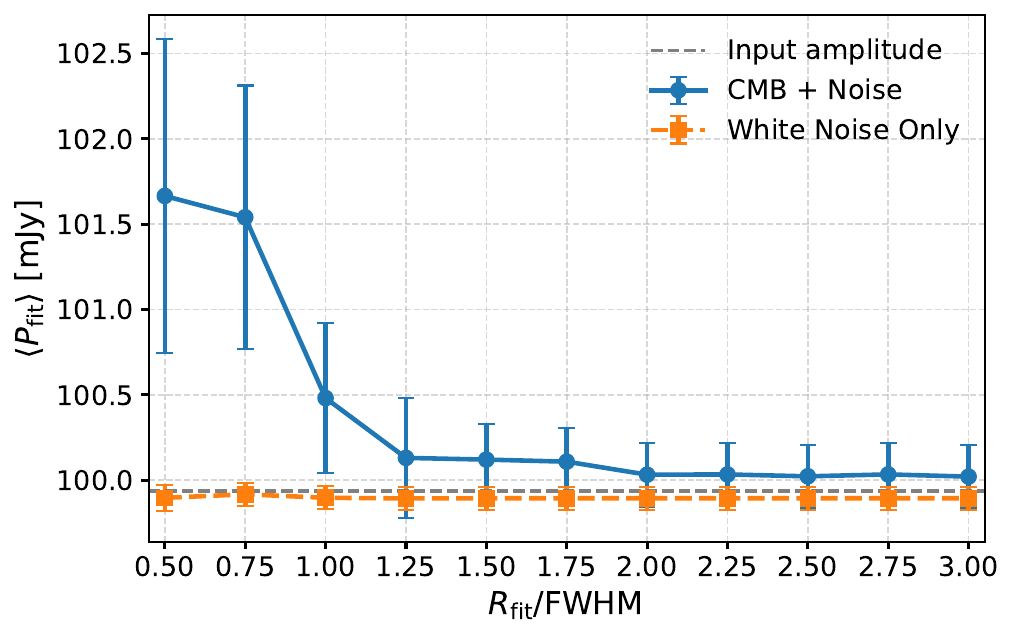}
    \end{subfigure}
    \hfill
    \begin{subfigure}{0.48\textwidth}
        \includegraphics[width=\linewidth]{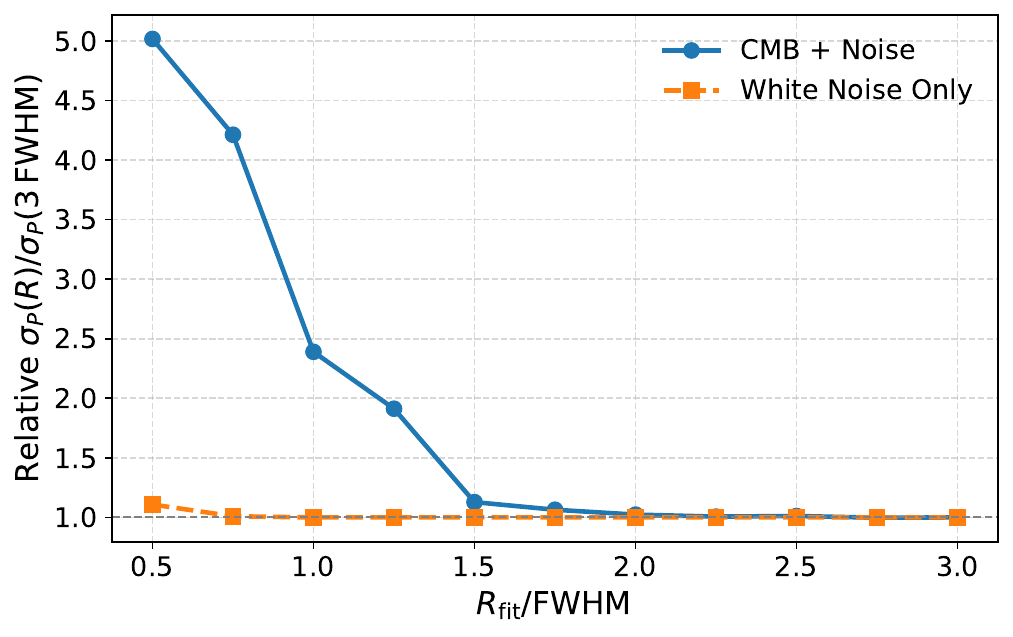}
    \end{subfigure}
    \caption{
    Sensitivity of the recovered point-source amplitude to the fitting radius.
    \textbf{Left:} mean recovered amplitude $\langle P_{\mathrm{fit}}\rangle$ with $1\sigma$ uncertainties for the {\it CMB+noise} (blue) and {\it white-noise-only} (gray) cases.
    \textbf{Right:} relative statistical uncertainty $\sigma_P(R)/\sigma_P(3\,\mathrm{FWHM})$, normalized to its value at $R_{\mathrm{fit}}=3\,\mathrm{FWHM}$.
    The amplitude converges by $R_{\mathrm{fit}}\simeq1.5\,\mathrm{FWHM}$, and while the total uncertainty continues to decrease by $\sim10\%$ up to $2\,\mathrm{FWHM}$ in the presence of CMB covariance, the improvement is minor, validating our adopted choice of $1.5\,\mathrm{FWHM}$.}
    \label{fig:rfit_compare}
\end{figure}

To validate this choice, we varied $R_{\mathrm{fit}}$ from $0.5$ to $3\,\mathrm{FWHM}$ for both {\it CMB+noise} and {\it white-noise-only} simulations (figure~\ref{fig:rfit_compare}) at 155GHz. For the {\it CMB+noise} case, the recovered amplitudes converge to the input value for $R_{\mathrm{fit}}\gtrsim 1.25\,\mathrm{FWHM}$, and the statistical uncertainty decreases rapidly up to $R_{\mathrm{fit}}\simeq 1.5\,\mathrm{FWHM}$. Beyond this point the improvement becomes marginal: increasing $R_{\mathrm{fit}}$ from $1.5$ to $2.0\,\mathrm{FWHM}$ reduces the uncertainty by only $\sim10\%$, while the number of included pixels grows as $O(p^3)$, substantially increasing the computational cost.

In the {\it white-noise-only} case, the uncertainty decreases quickly at small radii—reflecting the inclusion of additional independent noise samples—and then reaches a stable plateau by $\sim1\,\mathrm{FWHM}$. This plateau persists at all larger radii, confirming that uncorrelated noise does not benefit much from enlarging the aperture once the beam is fully sampled. The mild residual dependence seen in the {\it CMB+noise} case therefore originates from correlated sky variance rather than from the fitting method itself.

Given that $1.5\,\mathrm{FWHM}$ encloses nearly all beam power, reaches the convergence regime of the recovered amplitude, and provides a favorable balance between statistical precision and computational efficiency, we adopt it as a practical and near-optimal choice for the fitting radius. 

Second, we adopt $R_{\mathrm{neigh}} = 3\,\mathrm{FWHM}$. This choice is motivated by the geometry of the fitting aperture: with a fitting radius of $R_{\mathrm{fit}} = 1.5\,\mathrm{FWHM}$, two point sources separated by $3\,\mathrm{FWHM}$ have fitting windows that touch exactly at their $1.5\,\mathrm{FWHM}$ boundaries. In this configuration, the profile of a neighboring source does not appreciably enter the central fitting region, and its influence on the recovered amplitude becomes negligible. Although the exact level of contamination can depend on the flux ratio between sources, tests under our simulation settings confirm that extending the neighbor search to $3\,\mathrm{FWHM}$ is sufficient to include all sources that could affect the joint-fit model. Increasing $R_{\mathrm{neigh}}$ beyond this value does not produce any noticeable change in the fitted parameters.

\begin{figure}[!htb]
    \centering
    \begin{subfigure}{0.47\textwidth}
        \includegraphics[width=\linewidth]{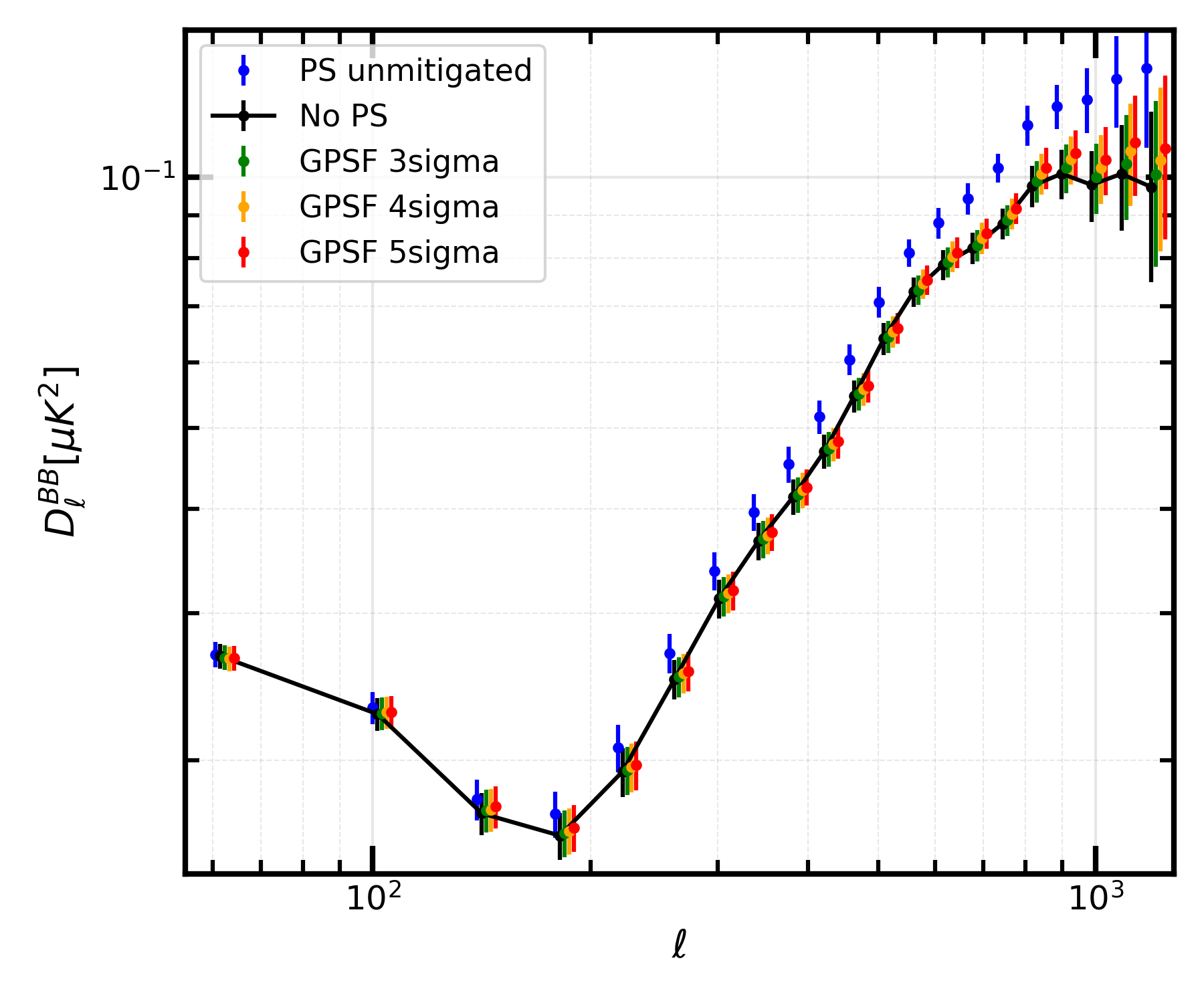}
        \caption{Noise-debiased $B$-mode power spectrum.}
    \end{subfigure}
    \hfill
    \begin{subfigure}{0.47\textwidth}
        \includegraphics[width=\linewidth]{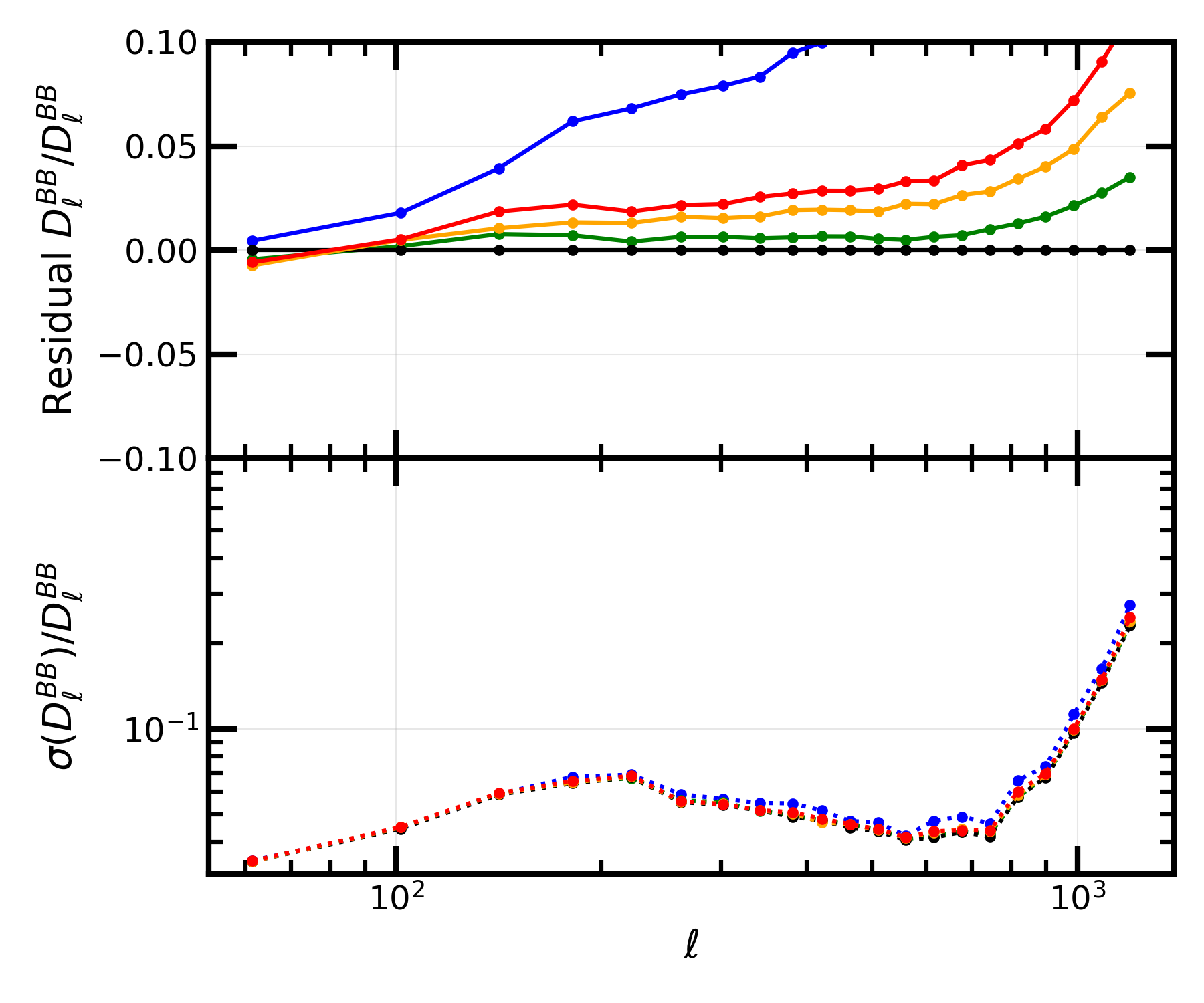}
        \caption{Relative residual and standard deviation.}
    \end{subfigure}
    \caption{
    Sensitivity of polarized flux significance threshold $k_\sigma$.
    (\textbf{a})~Noise-debiased $B$-mode power spectra at 155\,GHz with standard deviations estimated from 200 realizations (CMB+noise).
    (\textbf{b})~Relative residuals and relative standard deviations of each $k_\sigma$ configuration.
    The case {\it No PS} serves as the reference. The relative residual is defined as the difference between the mean power spectrum of each method and that of the reference, normalized by the reference mean.
    The relative standard deviation is computed as the standard deviation of each method divided by its own mean.}
    \label{fig:ksigma_compare}
\end{figure}

Finally, the polarized flux significance threshold $k_{\sigma}=3$ was selected as a conservative balance between completeness and reliability.
We verified that values between $k_{\sigma}=3$ and $5$ change the recovered amplitudes by a few percent in relative residual (see figure~\ref{fig:ksigma_compare}).

\section{Map preprocessing}\label{appendix:map_preprocess}

Before NILC, it is important to highlight two preprocessing steps on the frequency maps that require special attention: EB leakage correction and smoothing.

\subsection{EB leakage correction}\label{sec:lkg_correct}
EB leakage occurs when performing a spherical harmonic transformation on a partial sky, as the orthogonality of spherical harmonics is no longer satisfied. For ground-based CMB observations, such leakage is inevitable due to limited sky coverage, introducing bias in the estimation of $r$. Therefore, correcting for EB leakage is essential.

Recent studies have demonstrated that the \textit{Recycling} method provides an effective, blind estimation and correction of EB leakage~\cite{Liu:2018kut}. This method, designed for partial-sky observations, directly reconstructs a pure $B$-mode map.

The recycling method follows these steps:

\begin{enumerate}
    \item From the original simulated partial-sky $\mathrm{P} = (Q, U)$ maps, compute the spherical harmonic coefficients $a_{\ell m}^E$ and $a_{\ell m}^B$. Convert $(a_{\ell m}^E, 0)$ into $\mathrm{P}_E^{\prime} = (Q_E, U_E)^{\prime}$ and $(0, a_{\ell m}^B)$ into $\mathrm{P}_B^{\prime} = (Q_B, U_B)^{\prime}$, where the latter contains the contaminated $B$-mode contribution.

    \item Construct the EB leakage template by applying a mask to $\mathrm{P}_E^{\prime} = (Q_E, U_E)^{\prime}$, transforming it into $a_{\ell m}^{B\prime}$, and then inverting $(0, a_{\ell m}^{B\prime})$ into $\mathrm{P}_B^{\prime\prime} = (Q_B, U_B)^{\prime\prime}$.

    \item Perform a linear fit between $\mathrm{P}_B^{\prime\prime} = (Q_B, U_B)^{\prime\prime}$ and $\mathrm{P}_B^{\prime} = (Q_B, U_B)^{\prime}$, and denote the fitting coefficient as $a$.

    \item Compute the cleaned map:
    \begin{equation}
        \mathrm{P}^{\mathrm{clean}} = \mathrm{P}_B^{\prime} - a \mathrm{P}_B^{\prime\prime}.
    \end{equation}
\end{enumerate}

Once the cleaned polarization maps are obtained, the pure $B$-mode map can be reconstructed using a spherical harmonic transformation. This cleaned $B$-mode map represents the best blind estimate~\cite{2019JCAP...04..046L}, effectively minimizing errors caused by EB leakage. Any further refinement would require additional knowledge of the unobserved sky regions.

\subsection{Smoothing}\label{sec:smooth}
Before NILC, the input frequency maps are smoothed to a common angular resolution—17 arcmin in our data simulation—to ensure that all maps share the same effective beam size. This step guarantees consistency of the CMB signal across scales, as required by ILC. The smoothing is performed in harmonic space by first deconvolving the maps’ original beams and then convolving them to a uniform 17 arcmin FWHM, which can be expressed as:
\begin{align}
a_{\ell m}^{out} = \frac{b_{\ell}^{out}}{b_{\ell}^{in}} a_{\ell m}^{in}.
\end{align}
To mitigate the deconvolution artifacts near the mask edges, we apply a C1 apodization with a 3-degree taper prior to deconvolution.

\section{Needlets internal linear combination}\label{appendix:nilc}
We apply the NILC algorithm, a blind component separation method, to reconstruct the original CMB signal. NILC is a localized, multi-scale variant \textit{Internal Linear Combination} (ILC) method. In this work, NILC is performed on the \textit{B-mode} polarization maps for all five cases. The NILC code developed for this work, \texttt{openilc}\footnote{\url{https://github.com/dreamthreebs/openilc}}, is publicly accessible on the internet.

The ILC is a blind component separation technique commonly used in CMB data analysis. It extracts the CMB signal from multi-frequency observations by forming a linear combination of the input maps with weights $\mathbf{w}$ that minimize the variance of the resulting map. Specifically, the cleaned CMB estimate at sky direction $\hat{n}$ is given by
\begin{align}
\hat{s}(\hat{n}) = \mathbf{w}^T \mathbf{x}(\hat{n}),
\end{align}
where $\mathbf{x}(\hat{n})$ is the vector of observed maps at different frequencies. The variance of the cleaned map is
\begin{align}
\sigma_{\hat{s}}^2 = \mathbf{w}^T \mathbf{C} \mathbf{w},
\end{align}
which is minimized under the constraint that the combination preserves the CMB signal:
\begin{align}
\mathbf{w}^T \mathbf{a} = 1,
\end{align}
where $\mathbf{C}$ is the frequency-frequency covariance matrix of the observations, and $\mathbf{a}$ is the frequency response vector of the CMB, typically taken as a vector of ones in thermodynamic units. The solution to this constrained optimization yields the ILC weights:
\begin{align}
\mathbf{w} = \frac{\mathbf{C}^{-1} \mathbf{a}}{\mathbf{a}^T \mathbf{C}^{-1} \mathbf{a}}.
\end{align}

The NILC method follows the same principle as ILC but performs the variance minimization in the needlet domain, which provides localization both in scale and on the sky. Each frequency map $x_\nu(\hat{n})$ is first decomposed into needlet coefficients:
\begin{align}
\beta_{j,k}^{(\nu)} = \int_{\mathcal{S}^2} x_\nu(\hat{n}) \psi_{j,k}(\hat{n}) \, d\Omega,
\end{align}
where $\psi_{j,k}(\hat{n})$ are the needlet basis functions indexed by scale $j$ and position $k$. The collection of needlet coefficients across all frequencies forms the vector
\begin{align}
\boldsymbol{\beta}_{j,k} = [\beta_{j,k}^{(1)}, \beta_{j,k}^{(2)}, \dots, \beta_{j,k}^{(N)}]^T.
\end{align}

At each scale $j$ and location $k$, a local covariance matrix is estimated as
\begin{align}
\mathbf{C}_{j,k} = \langle \boldsymbol{\beta}_{j,k} \boldsymbol{\beta}_{j,k}^T \rangle,
\end{align}
where the average is taken over a local neighborhood of needlet coefficients. This localization enables the NILC weights to adapt to spatially varying foregrounds and noise.

The optimal NILC weights are computed in the same form as in standard ILC, but using the local covariance matrix:
\begin{align}
\mathbf{w}_{j,k} = \frac{\mathbf{C}_{j,k}^{-1} \mathbf{a}}{\mathbf{a}^T \mathbf{C}_{j,k}^{-1} \mathbf{a}}.
\end{align}
These weights are applied to the needlet coefficients to produce the cleaned CMB estimate at each scale and position:
\begin{align}
\beta_{j,k}^{\mathrm{CMB}} = \mathbf{w}_{j,k}^T \boldsymbol{\beta}_{j,k}.
\end{align}
The final cleaned CMB map is reconstructed by summing over all needlet scales and positions:
\begin{align}
\hat{s}(\hat{n}) = \sum_{j,k} \beta_{j,k}^{\mathrm{CMB}} \psi_{j,k}(\hat{n}).
\end{align}

\section{Power spectra at 95, 215, 270GHz bands}\label{app:95215270}

\begin{figure}[bpth]
    \centering
    \subfloat{\includegraphics[width=0.47\textwidth]{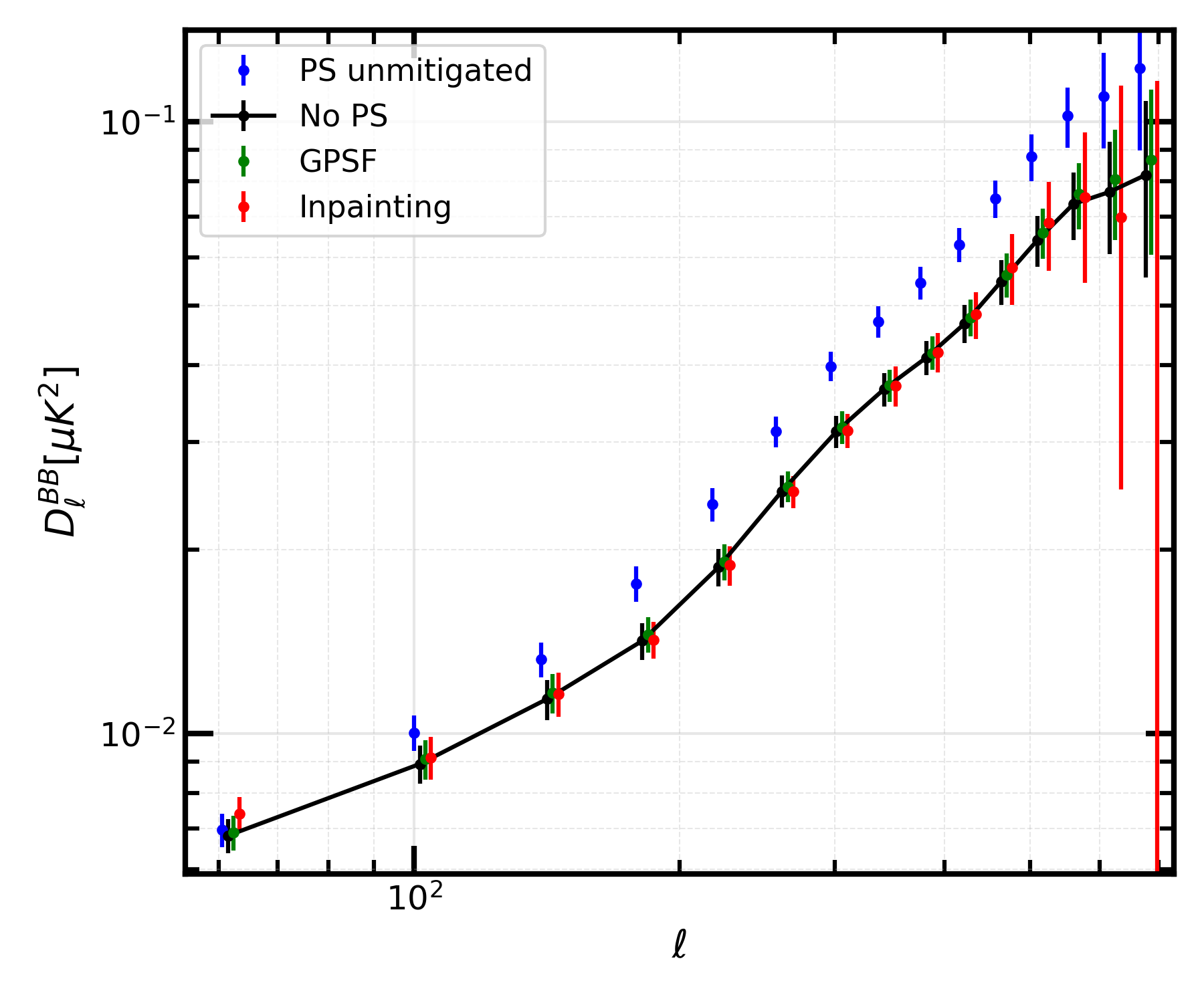}}\hfill
    \subfloat{\includegraphics[width=0.47\textwidth]{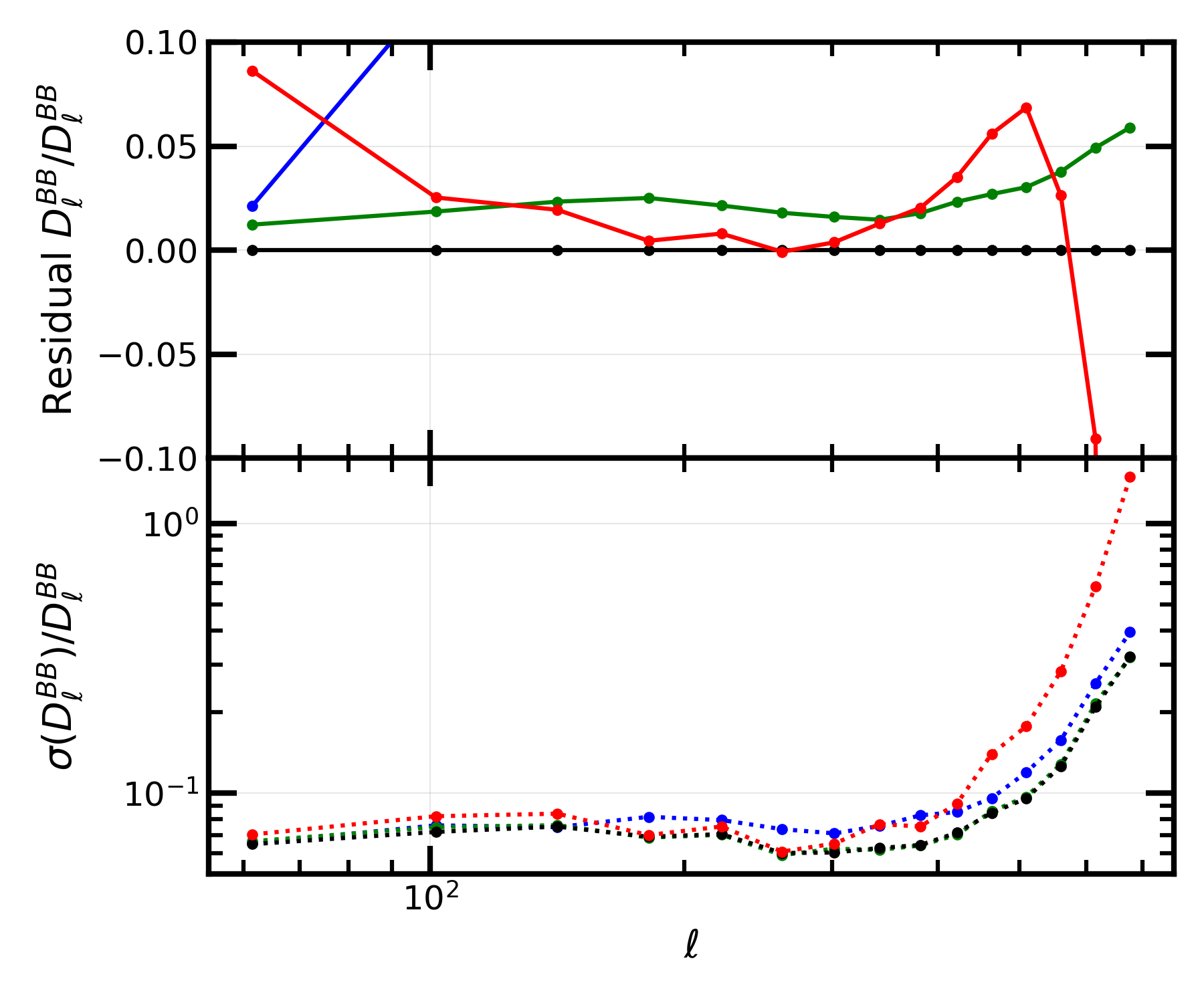}}\\[0.5ex]
    \subfloat{\includegraphics[width=0.47\textwidth]{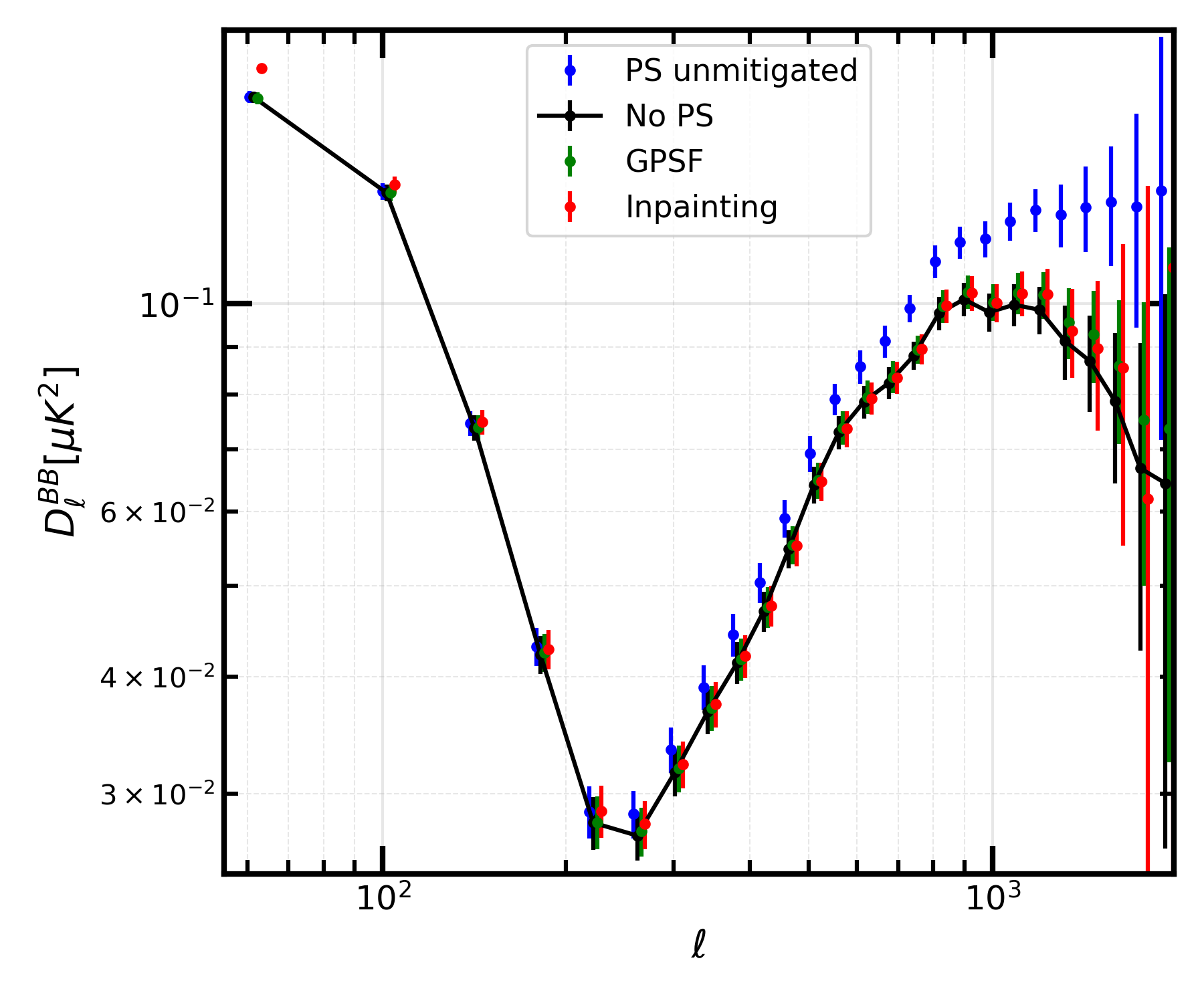}}\hfill
    \subfloat{\includegraphics[width=0.47\textwidth]{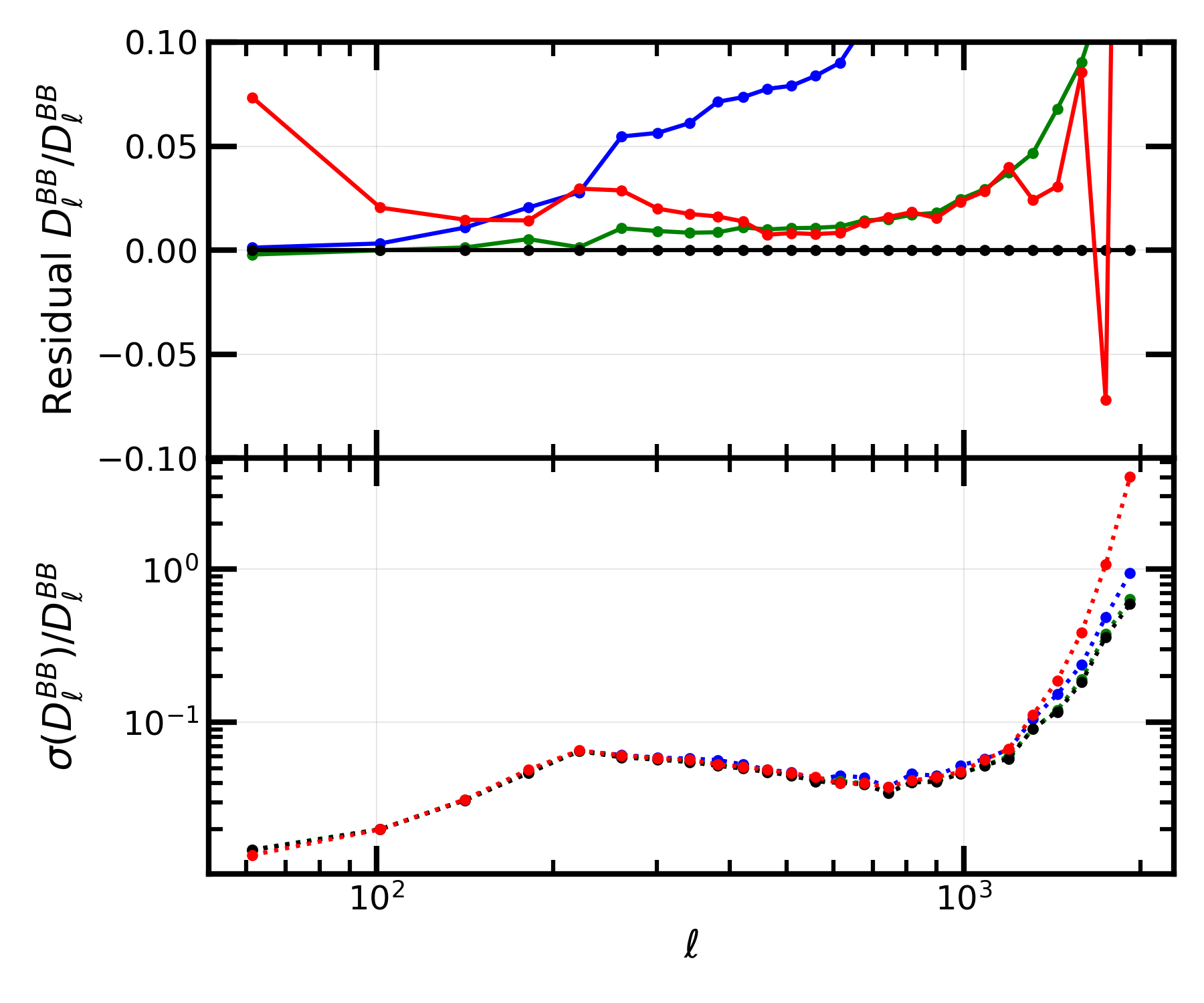}}\\[0.5ex]
    \subfloat{\includegraphics[width=0.47\textwidth]{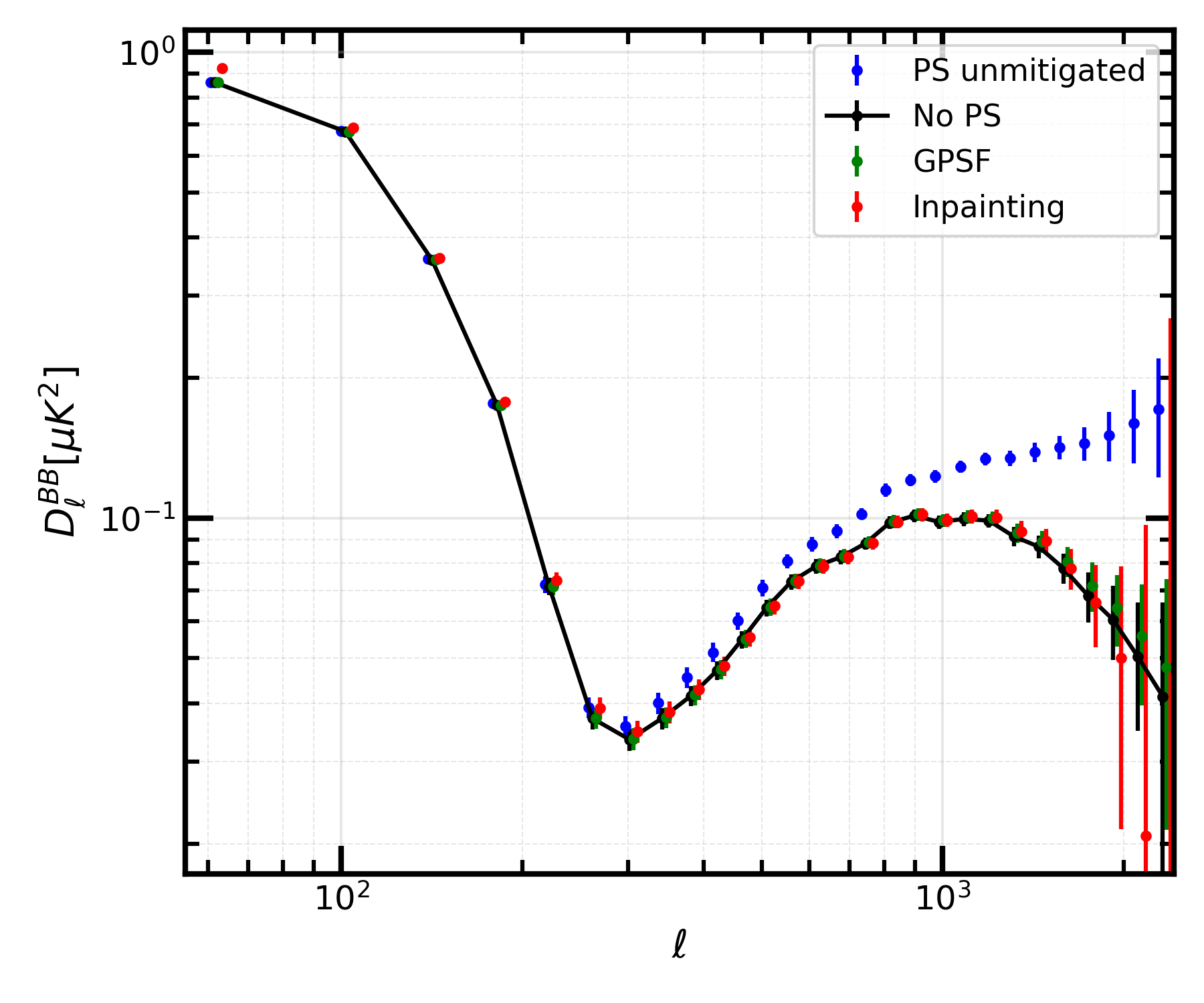}}\hfill
    \subfloat{\includegraphics[width=0.47\textwidth]{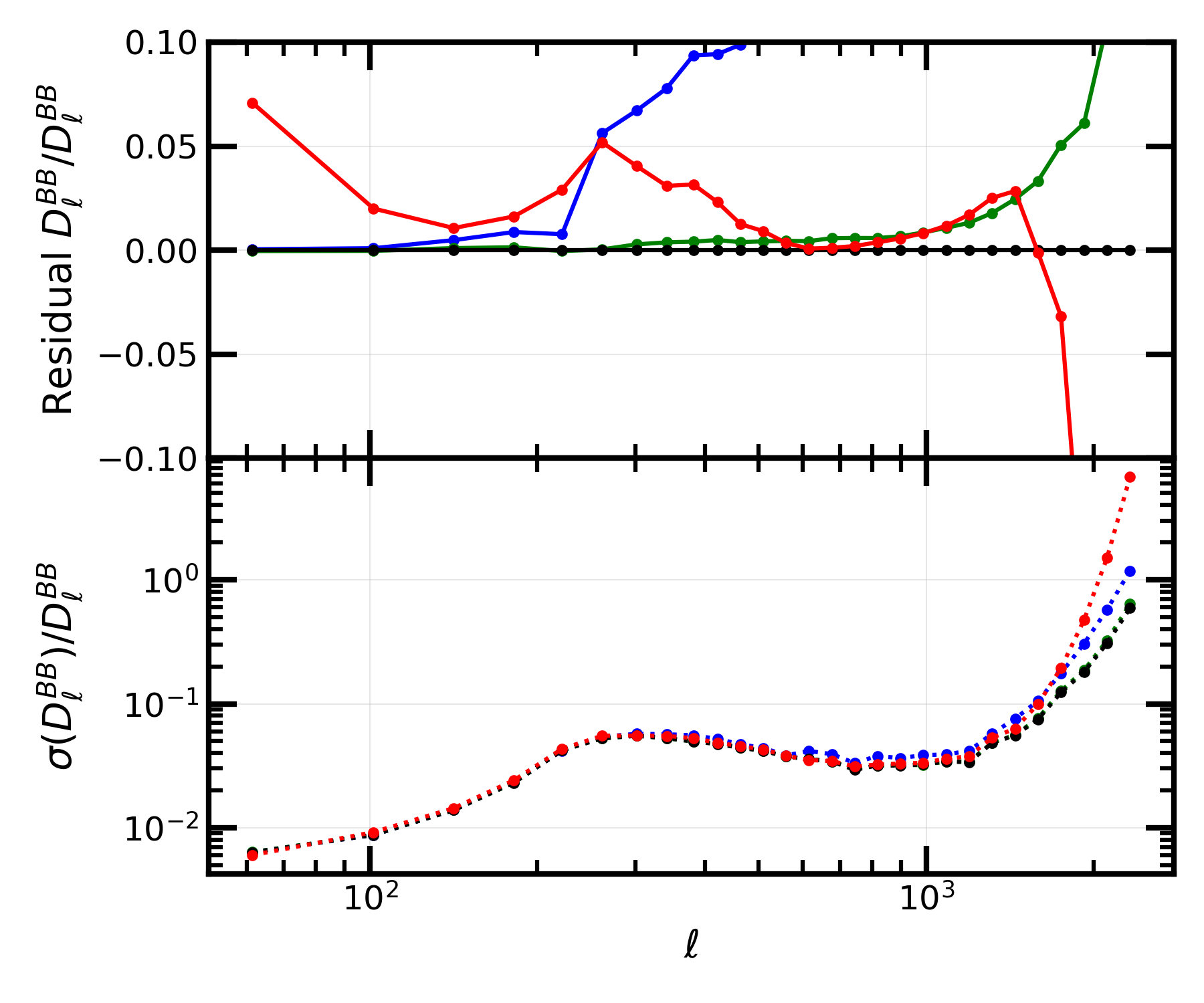}}
    
    \caption{Noise-debiased B-mode power spectra and corresponding residual analysis for different point-source mitigation methods. From top to bottom, the results correspond to 95, 215, and 270\,GHz. In each row, the left panel presents the noise-debiased B-mode power spectrum with standard deviations estimated from 200 realizations (accounting for CMB and noise), while the right panel displays the relative residuals and relative standard deviations. The \textit{No PS} case serves as the reference.}
    \label{fig:each_freq_cl_other}
\end{figure}

For the completeness of section \ref{sec:each_freq}, we supplement the power spectra from observations at 95 GHz, 215 GHz, and 270 GHz for each case in figure~\ref{fig:each_freq_cl_other}.
In these bands, diffuse foregrounds dominate the spectra at low multipoles ($\ell$), while point-source contamination becomes increasingly prominent at high $\ell$, as indicated by the differences between \textit{No PS} and \textit{PS unmitigated} case. 
In terms of comparing the relative residuals and relative standard deviations,
\textit{Inpainting} consistently reduces the point-source contribution on intermediate multipole ranges. and
exhibits larger residuals at both low and high $\ell$, along with a increased standard deviation at high $l$. The \textit{GPSF} method provides relatively stable performance across the full multipole range and across the 95 to 270\,GHz frequency bands.

\section{Bias analysis of GPSF}\label{app:bias}       
\begin{figure}[htbp]
    \centering
    \includegraphics[width=0.6\textwidth]{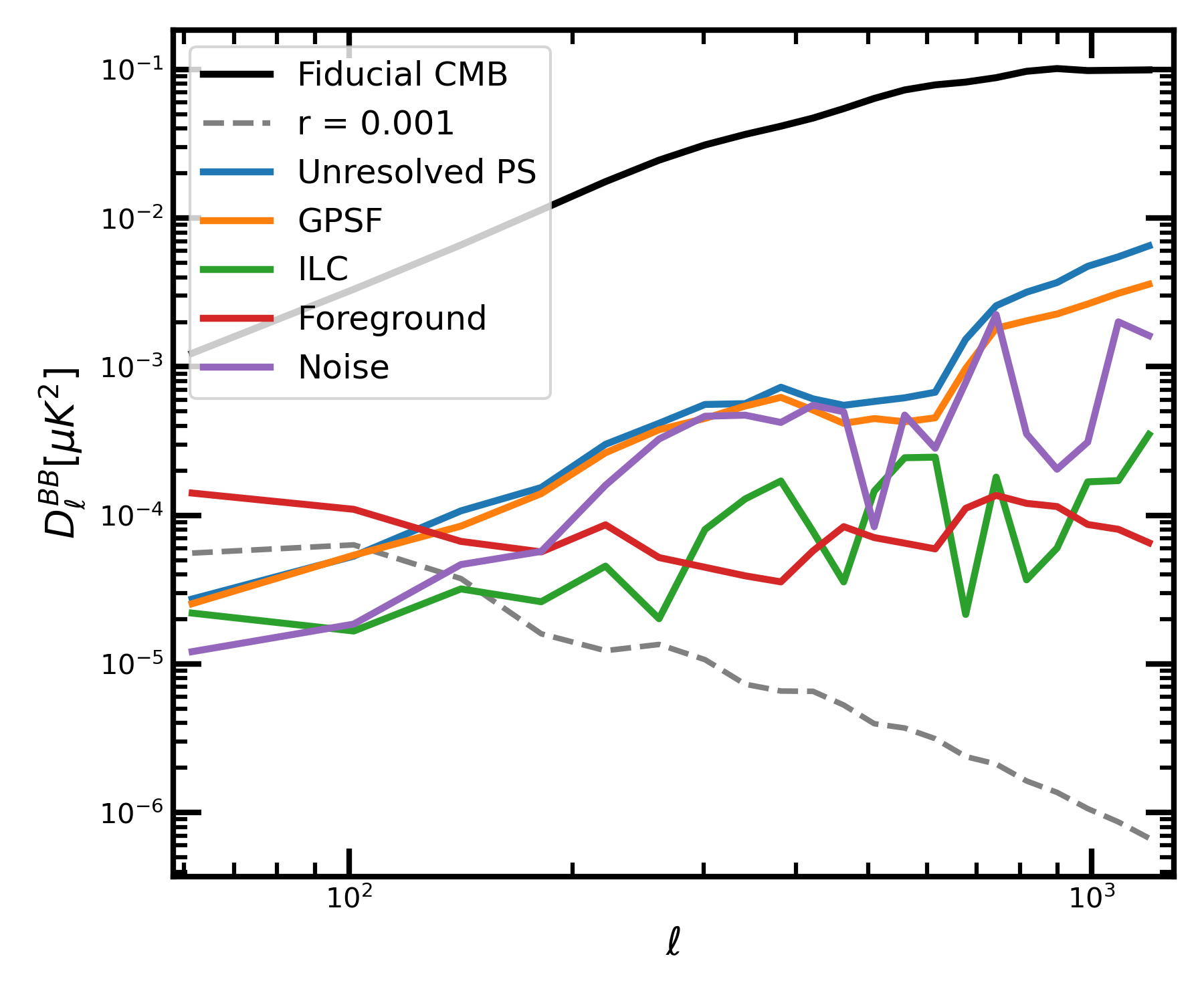}
    \caption{BB angular power spectra of residuals of \textit{GPSF} pipeline, compared with the fiducial CMB signal. The ILC bias is plotted in absolute amplitude.}
    \label{fig:bias}
\end{figure}
\begin{figure}[htbp]
    \centering
    \includegraphics[width=0.6\textwidth]{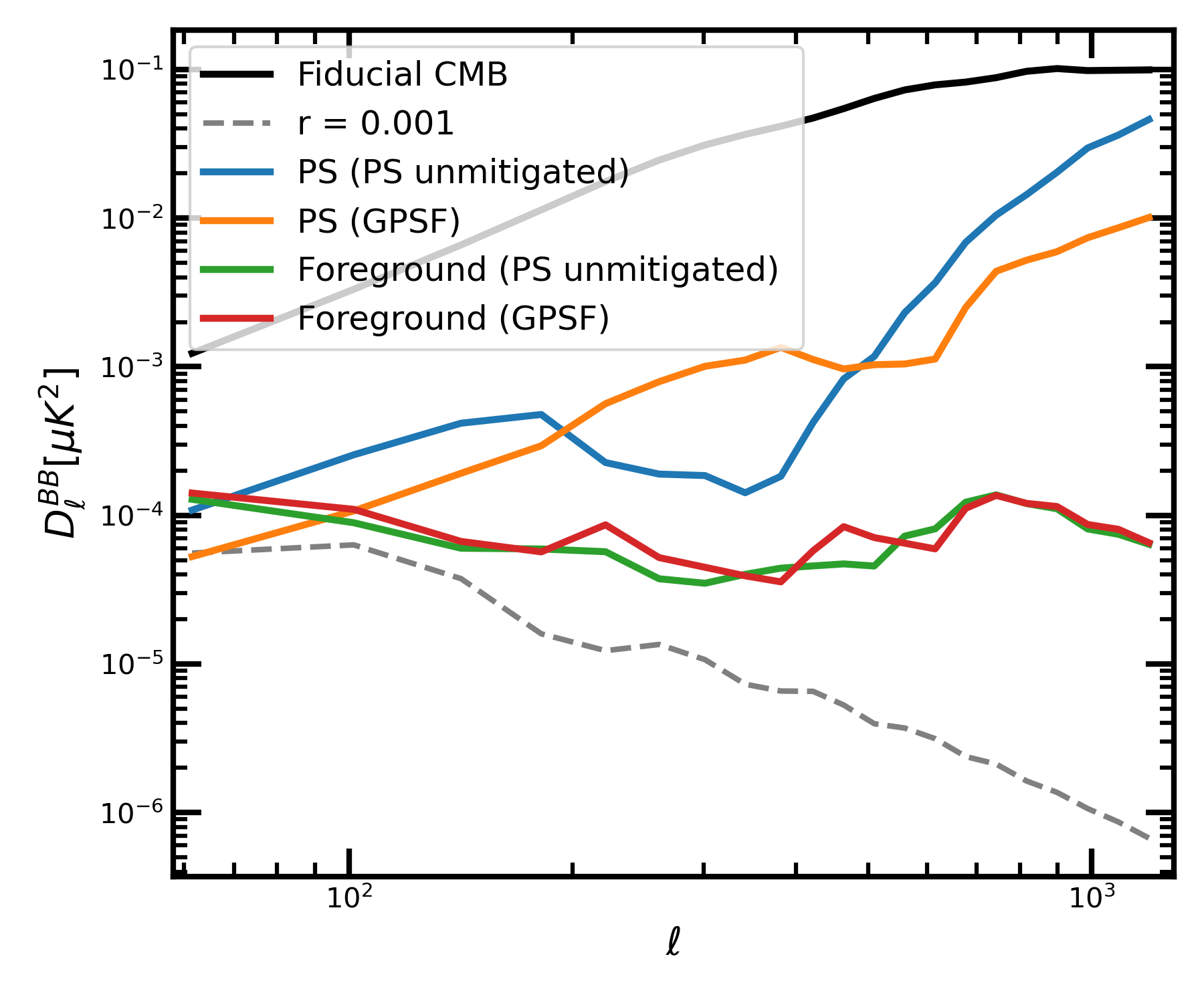}
    \caption{Diffuse foreground and point-source residuals from the \textit{GPSF} and \textit{PS unmitigated} pipelines, compared with the fiducial CMB signal.}
    \label{fig:cpr_bias}
\end{figure}

To gain a deeper understand the origin of residual contamination after NILC and its impact on cosmological parameter inference for GPSF method, 
we analyze the contributions from individual components in the cleaned CMB maps, which include diffuse foregrounds, instrumental noise, unresolved point sources, and the residuals arising from point-source removal using GPSF. In figure~\ref{fig:bias}, we presents a comparison of the BB angular power spectra of these residuals, highlighting the residual contributions from different components at various scales.

The residuals from diffuse foregrounds and instrumental noise originate from incomplete suppression during the NILC component separation process. Similarly, the unresolved point source residuals arise from faint sources that remain in the map because their emission is too weak or spatially complex to be effectively removed by NILC. In contrast, the GPSF removal residuals stem from the point-source fitting process itself: the fitted amplitudes may partially pick up contributions from nearby foregrounds, CMB or noise, leading to small residuals that depend on the local background structure.


As shown in figure~\ref{fig:bias}, the power-spectrum residuals from unresolved point sources---as well as the residuals left after GPSF removal of resolved sources---is small but non-negligible across all multipoles. Their impact is subdominant compared with the diffuse foreground residuals, which dominate at low multipoles ($\ell < 100$) and constitute the main source of bias in the recovered tensor-to-scalar ratio~$r$.
We also quantify the NILC bias using CMB+noise–only Monte Carlo simulations. As illustrated in figure~\ref{fig:bias}, the ILC bias in the GPSF pipeline is at the level of approximately $0.3\%$ of the fiducial CMB $B$-mode power spectrum. This bias is also subdominant relative to diffuse foreground residuals and therefore has only a minimal impact on the recovered value of $r$.

We further compare the diffuse foreground and point-source residuals\footnote{For \textit{GPSF}, the point-source residual includes both unresolved point sources and GPSF removal residuals. For \textit{PS unmitigated}, the point-source residual includes contributions from resolved and unresolved point sources.} between the \textit{GPSF} pipeline and the \textit{PS unmitigated} pipeline, as shown in figure~\ref{fig:cpr_bias}. The results indicate that GPSF reduces point-source residuals by approximately 50\% at low ($\ell < 240$) and high multipoles ($\ell > 500$), while slightly increasing the residuals in the intermediate range ($240 < \ell < 500$). In contrast, the diffuse foreground residuals remain comparable between the two pipelines, indicating that GPSF has little impact on diffuse foreground suppression. Taken together, these findings suggest that incorporating GPSF into the NILC pipeline can effectively mitigate point-source contamination on large and small angular scales, without compromising the suppression of diffuse foregrounds, and with only a modest trade-off in the intermediate multipole regime.

\end{document}